\documentclass[12pt]{article}
\usepackage{epsfig,amsfonts,amsbsy,ifthen}
\makeatletter
\@addtoreset{equation}{section}
\makeatother
\renewcommand{\theequation}%
{\thesection.\arabic{equation}}
\newcounter{topic}[subsubsection]

\newcommand{\topic}[1]%
{\par\bigskip\par%
\noindent\refstepcounter{topic}(\alph{topic})\ {\em #1:}\/\ }
\newcommand{\ret}{\nonumber\\}

\newcommand{\Rem}{\bigskip\noindent{\em Remark\/}:\ }

\newcommand{\sqbk}[1]{\left[#1\right]}
\newcommand{\cbk}[1]{\left\{#1\right\}}
\newcommand{\bkt}[1]{\left\langle#1\right\rangle}
\newcommand{\sbkt}[1]{\langle#1\rangle}
\newcommand{\set}[2]{\left\{#1\,\Bigl|\,#2\right\}}
\newcommand{\partialD}[1]%
{\frac{\partial}{\partial #1}}
\newcommand{\partialDD}[1]%
{\frac{\partial^2}{\partial #1^2}}
\newcommand{\partialf}[2]%
{\frac{\partial #1}{\partial #2}}
\makeatletter
\long\def\@makecaption#1#2{{\small
\advance\leftskip1cm
\advance\rightskip1cm
\vskip\abovecaptionskip
\sbox\@tempboxa{#1: #2}%
\ifdim \wd\@tempboxa >\hsize
 #1: #2\par
\else
\global \@minipagefalse
\hb@xt@\hsize{\hfil\box\@tempboxa\hfil}%
\fi
\vskip\belowcaptionskip}}
\makeatother
\newcommand{\eqref}[1]{(\ref{#1})}
\newcommand{\sumtwo}[2]%
{\mathop{\sum_{#1}}_{#2}}
\newcommand{\sumthree}[3]%
{\mathop{\mathop{\sum_{#1}}_{#2}}_{#3}}
\newcommand{\sumfour}[4]%
{\mathop{\mathop{\mathop{\sum_{#1}}_{#2}}_{#3}}_{#4}} 
\newcommand{\mintwo}[2]%
{\mathop{\min_{#1}}_{#2}}
\newcommand{\minthree}[3]%
{\mathop{\mathop{\min_{#1}}_{#2}}_{#3}}
\newcommand{\minfour}[4]%
{\mathop{\mathop{\mathop{\min_{#1}}_{#2}}_{#3}}_{#4}} 
\newcommand{\calS}{{\cal S}}
\newcommand{\calU}{{\cal U}}
\newcommand{\la}{\lambda}
\newcommand{\La}{\Lambda}
\newcommand{\rb}{\boldsymbol{r}}
\newcommand{\xb}{\boldsymbol{x}}
\newcommand{\yb}{\boldsymbol{y}}
\newcommand{\ub}{\boldsymbol{u}}
\newcommand{\vb}{\boldsymbol{v}}
\newcommand{\ob}{\boldsymbol{o}}
\newcommand{\eb}{\boldsymbol{e}}
\newcommand{\etab}{\boldsymbol{\eta}}
\newcommand{\zetab}{\boldsymbol{\zeta}}
\newcommand{\etax}{\eta_{\xb}}
\newcommand{\etay}{\eta_{\yb}}
\newcommand{\etabx}{\etab^{\xb}}
\newcommand{\zetax}{\zeta_{\xb}}

\newcommand{\zetabx}{\zetab^{\xb}}
\newcommand{\etabxy}{\etab^{\xb,\yb}}
\newcommand{\xiL}{\xb\in\La}
\newcommand{\HL}{H_\Lambda}

\newcommand{\dHxy}{\HL(\etab)-\HL(\etabxy)}
\newcommand{\cTEL}{c^{(T,E)}_\Lambda}
\newcommand{\deltaT}{\Delta T}

\begin{document}

\noindent
{\Large \bf 
Steady state thermodynamics
}
\setcounter{footnote}{1}
\footnotetext{
December 20, 2005.
To appear in J. Stat. Phys.
Archived as {\tt cond-mat/0411052}.
}

\bigskip
\noindent
Shin-ichi Sasa\footnote
{Department of Pure and Applied Sciences, 
University of Tokyo, 
Komaba, Tokyo 153-8902, Japan
(electronic address: 
\tt
sasa\makeatletter @\makeatother jiro.c.u-tokyo.ac.jp)
}
and
Hal Tasaki\footnote{
Department of Physics,
Gakushuin University,
Mejiro, Toshima-ku, Tokyo 171-8588,
Japan
(electronic address: 
\tt
hal.tasaki\makeatletter @\makeatother gakushuin.ac.jp)
}

\bigskip
\begin{quotation}
The present paper reports our attempt to search for a new universal framework in nonequilibrium physics.
We propose a thermodynamic formalism 
that is expected to apply to a large class of nonequilibrium
steady states including a heat conducting fluid,
a sheared fluid, and an electrically conducting fluid.
We call our theory steady state thermodynamics (SST)
after Oono and Paniconi's original proposal.
The construction of SST
is based on a careful examination of how the basic notions
in thermodynamics should be modified in nonequilibrium steady
states.
We define all thermodynamic quantities through operational 
procedures which can be (in principle) realized experimentally.
Based on SST thus constructed, we make some nontrivial
predictions, including an extension of Einstein's formula on 
density fluctuation, an extension of the minimum work principle,
the existence of a new osmotic pressure of a purely nonequilibrium
origin, and a shift of coexistence temperature.
All these predictions may be checked experimentally to test SST for
its quantitative validity.
\end{quotation}

\tableofcontents

\section{Introduction}
\label{s:in}

\subsection{Motivation and the goal of the paper}
\label{s:intro}
Construction of statistical mechanics that applies to  
nonequilibrium states 
has been a challenging open problem in theoretical physics. 
By statistical mechanics, we mean a universal theoretical
framework that enables one to precisely characterize 
 states of a given system, and to compute
(in principle) arbitrary macroscopic quantities.
Nobody knows what the desired nonequilibrium statistical 
mechanics should look like.
Indeed it seems highly unlikely that there is 
statistical mechanics that applies to {\em any}\/ nonequilibrium
systems.
A much more modest (but still extremely ambitious)
goal is to look for a theory that applies to nonequilibrium steady
states, which are out of equilibrium 
but have no macroscopically observable time dependence.
There may be a chance that 
probability distributions for nonequilibrium steady states can be 
obtained from a general principle, 
analogous to the  equilibrium statistical mechanics.
Our ultimate goal is to find such a principle, 
but the goal (if any) is still very far away.

We wish to recall the history of equilibrium statistical mechanics.
When Boltzmann, Gibbs, and others constructed statistical
mechanics, the formalism of thermodynamics played a fundamental
role as a theoretical guide.
In particular, Gibbs seems to have intentionally 
sought for a probability distribution which most naturally
recovers some of the thermodynamic relations.

In our attempt toward nonequilibrium statistical mechanics, 
we too would like to start from the level of 
phenomenology and look for a possible thermodynamics.
By a thermodynamics, we mean a rigid mathematical structure 
consisting of mathematical relations among certain quantities in 
a physical system.
The mathematical structure of thermodynamics is clearly
and abstractly explained, for example, in \cite{Wightman79,OonoPaniconi98,LiebYngvason99}.
The conventional thermodynamics for equilibrium systems is a
typical and no doubt the most important example of thermodynamics,
but it is not the only example 
(see, for example, section 3 of 
\cite{OonoPaniconi98} and Appendix 1. A1 of \cite{Gallavotti99}).

Then it makes sense to look for a thermodynamics in a physical context
other than equilibrium systems.
We wish to do that for nonequilibrium steady states.
If it turns out that 
there is no sensible thermodynamics for nonequilibrium
steady states, then we should give up seeking for 
 statistical mechanics.
If there is a thermodynamics, on the other hand,
then we can start
looking for statistical mechanics which is
consistent with the thermodynamics.
Our goal in the present paper is to propose a thermodynamics
for nonequilibrium steady states, and to convince the readers that 
our proposal is essentially the unique possible thermodynamics.

\bigskip

The standard theory of nonequilibrium 
thermodynamics (see section~\ref{s:NETD})
is based on the local equilibrium hypothesis, which roughly 
asserts that each small part of a nonequilibrium state can be regarded 
as a copy of a suitable equilibrium state.
But such a description seems 
insufficient for general 
nonequilibrium steady states, especially when the
``degree of nonequilibrium'' is not small.

Consider, for example, a system 
with  steady heat flow. 
It is true that quantities like the temperature
and the density become essentially constant within a sufficiently small 
portion of the system. But no matter how small the portion is, there always 
exists a heat flux passing through it and hence the local state  is 
{\em not}\/ isotropic. 
It is quite likely that the pressure tensor, for example,
becomes anisotropic, and the equation of state is consequently modified.
Then the local state cannot be identical
to an equilibrium state, but should be described rather 
as a {\em local steady state}\/.

There has been some attempts to formulate thermodynamics
for nonequilibrium steady states by going beyond
local equilibrium treatments.
See section~\ref{s:EIT}.
Among these attempts, we regard 
the steady state thermodynamics (SST)
proposed by Oono and Paniconi \cite{OonoPaniconi98}
to be most sophisticated and
promising.
The basic strategy of Oono and Paniconi
is to seek for a universal thermodynamic formalism
respecting general mathematical structure of thermodynamics
and operational definability of thermodynamic quantities.
As far as we know, no other proposals of nonequilibrium thermodynamics
follow such logically strict rules.
Oono and Paniconi's SST, however, is still too abstract to be
tested empirically.

In the present paper, we follow the basic strategy of Oono and 
Paniconi's, but try to construct much more concrete theory
which leads to nontrivial predictions.
Our strategy in the present work may be summarized as follows.
\begin{itemize}
\item
Concentrate on some typical examples
(i.e., a heat conducting fluid, a sheared fluid, and an electrically
conducting fluid)
of nonequilibrium steady states, always trying to elucidate
universal aspects of the problem.
\item
Examine carefully how the basic notions of thermodynamics
(for example, scaling, extensivity/intensivity, and operations to systems)
should be modified in nonequilibrium steady states.
\item
Define every thermodynamic quantity through a purely operational 
procedure which can be realized experimentally.
\item
Make concrete predictions which may be checked experimentally to test
our theory for its quantitative validity.
\end{itemize}
As a result, our theory has no direct logical connection with 
Oono and Paniconi's SST.
But we keep the name SST to indicate that we share the
basic philosophy with them.

Our theory is of course based on some phenomenological assumptions,
the biggest one being the assumption that there exists a sensible
thermodynamics.
Although we are confident about theoretical consistency of our 
SST, its validity must ultimately be tested empirically.

If we restrict ourselves to certain idealized (but still nontrivial) theoretical models, we can demonstrate that the formalism of SST is indeed realized.
We shall present such model dependent results as Appendices.
The most complete ``existence proof'' is the results in Appendix~\ref{s:spn} about the driven lattice gas, a standard stochastic model for nonequilibrium steady states.
For a sheared fluid with a ``weak coupling'', we also recover a significant part (but, not the whole) of SST as we describe in Appendix~\ref{s:WCS}.

Of course we have no intention to claim that our SST should cover nonequilibrium states in general.
Systems with explicit macroscopic time-dependence are out of consideration from the beginning.
Systems which are too unstable to maintain stable thermodynamics cannot be treated.
Moreover, since we make  a full use of the pressure, model systems (such as chains of oscillators) which do not possess well-defined pressure do not fit into our scheme.
We nevertheless hope that our formalism covers a generic and nontrivial class of nonequilibrium steady states.

\bigskip

The organization of this long paper may easily be read off from the
table of contents.
After discussing necessary materials from equilibrium physics
in section~\ref{s:eq},
we carefully describe our assumptions, and construct
steady state thermodynamics step by step
through sections~\ref{s:ss} to \ref{s:fu}.
To help the readers, the beginning of each of these sections contains a brief summary of the section.

Before going into this massive main body, the reader is
invited to take a look at the next section~\ref{s:QT},
where we offer a very quick tour of our construction
and predictions.
In addition, we compare our approach with some
of the existing attempts in section~\ref{s:EA},
discuss possible experimental verifications in
section~\ref{s:exp}, and answer some of the
``frequently asked questions'' in
section~\ref{s:faq}.
Appendices, which treat model dependent results, may be studied
independently after reading the introductory section~\ref{s:QT}.

We should better stress here that the present paper does {\em not}\/ report
a standard scientific research in which one provides answers to well established problems.
We report our (admittedly ambitious) attempt to search for a novel universal framework that describes Nature.
We thus take a nonstandard approach where we proceed step by step, stating each assumption carefully, examining its consistency, and discussing the consequences.
We have tried our best to make the presentation as transparent as possible, not hiding any subtle points.

\subsection{A quick look at steady state thermodynamics (SST)}
\label{s:QT}
To give the reader a rough idea about what our 
steady state thermodynamics is all about, 
we shall here outline (rather superficially)
our construction and predictions
in a single example of a sheared fluid.
Every step illustrated here will be examined and explained carefully
in latter sections of the paper.
In particular we will thoroughly  discuss in the latter sections why 
we believe that the present construction is essentially the 
unique way toward a sensible thermodynamics
for nonequilibrium steady states.

\subsubsection{Nonequilibrium steady state in a sheared fluid}
\label{s:QTs}
\begin{figure}
\centerline{\epsfig{file=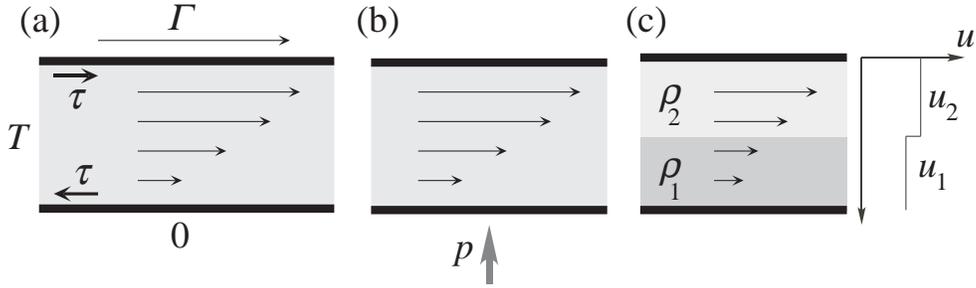,width=13cm}}
\caption[dummy]{
(a) The nonequilibrium steady states of a sheared fluid.
The velocity gradient is maintained by shear forces 
exerted on the fluid by the upper and the lower walls. 
The two walls exert exactly the opposite forces, whose
magnitudes are $\tau$, on the fluid.
The steady state is parameterized as 
$(T,\tau;V,N)$.
(b)~The pressure is determined by measuring the vertical mechanical
force exerted on the wall.
(c)~The chemical potential is determined by adding
external potential to the system.
}
\label{f:QTss}
\end{figure}

Suppose that $N$ moles of fluid is contained in  a box with the cross
section area $A$ and height $h$, and kept at a constant temperature
$T$ with the aid of an external heat bath.
To make the state nonequilibrium, the upper wall of the box
is moved horizontally with a constant speed $\Gamma$
while the lower wall is kept at rest\footnote{
It is convenient to imagine that periodic boundary conditions
are imposed in the horizontal directions.
Experimentally, one should modify the geometry (say, into a ring shape)
to keep on moving the wall.
}.
We suppose that the walls are ``sticky'', and the fluid will reach, after a
sufficiently long time, a nonequilibrium steady state with a velocity
gradient as in Fig.~\ref{f:QTss}~(a).
We denote by $\tau$ the total horizontal force that the upper wall 
exerts on the fluid.
The steadiness implies that the lower wall exerts exactly
the opposite force on the fluid.
Clearly the shear force $\tau$ measures the 
``degree of nonequilibrium'' of the steady state.

We shall parameterize the nonequilibrium steady state as
$(T,\tau;V,N)$  where $V=Ah$ is the volume.
As in the equilibrium thermodynamics (see section~\ref{s:td}),
we frequently consider scaling, decomposition, and combination
of steady states.
In doing so, we always use the convention to fix the cross section
area $A$ constant, and vary only the height $h$.

In this convention of scaling, $T$ and $\tau$
are identified as intensive variables,
while $V$ and $N$ as extensive variables.
These identifications are fundamental in our construction of SST.

We stress that the above convention of scaling and the choice of
thermodynamic variables are results of very careful examination of 
general structures of thermodynamics and the characters specific to
nonequilibrium steady states.
These points are discussed in sections~\ref{s:ss}
and \ref{s:bf}.

\subsubsection{Pressure and chemical potential}
\label{s:QTpc}
We now fix the two intensive parameters $T$ and $\tau$,
and determine the pressure $p(\rho)$ and the chemical
potential $\mu(\rho)$ as functions of the density $\rho=V/N$.
We insist on determining these quantities in a purely operational 
manner, only using procedures that can be realized experimentally.
This is the topic of section~\ref{s:op}.

The pressure $p(\rho)$ is simply defined as the mechanical pressure
on the lower or the upper wall as in Fig.~\ref{f:QTss}~(b).
In other words we concentrate on the vertical component of the pressure.

The measurement of  the chemical potential $\mu(\rho)$ requires 
extra cares.
We (fictitiously) divide the system into half along a horizontal plane,
and apply a potential which is equal to $u_1$ in the lower half and equal
to $u_2$ in the upper half.
We denote by $\rho_1$ and $\rho_2$ the densities 
in the lower and the upper parts, respectively, in the steady state
under the potential.
We shall define the SST chemical potential $\mu(\rho)$ as
a function which satisfies
\begin{equation}
\mu(\rho_1)+u_1=\mu(\rho_2)+u_2,
\label{e:QTmu}
\end{equation}
for any $u_1$ and $u_2$.

Note that this only determines the difference
of $\mu(\rho)$.
There remains a freedom to add an arbitrary constant to $\mu(\rho)$.
In other words, we have determined the $V$, $N$ dependence of
the chemical potential, but not $T$, $\nu$ dependence.

An essential point of these definitions is that the
Maxwell relation
\begin{equation}
\partialf{p(\rho)}{\rho}=\rho\partialf{\mu(\rho)}{\rho}
\label{e:QTMax}
\end{equation}
can be shown to hold in general.

\subsubsection{Helmholtz free energy}
\label{s:QTf}
Since we have determined the pressure and the chemical potential,
we can introduce and investigate the SST free energy.
This is done in section~\ref{s:f}.
We define the specific free energy
through the
Euler equation as
\begin{equation}
f(\rho)=-p(\rho)+\rho\,\mu(\rho).
\label{e:QTf}
\end{equation}
The extensive free energy is obtained as
$F(T,\tau;V,N)=V\,f(N/V)$.
We have thus operationally determined the
$V$, $N$ dependence of $F(T,\tau;V,N)$ for each
$T$ and $\tau$.

We make three predictions which involve the 
$V$, $N$ dependence of the free energy.
The first two of these phenomenological conjectures can be verified 
by making plausible assumption about contact, as we shall see in Appendix~\ref{s:WCS}.
In a class of stochastic processes treated in 
Appendix~\ref{s:spn}, all the three conjectures are derived.

\begin{figure}
\centerline{\epsfig{file=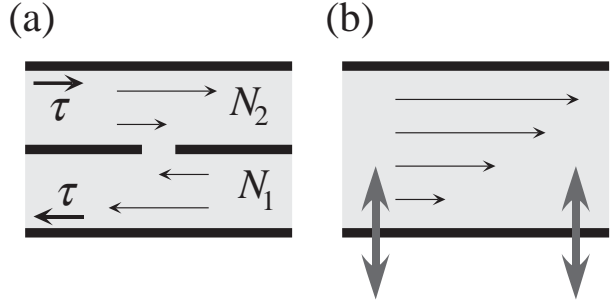,width=8cm}}
\caption[dummy]{
Two basic applications of the SST free energy $F(T,\tau;V,N)$.
(a)~We conjecture that  Einstein's formula for density fluctuation
extends to nonequilibrium steady states, 
provided that the two regions
are separated by a  wall with a window in it.
(b)~The minimum work principle is conjectured to hold when the agent is 
only allowed to move horizontal walls vertically.
}
\label{f:QTF}
\end{figure}

The first prediction is an extension of Einstein's formula on
macroscopic density fluctuation.
Consider the steady state $(T,\tau;2V,2N)$ with 
$2N$ moles of fluid in a box with volume $2V$.
We divide the system into two identical parts
with volumes $V$ by a horizontal wall with a small window\footnote{
See Appendix~\ref{s:WCS} for details about the window.
} in it as in 
Fig.~\ref{f:QTF}~(a).
We fix the wall in the middle, and apply the same shear force $\tau$ to the 
upper and the lower walls to maintain the constant shear in the whole system.
In this way the two parts are coupled weakly and exchange fluid molecules.
Let $N_1$ and $N_2$ be the amounts of 
fluid in the lower and the upper parts, respectively.
Although both $N_1$ and $N_2$ should be equal to 
$N$ in the average, one always observes a fluctuation in
a finite system.
Our conjecture is that the probability $\tilde{p}(N_1,N_2)$
of observing $N_1$ and $N_2$ moles of fluid in the two parts
is given by
\begin{equation}
\tilde{p}(N_1,N_2)\propto
\exp[-\frac{1}{k_{\rm B}T}\{
F(T,\tau;V,N_1)+F(T,\tau;V,N_2)
\}],
\label{e:QTpt}
\end{equation}
where $k_{\rm B}$ is the Boltzmann constant.
Unlike the corresponding relation in equilibrium, this relation is expected to
hold only when the two regions are separated by the horizontal wall with a window.

The second prediction is the fluctuation-response relation for time-dependent processes that takes place in the same setting as above.
We shall leave details to sections~\ref{s:df}, \ref{s:WCSL}, and \ref{s:DLGLR}.

The third prediction is an extension of the minimum work principle, a
version of the second law of thermodynamics.
Suppose that an outside agent moves one of the horizontal walls of 
the box vertically, always keeping the wall horizontal as in 
Fig.~\ref{f:QTF}~(b).
We assume that $T$ and $\tau$ are kept constant during the operation.
We denote by $V$ and $V'$ the initial and the final volumes, respectively.
Denoting by $W$ the total mechanical work done by the agent, we 
can write the conjectured minimum work principle 
for nonequilibrium steady states as
\begin{equation}
W\ge F(T,\tau;V',N)-F(T,\tau;V,N),
\label{e:QTmwp}
\end{equation}
which has exactly the same form as the corresponding equilibrium relation.
An essential difference is that we here severely restrict allowed operations.

\subsubsection{Flux-induced osmosis and shift of coexistence temperature}
\label{s:QTc}
We shall now determine the SST free energy $F(T,\tau;V,N)$
completely and make further conjectures.
This is the topic of section~\ref{s:fu}.

\begin{figure}
\centerline{\epsfig{file=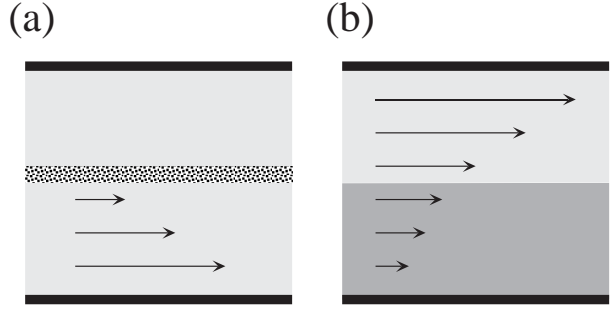,width=8cm}}
\caption[dummy]{
(a)~A nonequilibrium steady state with a finite shear is in contact with
an equilibrium state without a shear.
The two states are separated by a porous wall that allows fluid to pass thorough.
The top wall and the porous wall are at rest while the bottom wall has a constant horizontal velocity.
This setup is used to determine the chemical potential and the free energy
completely.
SST leads to a conjecture that the (vertical) pressure in the steady state
is always larger than that in the equilibrium state.
(b)~A nonequilibrium steady state with a phase coexistence.
We conjecture that the coexistence temperature $T_{\rm c}(p,\tau)$
deviates from that in the equilibrium.
}
\label{f:QTc}
\end{figure}

The key idea is to consider the setting in
Fig.~\ref{f:QTc}~(a),
where a nonequilibrium steady state $(T,\tau;V,N)$
with a finite shear is in contact with an equilibrium state
$(T,0;V',N')$ via a porous wall.
Since the two states can exchange fluid,
we require that
\begin{equation}
\mu(T,\tau;V,N)=\mu(T,0;V',N')
\label{e:QTmm}
\end{equation}
as in equilibrium thermodynamics.
Since $\mu(T,0;V',N')$ is an already known equilibrium quantity,
we use \eqref{e:QTmm} as the definition of the 
SST chemical potential.

Now that the chemical potential has been fully determined,
we can also determine the SST free energy $F(T,\tau;V,N)$
through \eqref{e:QTf}, including its dependence on $T$ and $\tau$.
Then we can define the SST entropy
\begin{equation}
S(T,\tau;V,N)=-\partialD{T}F(T,\tau;V,N),
\label{e:QTSP1}
\end{equation}
and a new extensive quantity
\begin{equation}
\Psi(T,\tau;V,N)=-\partialD{\tau}F(T,\tau;V,N).
\label{e:QTSP2}
\end{equation}
We call $\Psi(T,\tau;V,N)$ the nonequilibrium order parameter,
since we can show
(under the assumption about concavity of $F(T,\tau;V,N)$ in $\tau$)
that $\Psi(T,0;V,N)=0$  and 
$\Psi(T,\tau;V,N)=-\Psi(T,-\tau;V,N)\ge0$ if
$\tau\ge0$.

The nonequilibrium order parameter $\Psi(T,\tau;V,N)$
characterizes two important phenomena, which are intrinsic to
nonequilibrium steady states.

The first phenomenon takes place in the setting of Fig.~\ref{f:QTc}~(a).
Suppose one fixes the pressure $p_{\rm eq}$
of the equilibrium part, and changes the shear force $\tau$.
Then we can show that the pressure $p_{\rm ss}$ 
of the steady state satisfies
\begin{equation}
\partialf{p_{\rm ss}}{\tau}
=
\frac{\Psi(T,\tau;V,N)}{V}.
\label{e:QTFIO}
\end{equation}
Sine $p_{\rm ss}=p_{\rm eq}$ when $\tau=0$,
this (and the knowledge about the sign of $\Psi$)
implies that 
$p_{\rm ss}\ge p_{\rm eq}$  in general.
We expect that $p_{\rm ss}>p_{\rm eq}$ holds for $\tau\ne0$.
The steady state always has a higher pressure than the 
equilibrium state.
We call this pressure difference the flux-induced osmosis
(FIO).
Note that FIO can never be predicted within the standard local
equilibrium treatments.

To see the second phenomenon, suppose that two phases
(such as gas and liquid) coexist within a steady state
as in Fig.~\ref{f:QTc}~(b).
We denote by $T_{\rm c}(p,\tau)$
the temperature at which the coexistence takes place
when the pressure and the shear force are
fixed at $p$ and $\tau$, respectively.
We can then show that
\begin{equation}
\partialf{T_{\rm c}(p,\tau)}{\tau}=-
\frac{\Psi_{\rm g}-\Psi_{\ell}}{S_{\rm g}-S_{\ell}},
\label{e:QTTc}
\end{equation}
where $S_{\rm g,\ell}$ and $\Psi_{\rm g,\ell}$
are the entropy and the nonequilibrium order parameter
in the gas and the liquid phases, respectively.
This means that in general the coexistence temperature $T_{\rm c}(p,\tau)$
in a nonequilibrium steady state is different from that in equilibrium.
This, again, is a truly nonequilibrium phenomenon.
Applied to the phase coexistence between a fluid and a solid phases, the same argument yields $T_{\rm c}(p,\tau)<T_{\rm c}(p,0)$.
Thus the shear induces melting.

It is important that the same quantity $\Psi$ plays the essential
roles in the above two phenomena.
This means that we can test  the quantitative validity of SST
through purely experimental  studies.

\subsection{Existing approaches to nonequilibrium steady states}
\label{s:EA}
In the present section, we briefly discuss some of the existing 
approaches to nonequilibrium steady states, and 
see how they are (or how they are not) related to our own approach 
of SST. We note that the aim here is not to give an exhaustive 
and balanced review of the field, but to place our new work in 
the context of (necessarily biased) summary of nonequilibrium 
thermodynamics and statistical mechanics.

\subsubsection{Phenomenological theories in the linear nonequilibrium regime}
\label{s:PhLin}
Probably the best point to start this  discussion on nonequilibrium 
physics is Einstein's celebrated  work on the Brownian motion.
We have no intention of going deeply into the work, but wish to 
mention that Einstein's formula
\begin{equation}
D=k_{\rm B}T\,\mu,
\label{e:Brownian}
\end{equation}
derived in \cite{Einstein05a,Einstein05b} represents a deep fact 
that the transport coefficient (the mobility $\mu$) in a driven 
nonequilibrium state is directly related to the diffusion constant 
$D$, which characterizes fluctuation in the equilibrium state.

\topic{Onsager's theory}\label{s:Ons}
Such a relation between equilibrium fluctuation and nonequilibrium 
transport was stated as a fundamental principle of (linear) nonequilibrium
physics by Onsager. In his famous paper on the reciprocal 
relations \cite{Onsager31a}, he formulated 
the {\em regression hypothesis}\/ which asserts that 
``the average regression of fluctuations (in equilibrium) will 
obey the same laws as the corresponding macroscopic irreversible 
processes'' \cite{Onsager31b}.
From the regression hypothesis and microscopic reversibility of 
underlying mechanics, Onsager \cite{Onsager31a,Onsager31b}  
derived the reciprocal relations for transport coefficients.
Since the reciprocal relations are established experimentally, this 
provides a strong support to the regression hypothesis, at least 
in the linear nonequilibrium regime. 
It is fair to say that, as far as nonequilibrium steady states in 
linear transport regime are concerned, Onsager constructed a beautiful 
phenomenology with sound theoretical and empirical bases.

Onsager's theory is essentially related to (at least) three subsequent developments in nonequilibrium physics that we shall discuss in the following.

\topic{Linear response relations}
\label{s:LRR}
A series of formulae that express various transport coefficients 
in terms of time-dependent equilibrium correlation functions
was found in various contexts, the first example being that by Nyquist \cite{Nyquist28}, who precedes Onsager. 
These formulae are now known under the generic name {\em linear response relations}\/.
See, for example, \cite{KuboTodaHashitsume85,Nakano93}.
We believe that the conceptual basis of these relations should 
be sought in a certain form of regression hypothesis, i.e., quantitative 
correspondence between nonequilibrium transport and equilibrium fluctuation.

\topic{Variational principles}
\label{s:varational}
The second development is the establishment of variational principles
which relate currents to the corresponding forces in the linear response regime.
The simplest version of such principles, called the {\em principle of the 
least dissipation of energy}\/, is obtained as a direct consequence of 
the reciprocal relations \cite{Onsager31a,Onsager31b}.
Another type of variational principle 
attempting to characterize nonequilibrium steady states is called the {\em principle of minimum entropy production}\/ \cite{Prigogine67}. 
It is understood that  all the correct variational principles 
in linear transport regime are based on the Onsager-Machlup theory 
\cite{Hashitsume52,OnsagerMachlup53,Hashitsume56} which concerns a large deviation functional 
for the history of fluctuations. 
See, for example, \cite{Ono61}.  

\topic{Nonequilibrium thermodynamics}
\label{s:NETD}
Flux-force relations with the reciprocity constitute fundamental
ingredients of the standard theory known as
nonequilibrium thermodynamics, which  provides a 
macroscopic description of a system which slightly deviates from 
equilibrium \cite{Prigogine67,GrrotMazur62}.
A fundamental assumption in this approach is that 
a small portion of the system in the nonequilibrium state can be 
regarded as a {\em local equilibrium state}\/ in the sense that 
all the thermodynamic relations in equilibrium (not only universal 
relations, but also equations of states specific to each system) 
are valid without modifications. One then allows macroscopic 
thermodynamic variables to vary slowly in space and time, assuming 
that there takes place linear transport 
according to a given set of transport coefficients.

\topic{Relation to SST}
\label{s:PTSST}
We wish to see how our SST is related to these theories.
In short, we see no direct logical connection 
for the moment.
All of these theories are essentially limited to linear transport regime 
with very small ``degree of nonequilibrium'', while SST is designed 
to apply to any nonequilibrium steady states.
The variational principles mentioned above
attempt to characterize the steady state itself, while the SST free energy mainly describes 
the response of nonequilibrium steady states to external operations 
(such as the change of the volume) under a fixed degree of nonequilibrium.
We can say that, at least for the moment, SST covers aspects complimentary 
to that dealt with the above theories. It would be very interesting 
to incorporate Onsager's and related phenomenology into SST, but we do not yet see 
how this can be accomplished.

We have already stressed in section~\ref{s:intro} the difference between 
the nonequilibrium thermodynamics and our SST. Our main motivation is to 
construct thermodynamics that applies to systems very far from equilibrium.
We must abandon the description in terms of local equilibrium states, and 
replace it with that in terms of local steady states\footnote{
It should be noted that, in the present work, we are concentrating on 
characterizing local steady states, and not yet considering spatial and 
temporal variation of macroscopic variables. It is among our future plan 
to patch together local steady states to describe non-uniform nonequilibrium 
states.
}.

\subsubsection{Approaches from microscopic dynamics}
\label{s:MD}
It is a natural idea to realize and characterize nonequilibrium steady 
states by using equilibrium states and microscopic (classical or quantum) dynamics.
Suppose that we are interested in a heat conducting steady state.
We prepare an arbitrary (macroscopic) subsystem, and couple it to 
two ``heat baths'' which are much larger than the subsystem.
The two heat baths are initially in thermal equilibria with different 
temperatures. We then let the whole system evolve according to the 
microscopic equation of motion. 
After a sufficiently long 
(but not too long) time, the subsystem is expected to reach a steady 
heat conducting state. By projecting only onto the subsystem, we get 
the desired nonequilibrium steady state. If such a projection can be 
executed for general systems, there is a chance that we can extract 
a universal description for nonequilibrium steady states.

Of course the procedure described above is in general too difficult 
to be carried out literally  even in the linear response regime.
We shall see two approximate calculation schemes within the conventional statistical mechanics (which are (a) and (b)), and some of more mathematical approaches (which are (c), (d), and (e)).

If and when these theories provide us with concrete information about the structure of nonequilibrium steady states and their response to external operations,  we can (and should) check the consistency between such predictions and those obtained from SST.
For the moment most of the known results are rather formal, and we  do not find any concrete results which should be compared with SST.

\topic{Linear response theory}
\label{s:LRT}
Probably the most well-known of such schemes is the linear response theory \cite{KuboTodaHashitsume85,Nakano93}. 
Although this theory is sometimes referred to as  a ``microscopic (or rigorous) derivation'' of linear response relations
(see, for example, \cite{KuboTodaHashitsume85}), it is after all a formal perturbation theory about the equilibrium state, and does not deal with the intrinsic characterization of nonequilibrium steady states.
As far as we understand, certain phenomenological principle must be invoked 
to justify such a derivation.

\topic{Methods based on the Liouville equation}
\label{s:Liouville}
In classical mechanics, the Liouville equation can be a starting point 
for microscopic considerations. 
An example is the derivation of the 
non-linear response relation of \cite{YamadaKawasaki68}, 
which leads to the Kawasaki-Gunton formula \cite{KawasakiGunton73} for 
a nonlinear shear viscosity and normal stresses.
Another example is the establishment of the existence of  
long range spatial correlations of fluctuations 
in nonequilibrium steady states \cite{DorfmanKirkpatrickSengers94}.
These results were obtained by employing the projection 
operator method pioneered by Zwanzig 
\cite{Zwanzig61} and  Mori \cite{Mori65}.
Furthermore, through 
a formal argument based on the Liouville equation,
McLennan \cite{McLennan90} and Zubarev \cite{Zubarev74} proposed a measure 
that describes (or is claimed to describe) nonequilibrium states.

Although the derivations of these results involve (often uncontrolled) 
assumptions, the nonlinear response relation, the Kawasaki-Gunton formula, 
and the power-law decay of spatial correlations are believed to be physically sound,
since they can also be derived in simple manners from phenomenological 
considerations.  See \cite{Crooks00} for the 
nonlinear response relation, \cite{WadaSasa03} 
for the Kawasaki-Gunton formula, and \cite{DorfmanKirkpatrickSengers94} 
for the long range correlations.  
The measure proposed by McLennan and Zubarev is supported by neither a controlled theory nor a phenomenological argument.
It is therefore difficult to judge its physical validity and usefulness.

\topic{Weak coupling limit}
\label{s:weakcoupling}
In the weak coupling limit of quantum systems, 
the procedure of projection can be executed rigorously \cite{SpohnLebowitz78}.
Relaxation to the steady state, the reciprocal relations in linear transport, 
and the principle of minimum entropy production are established.
In this study, however, explicit forms of nonequilibrium steady states 
are not obtained.

\topic{$\rm C^*$ algebraic approaches}
\label{s:C*}
There is a series of works in which heat baths are modeled by infinitely 
large systems of ideal gases, and the time evolution is discussed by using 
the C$^*$ algebraic formalism. See, for example, \cite{JaksicPillet02}.
As far as we understand, the results obtained in this direction mainly 
focus on what happens when more than two baths are put into contact, 
rather than what happens in the subsystem where transport is taking place.
We still do not get much information about the structure of nonequilibrium 
steady states from these works.

\topic{Chain of anharmonic oscillators}
\label{s:chain}
A standard model for heat conduction in classical mechanics is the chain of 
coupled anharmonic oscillators whose two ends are attached to two heat baths 
with different  temperatures. 
From numerical simulations (see, for example, \cite{HuLiZhao00}) it is 
expected that the model exhibits a ``healthy'' heat conduction, i.e., 
obeys the Fourier law.
Mathematically, basic results including the existence, 
uniqueness and 
mixing property of the nonequilibrium steady states are proved under 
suitable conditions \cite{EckmannPilletReyBelle99,EckmannHairer00}, 
but no concrete information about the structure of the heat conducting 
state is available. Recently a new perturbative method for the 
nonequilibrium steady state of this model was developed 
\cite{LefevereSchenkel04}.


\subsubsection{Approaches from meso-scale models}
\label{s:SD}
We turn to approaches to nonequilibrium steady states that employ a class of models which are neither microscopic (as in mechanical treatments) nor macroscopic (as in thermodynamic treatments).
The class, which may be called mesoscopic, includes the Boltzmann 
equation, the nonlinear Langevin equations for slowly varying macroscopic
variables, and the driven lattice gas.

\topic{Boltzmann equation}
\label{s:Boltz}
The method developed by Chapman and  Enskog \cite{ChapmanCowling39} 
enables one to explicitly compute perturbative solutions of the 
Boltzmann equation. 
Expecting that the Boltzmann equation, which  was originally introduced to describe relaxation to equilibrium, 
may be extended to study nonequilibrium phenomena\footnote{
The Boltzmann equation can be derived from the BBGKY hierarchy in 
a low density limit around the (spatially uniform) equilibrium state. 
(See \cite{Lanford75} for the mathematical justification of the derivation.) 
We have to keep in mind, however, that there is a logical possibility that 
correction terms to the Boltzmann equation appear in the truncation 
process from the BBGKY hierarchy when the spatial non-uniformity of the 
states are taken into account.
}, nonequilibrium stationary 
distribution functions have been calculated.
Recently, for example, a systematic calculation for heat conducting nonequilibrium  
steady states was performed \cite{KimHayakawa03a,KimHayakawa03b}. 
Such a study reveals detailed properties of the nonequilibrium 
steady states, and may become an  important guide in construction of 
phenomenology and statistical mechanical theory. 
The relation of this result to SST will
be discussed in section~\ref{s:FIO}.
 As for recent progress in this direction, 
see references in  \cite{KimHayakawa03a,KimHayakawa03b}.

\topic{Nonlinear Langevin model for macroscopic variables}
\label{s:NLM}
Nonlinear Langevin models for macroscopic variables were useful 
to study anomalous behavior of transportation coefficients at the
critical point \cite{HohenbergHalperin77}. 
The shift of the critical temperature under the influence of shear flow 
as well as the corresponding critical exponents were calculated 
by analyzing the so called model H with the steady shear flow \cite{OnukiKawasaki79}. 

Such an approach might produce correct results for universal quantities (such as the critical exponents) which are insensitive to minor details of models.
It is questionable, however, whether a non-universal quantity like the critical temperature shift can be properly dealt with.
Results from a model calculation may be always improved by making the model more and more complicated, but  such a process of improvement seems endless.
If the formulation of SST is true, on the other hand, the shift of coexistence temperatures
should be related to other measurable quantities through the (conjectured) extended Clapeyron
law, which is expected to be universal\footnote{
Needless to say, 
thermodynamic phases may be in principle determined from  microscopic descriptions
when and if statistical 
mechanics for nonequilibrium steady states is constructed.
}.

\topic{Driven lattice gas}
\label{s:DLGO}
Given the history that the lattice gas models (equivalently, the Ising model)
was the paradigm model in the study of equilibrium phase transitions, it is 
natural that various stochastic models of lattice gases for nonequilibrium 
states were studied. See, for example, \cite{Spohn91}.
The simplicity of these models made it possible to resolve some delicate 
issues rigorously, a notable examples being the long-range correlations 
\cite{Spohn83} and the anomalous current fluctuation \cite{PraehoferSpohn02}.

A standard nontrivial model is the driven lattice 
gas \cite{KatzLebowitzSpohn84}, in which particles on lattice are subject 
to hard core on-site repulsion, nearest neighbor interaction, and a constant 
driving force. Many results, both theoretical and numerical, have been 
obtained \cite{Spohn91,SchimttmannZia95}, but the structure of the 
nonequilibrium steady state is still not very well understood except for 
some partial results including the recent perturbation expansion 
\cite{LefevereTasaki04}. In \cite{EynkLebowitzSpohn96}, hydrodynamic limit 
and fluctuation was studied for the nonequilibrium steady state in the 
driven lattice gas. Possibility of thermodynamics of driven lattice gas 
being ``shape-dependent'' was pointed out in \cite{AlexanerEyink98}.
In SST, such a shape-dependence is properly taken into account in the basic formalism.

For us the driven lattice gas provides a very nice ``proving ground'' for 
various proposals and conjectures of SST. Some of our discussions in the 
present paper are based on earlier numerical works by Hayashi and Sasa 
in \cite{HayashiSasa03}. In  Appendix~\ref{s:spn} of the present 
paper, we also discuss theoretical results  about SST realized in  driven 
lattice gases.

In spite of all these interesting works, we always have to keep in mind 
that physical basis of these stochastic lattice models are still unclear.
As for the stochastic dynamics near equilibrium, it is well appreciated 
that the detailed balance condition (which was indeed pointed out in 
Onsager's work on the reciprocal relations \cite{Onsager31a}) is the 
necessary and sufficient condition to make the model physically meaningful.
As for dynamics far away from equilibrium, we still do not know of any 
criteria that should replace the detailed balance condition.

\subsubsection{Recent progress}
\label{s:RP}

In the last decade, there have been some progress in new directions 
of study on nonequilibrium steady states. 
They are fluctuation theorem, 
additivity principle, and dynamical fluctuation theory. 
We shall briefly review
them and comment on the relevance to SST.

\topic{Fluctuation theorem}
\label{s:FTspG}
In a class of chaotic dynamical systems, a highly nontrivial symmetry in 
the entropy production rate, now known by the name 
{\em fluctuation theorem}\/, was found 
\cite{EvansCohenMorris93,GallavottCohen95}.
The fluctuation theorem was then extended to nonequilibrium steady states 
in various systems.  See \cite{Kurchan98,LebowitzSpohn99,Maes99}.

Now it is understood that the essence of the fluctuation theorem 
lies in the fact that the relevant nonequilibrium steady states are described 
by Gibbs measures for space-time configurations \cite{Maes99}.
It is known that nonequilibrium steady states that are modeled by a 
class of chaotic dynamical system \cite{Gallavotti99} or by a class 
of stochastic processes \cite{LebowitzSpohn99,LebowitzMaesSpeer90} 
are described by space-time Gibbs measures. But it is not yet clear 
if the description in terms of a space-time Gibbs measure is universally
valid.

A more important question is whether a space-time description is really
necessary for nonequilibrium steady states. One might argue that any 
nonequilibrium physics should be described in space-time language, since
the time-evolution must play a crucial role. On the other hand, one may 
also expect that the temporal axis is redundant for the description of 
nonequilibrium {\em steady}\/ states since nothing depends on time.

Our formalism of SST is based on the assumption that one can construct 
a consistent macroscopic phenomenology without explicitly dealing with 
the temporal axis. If a space-time description is mandatory for 
nonequilibrium physics, our attempt should reveal its own failure as 
we pursue it. So far we have encountered no inconsistencies.

\topic{Additivity principle}
\label{s:DLS}
Recently Derrida, Lebowitz, and Speer obtained exact large deviation 
functionals for the density profiles in the nonequilibrium steady states
of the one dimensional
lattice gas models (the symmetric exclusion process 
\cite{DerridaLebowitzSpeer01,DerridaLebowitzSpeer02a} and the asymmetric 
exclusion process \cite{DerridaLebowitzSpeer02b,DerridaLebowitzSpeer03})
attached to two particle baths with different chemical potentials.
In the equilibrium states, the corresponding large deviation functional 
coincides with the thermodynamic free energy.
Moreover their large deviation
functional satisfies a very suggestive variational principle named
{\em additivity principle}\/.
It was 
further proposed \cite{BodineauDerrida04} that, in a large class of one dimensional models, the large deviation functional for current satisfies a similar additivity principle.

It would be quite interesting if these large deviation functionals could 
be related to the SST free energy that we construct operationally.
Unfortunately we still do not see any explicit relations.
A difficulty comes from the restriction to one dimensional lattice systems, 
where it is not easy to realize macroscopic operations which are essential 
in our construction.
It is thus of great interest whether the additivity principles
can be extended to higher dimensions.

\topic{Dynamical fluctuation theory}
\label{s:DFT}
Bertini, De Sole, Gabrielli, Jona-Lasinio, and Landim \cite{Bertinietal01,Bertinietal02} re-derived the above mentioned large deviation functional by analyzing the model 
of fluctuating hydrodynamics. 
In \cite{Bertinietal01,Bertinietal02}, the large deviation functional of the density profile 
is obtained through the history minimizing an action functional for 
spontaneous creation of a fluctuation.
When one is concerned with equilibrium dynamics, which has the detailed balance property, such a task can be accomplished essentially within the Onsager-Machlup theory \cite{OnsagerMachlup53}.
In nonequilibrium dynamics, where the detailed balance condition is explicitly violated, a modified version of the Onsager-Machlup theory had to be devised to 
derive  a closed 
equation\footnote{
Unfortunately, it is likely that the equations for the large deviation functional can be solved exactly only in special cases,
the model treated by Derrida, Lebowitz, and Speer
being an example.
} for the large deviation functional \cite{Bertinietal01,Bertinietal02}.
By using the equation, the possible form
of the evolution of fluctuations was determined, and a generalized type of 
fluctuation dissipation relation for nonequilibrium steady states was proposed.

In \cite{Bertinietal01,Bertinietal02}, the form of fluctuating hydrodynamics
must be assumed or derived from other microscopic models. 
We believe that the SST free 
energy, if it really exists, should be taken into account in this step.
It would be quite interesting if the fluctuation dissipation 
relation that they proposed is related to the generalized second law of our SST.

\subsubsection{Thermodynamics beyond local equilibrium hypothesis}
\label{s:EIT}
There of course have been a number of attempts to formulate 
nonequilibrium thermodynamics that goes beyond local equilibrium 
treatment\footnote{
Landauer \cite{Landauer75,Landauer94} made a deep criticism to
 thermodynamics and statistical mechanics for nonequilibrium 
states in general. He argued, correctly, that one cannot expect 
to fully characterize a nonequilibrium state by simply 
minimizing a local function of states like the energy or the free energy.
The main point of his argument is that a coupling between two different 
subsystems can be much more delicate and trickier than we are used to 
in equilibrium physics.

We can assure that our SST is perfectly safe from Landauer's criticism.
First of all, we ourselves have encountered the delicateness of 
variational principle in nonequilibrium steady states, and this 
observation led us to the (almost) unique choice of nonequilibrium 
thermodynamic variables.  
This point will be discussed in section~\ref{s:var}.
Delicateness of contact is another issue that we ourselves have 
realized (with a surprise) during the development of SST.
In section~\ref{s:am}, we shall argue that the contact between 
an equilibrium state and a nonequilibrium steady state may be very delicate.
}.
A considerable amount of works appear under the name 
{\em extended irreversible thermodynamics}
\/ \cite{JouCasasVazquezLebon88,JouCasasVazquezLebon01}.

In extended irreversible thermodynamics, thermodynamic functions 
with extra variables for the ``degree of nonequilibrium'' are 
considered, and thermodynamic relations are discussed for various 
systems. This is quite similar to what we shall do in our own SST.

As far as we have understood, however, the philosophies behind 
extended irreversible thermodynamics and our SST are very much different.
In the literature of extended irreversible thermodynamics, we do not 
find anything corresponding to our careful (and lengthy) 
discussions about the convention 
of scaling, the identification of  intensive and extensive variables, 
the (almost) unique choice of nonequilibrium variables, the fully 
operational construction of the free energy, or the proof of Maxwell relation.
We also notice that, in many works in the extended irreversible 
thermodynamics, different levels of approaches, such as macroscopic 
phenomenology, microscopic kinetic theory, and statistical 
mechanics (such as the maximum entropy method) are discussed simultaneously.
In our own approach to SST, in contrast, we have tried to completely 
separate thermodynamics from microscopic considerations, stressing 
what outcome we get (and we do not get) from purely macroscopic phenomenology.

Although it is impossible to examine all the existing literature, 
it is very likely that more or less the same comments apply to other 
approaches in similar spirit. Examples include \cite{MullerRuggeri98,
Keizer87,Eu98}.

To make the comparison more concrete, let us take a look at two examples.

The same problem of sheared fluid that we have briefly seen in 
section~\ref{s:QT} is treated, for example, in \cite{Eu89a,Eu89b}.
Although the final conclusion \cite{Eu89b} that shear induces melting 
is the same, everything else is just different.
The discussion in \cite{Eu89a,Eu89b} are essentially model dependent, 
while we try to derive universal thermodynamic relations.
Moreover the proposed thermodynamics in \cite{Eu89a,Eu89b} uses 
the shear velocity $\Gamma$ (more precisely the shear rate 
$\dot \gamma=\Gamma/h$ where $h$ is the distance between the 
upper and the lower walls) as the nonequilibrium variable.
But one of our major conclusions in the present work is that 
a thermodynamics with the variable $\Gamma$ (or $\dot \gamma$) 
has a pathological behavior.
Thus the analysis of \cite{Eu89a,Eu89b} can never be consistent with our SST.
Indeed it is our opinion that the introduction of the nonequilibrium 
entropy in \cite{Eu88}, which gives a foundation to the above works, 
is not well-founded. Analysis of sheared fluids in 
\cite{EvansHanley80a,EvansHanley80b} looks sounder to us, but still 
does not contain careful steps as in SST.

In \cite{DominguezJou95} the pressure in a heat conducting state is discussed.
This again is in a sharp contrast between our own discussion of a 
similar problem.
The work in \cite{DominguezJou95} is based on a formula of the Shannon 
entropy obtained from microscopic theories (kinetic theory and the 
maximum entropy calculation). We see no reason that the Shannon entropy 
gives meaningful thermodynamic entropy once the system is away from 
equilibrium. 
Operational meaning of the pressure is also unclear.
A gedanken experiment is proposed, but there seems to be no way of 
realizing this setting (even in principle) unless one precisely 
knows in advance the formula for the nonequilibrium pressure.
Our discussion, in contrast, starts from a completely operational 
definition of the pressure.
We also predict a shift of pressure due to nonequilibrium effects 
in section~\ref{s:FIO}, but as a universal thermodynamic relation.
Let us note in passing that the maximum entropy calculation 
(called information theory), on which \cite{DominguezJou95} and 
other related works rely (see also \cite{CasasVazquezjou94}), 
is found to produce results which are inconsistent with the 
Boltzmann equation \cite{KimHayakawa03b}.
 
To conclude, our SST is completely different in essentially all the 
aspects from the extended irreversible thermodynamics and other 
similar approaches.
 The only similarity is in superficial formalism, i.e., thermodynamic 
functions with extra variables.
It is our belief that our own approach achieves much higher standard 
of logical rigor, and has a better chance of providing truly powerful 
and correct description of nature.

\section{Brief review of equilibrium physics}
\label{s:eq}
Before dealing with nonequilibrium problems, we present a very
brief summary of thermodynamics (section~\ref{s:td}), statistical mechanics (section~\ref{s:sm}), and approaches based on
stochastic processes (section~\ref{s:sp}) for equilibrium systems.
The main purpose of the present section is to fix some notations and 
terminologies used throughout the paper,
to give some necessary background, and, most importantly,
to motivate our approach to steady state thermodynamics (SST).

\subsection{Equilibrium thermodynamics}
\label{s:td}
Equilibrium thermodynamics is a universal theoretical framework which applies
{\em exactly\/} to arbitrary macroscopic systems in equilibrium.

Here we restrict ourselves to the formalism of
equilibrium thermodynamics at a fixed temperature, since it is directly related
 to our approach to steady state thermodynamics.
 See, for example, \cite{Callen85} for relations between different formalisms
 of thermodynamics\footnote{
The most beautiful formalism of thermodynamics uses energy variable
instead of temperature.
See, for example, \cite{LiebYngvason99}.
}.

\subsubsection{Equilibrium states and operations}
\label{s:td1}
A fluid consisting of a single substance of amount\footnote{
The amount of substance $N$
is sometimes called the ``molar number'' since $N$ is usually measured
in moles.
} $N$ is contained in a 
container with volume $V$ and kept in an environment with a fixed temperature
$T$.
If we leave the system in this situation for a sufficiently long time, it reaches an equilibrium
state, where no observable macroscopic changes take place.
An equilibrium state of this system is known to be
uniquely characterized (at least in the macroscopic scale)
by the three macroscopic parameters $T$, $V$, and $N$.
We can therefore denote the equilibrium state symbolically as $(T;V,N)$.
Note that we have separated the intensive variable $T$ and the
extensive variables $V$, $N$ by a semicolon.
This convention will be used throughout the present paper.

In thermodynamics, various operations to equilibrium states play essential roles.
Let us review them briefly.

By gently inserting a thin wall into an equilibrium state, one can
{\em decompose\/} the state into two separate
equilibrium states.
This is symbolically denoted as
\begin{equation}
(T;V,N)\to(T;V_1,N_1)+(T;V_2,N_2),
\label{e:eqdec}
\end{equation}
where\footnote{
Note that $V_1$, $V_2$, $N_1$, $N_2$ in the right-hand side are
not arbitrary.
If one fixes (for example) $V_1$, $V_2$, then $N_1$, $N_2$ are determined
almost uniquely.
}
$V_1+V_2=V$ and $N_1+N_2=N$.
One may realize the inverse of this operation by attaching the two equilibrium
states together and removing the wall between them.
Another important operation is to put two or more equilibrium states together,
separating them by walls which do not pass fluids but are thermally conducting.
In this way we get an equilibrium state characterized by a single temperature
$T$ and more than one pairs of $(V,N)$.

Given an arbitrary $\la>0$, one can associate with an equilibrium
state $(T;V,N)$ its {\em scaled copy\/} as
\begin{equation}
(T;V,N)\to(T;\la V,\la N).
\label{e:eqsc}
\end{equation}
The scaled copy has exactly the same properties as the original state,
but its size has been scaled.

The intensive variable $T$ and the extensive variables $V$, $N$
show completely different behaviors under the operations
\eqref{e:eqdec} and \eqref{e:eqsc}.
One may roughly interpret that an intensive variable characterizes
a certain property of the environment of the system,
while an extensive variable measures an amount of the system.

\subsubsection{Helmholtz free energy}
\label{s:td2}
The Helmholtz free energy (hereafter abbreviated as ``free energy'')
$F(T;V,N)$ is a special thermodynamic function which carries essentially 
all the information regarding the equilibrium state $(T;V,N)$.

The free energy $F(T;V,N)$ is {\em concave\/}
 in the intensive variable $T$,
and is jointly {\em convex\/}\footnote{
\label{fn:conv}
A function $g(V,N)$ is jointly convex in $(V,N)$ if
$g(\la V_1+(1-\la)V_2,\la N_1+(1-\la)N_2)
\le\la g(V_1,N_1)+(1-\la)g(V_2,N_2)$
for any $V_1$, $V_2$, $N_1$, $N_2$, and $0\le\la\le1$.
} in the extensive variables
$V$ and $N$.
Corresponding to the decomposition \eqref{e:eqdec},
it satisfies the {\em additivity\/}
\begin{equation}
F(T;V,N)=F(T;V_1,N_1)+F(T;V_2,N_2),
\label{e:eqFad}
\end{equation}
and corresponding to the scaling \eqref{e:eqsc}, the {\em extensivity}
\begin{equation}
F(T;\la V,\la N)=\la\,F(T;V,N).
\label{e:eqFex}
\end{equation}

The free energy $F(T;V,N)$ appears in the second law of thermodynamics
in the form of the minimum work principle.
Consider an arbitrary mechanical operation
to the system executed by an external (mechanical) agent, 
a typical (and important) example being a change of volume caused
by the motion of a wall.
We assume that the system is initially in the equilibrium state 
$(T;V,N)$, and the 
operation is done in an environment with a fixed 
temperature $T$.
Sufficiently long time after the operation, the system will settle down to 
another equilibrium state $(T;V',N)$.
Let $W$ be the total mechanical work done by the agent during the whole
operation.
Then the {\em minimum work principle\/} asserts that the inequality
\begin{equation}
W\ge F(T;V',N)-F(T;V,N)
\label{e:eqmwp}
\end{equation}
holds for an arbitrary operation.
Note that the operation need not be gentle or slow.

Another important physical relation involving the free energy is 
the following formula about fluctuation in equilibrium.
Suppose that we have two systems of the same volume $V$
which are in a weak contact with each other which
allows fluid to move from one system to another slowly.
If the total amount of fluid is $2N$, the amount of substance
in each subsystem should be equal to $N$ in average.
But there always is a small fluctuation in the amount of substance.
Let $\tilde{p}(N_1,N_2)$ be the probability density that
the amounts of substance in the two subsystems are $N_1$ and $N_2$.
Then it is known that in equilibrium this probability behaves as
\begin{equation}
\tilde{p}(N_1,N_2)\propto
\exp\sqbk{-\frac{1}{k_{\rm B}T}\{F(T;V,N_1)+F(T;V,N_2)\}},
\label{e:eqpN12}
\end{equation}
where $k_{\rm B}$ is the Boltzmann constant.
This is the isothermal version of  Einstein's celebrated formula of
fluctuation.
See, for example, chapter XII of \cite{LandauLifshitz80}.

\subsubsection{Variational principles and other quantities}
\label{s:td3}
Let $V_1$, $V_2$, $N_1$, and $N_2$ be those in
the decomposition \eqref{e:eqdec}.
Then from the extensivity \eqref{e:eqFex} and 
the convexity of $F(T;V,N)$, one can show the variational relation\footnote{
The derivation is standard, but let us describe it for completeness.
Let $N=N_1+N_2$ and $\la=N_1/N$, and take arbitrary
$V_1'$, $V_2'$ with
$V_1'+V_2'=V$.
Then from the extensivity \eqref{e:eqFex}
and the convexity (see footnote~\ref{fn:conv}), we get
$F(T;V_1',N_1)+F(T;V_2',N_2)
=
\la\,F(T;{V_1'}/{\la},N)
+
(1-\la)\,F(T;{V_2'}/{(1-\la)},N)
\ge
F(T;V,N)$.
With the additivity \eqref{e:eqFad}, this
implies a variational principle \eqref{e:eqvar1}.
}
\begin{equation}
F(T;V_1,N_1)+F(T;V_2,N_2)
=
\mintwo{V_1',V_2'}{(V_1'+V_2'=V)}
\{F(T;V_1',N_1)+F(T;V_2',N_2)\},
\label{e:eqvar1}
\end{equation}
which corresponds to the situation in which the system is divided 
by a movable wall into two
parts with fixed amounts $N_1$ and $N_2$ of fluids.
The volumes
$V_1'$ and $V_2'$ of the two parts 
can vary within the constraint $V_1'+V_2'=V$, and finally settle to the equilibrium values $V_1$ and $V_2$, respectively.
If we define the {\em pressure\/} by
\begin{equation}
p(T;V,N)=-\partialf{F(T;V,N)}{V},
\label{e:eqp}
\end{equation}
the variational relation \eqref{e:eqvar1} leads to the condition
\begin{equation}
p(T;V_1,N_1)=p(T;V_2,N_2),
\label{e:eqpp}
\end{equation}
which expresses the mechanical balance between the two
subsystems (that have volumes $V_1$ and $V_2$, respectively).
This thermodynamic pressure coincides with the
pressure defined in a purely mechanical manner\footnote{
More precisely, the free energy is defined so that to ensure this 
coincidence.
}.

One can derive the similar variational relation 
\begin{equation}
F(T;V_1,N_1)+F(T;V_2,N_2)
=
\mintwo{N_1',N_2'}{(N_1'+N_2'=N)}
\{F(T;V_1,N_1')+F(T;V_2,N_2')\},
\label{e:eqvar2a}
\end{equation}
for the situation where the system is divided into two parts with
fixed volumes $V_1$, $V_2$, and the amounts $N_1'$ and $N_2'$  in the 
two parts may vary within the constraint $N_1'+N_2'=N$.
This leads to another balance condition
\begin{equation}
\mu(T;V_1,N_1)=\mu(T;V_2,N_2),
\label{e:eqmm}
\end{equation}
where
\begin{equation}
\mu(T;V,N)=\partialf{F(T;V,N)}{N}
\label{e:eqm}
\end{equation}
is the {\em chemical potential\/}.

To get a better insight about the chemical potential,
and to motivate our main definition of the chemical potential for 
SST (see section~\ref{s:cp}),
suppose that we apply a potential which is equal to $u_1$
in the subsystem with volume $V_1$
and is equal to $u_2$ in that with volume $V_2$.
Since the addition of a uniform potential $u$ simply changes
the free energy $F(T;V,N)$  to  $F(T;V,N)+uN$ ,
the variational relation in this case becomes
\begin{eqnarray}
&&F(T;V_1,N_1)+u_1N_1+F(T;V_2,N_2)+u_2N_2
\ret
&&=
\mintwo{N_1',N_2'}{(N_1'+N_2'=N)}
\{F(T;V_1,N_1')+u_1N_1'+F(T;V_2,N_2')+u_2N_2'\}.
\label{e:eqvar2}
\end{eqnarray}
Then the corresponding balance condition becomes
\begin{equation}
\mu(T;V_1,N_1)+u_1=\mu(T;V_2,N_2)+u_2
\label{e:eqmumu}
\end{equation}
which clearly shows that $\mu(T;V,N)$  is a kind of potential\footnote{
Note that  $\mu(T;V,N)$  is here defined by \eqref{e:eqm}.
It is also useful to consider the ``electrochemical potential''
 $\tilde\mu_u(T;V,N)=\mu(T;V,N)+u$, but we do not use it here.  
}.

These examples illustrate a very important role played by intensive
quantities in thermodynamics,
which role will be crucial to our construction of SST.
Suppose in general that one has an extensive variable
($V$ or $N$ in the present case) in the parameterization
of states.
Also suppose that two states are in touch with each other,
and each of them are allowed to change this extensive
variable under the constraint (like $V_1'+V_2'=V$ or
$N_1'+N_2'=N$) that the sum of the extensive variables
is fixed.
Then there exists an intensive quantity
(like $p$ or $\mu$) which is {\em conjugate}\/
to the extensive variable in question, and the condition for 
the two states to balance with each other is represented
by the equality (like \eqref{e:eqpp} or \eqref{e:eqmm}) of the
intensive quantity.
The product of the original extensive variable and the
conjugate intensive variable always has the dimension
of energy.

Finally we write down some of the useful relations which involve 
the pressure and the chemical potential.
From the extensivity \eqref{e:eqFex} of the free energy,
one gets the Euler equation
\begin{equation}
F(T;V,N)=-V\,p(T;V,N)+N\,\mu(T;V,N).
\label{e:eqEu}
\end{equation}
From the definitions \eqref{e:eqp} and \eqref{e:eqm},
one gets
\begin{equation}
\partialf{p(T;V,N)}{N}=-\partialf{\mu(T;V,N)}{V},
\label{e:eqMax1}
\end{equation}
which is one of the Maxwell relations.
Since the pressure and the chemical potential are intensive\footnote{
$p(T;\la V,\la N)=p(T;V,N)$ and $\mu(T;\la V,\la N)=\mu(T;V,N)$
for any $\la>0$.
},
one may define
$p(T,\rho)=p(T;1,N/V)=p(T;V,N)$ and
$\mu(T,\rho)=\mu(T;1,N/V)=\mu(T;V,N)$ 
with $\rho=N/V$.
Then the Maxwell relation \eqref{e:eqMax1} becomes
\begin{equation}
\partialf{p(T,\rho)}{\rho}=\rho\partialf{\mu(T,\rho)}{\rho}.
\label{e:eqMax2}
\end{equation}

\subsection{Statistical mechanics}
\label{s:sm}
Suppose that we are able to describe a macroscopic physical
system using a microscopic dynamics\footnote{
The description may not be ultimately microscopic.
A necessary requirement is that one can write down a reasonable
microscopic Hamiltonian.
}.
Let $\calS$ be the set of all possible microscopic states
of the system.
For simplicity we assume that $\calS$ is a finite set.
The system is characterized by the Hamiltonian $H(\cdot)$,
which is a real valued function on $\calS$.
For a state $s\in\calS$, $H(s)$ represents its energy.

The essential assertion of equilibrium statistical mechanics is  that
macroscopic properties of an equilibrium state can be reproduced by
certain probabilistic models.
An important example of such probabilistic models is the {\em canonical
distribution\/} in which the probability of finding the system in a state
$s\in\calS$ is given by
\begin{equation}
p_{\rm eq}(s)=\frac{e^{-\beta\,H(s)}}{Z(\beta)},
\label{e:can}
\end{equation}
where $\beta=(k_{\rm B}T)^{-1}$ is the inverse temperature, and
\begin{equation}
Z(\beta)=\sum_{s\in\calS}e^{-\beta\,H(s)}
\label{e:Z}
\end{equation}
is the partition function.
Moreover if we define the free energy as
\begin{equation}
F(\beta)=-\frac{1}{\beta}\log Z(\beta),
\label{e:FZ}
\end{equation}
it satisfies all the static properties of the free energy in thermodynamics,
including the convexity, and the variation properties.

The formula \eqref{e:eqpN12} about density fluctuation
holds automatically in the canonical formalism.
Let us see the derivation.
Consider a system describing a fluid, and denote by $\calS_N$ the
state space when there are $N$ molecules in the system.
We define $Z_N(\beta)=\sum_{s\in\calS_N}e^{-\beta H(s)}$
and $F(\beta,N)=-(1/\beta)\log Z_N(\beta)$.
Consider a new system obtained by weakly coupling two identical copies
of the above system, and 
suppose that the total number of molecules is fixed to $2N$.
Then the probability of finding a pair of states $(s,s')$
with $s\in\calS_{N_1}$, $s'\in\calS_{N_2}$, and $N_1+N_2=2N$
is
\begin{equation}
p_{\rm eq}(s,s')=\frac{e^{-\beta\{H(s)+H(s')\}}}{Z_{\rm tot}(\beta)},
\label{e:peqss}
\end{equation}
where $Z_{\rm tot}(\beta)$ is the partition function of the whole system.
Then the probability $\tilde{p}(N_1,N_2)$ of finding $N_1$ and $N_2$
molecules in the first and the second systems, respectively, is
\begin{eqnarray}
\tilde{p}(N_1,N_2)
&=&
\sumtwo{s\in\calS_{N_1}}{s'\in\calS_{N_2}}
p_{\rm eq}(s,s')
\ret
&=&
\frac{Z_{N_1}(\beta)\,Z_{N_2}(\beta)}{Z_{\rm tot}(\beta)}
\ret
&=&
\exp[-\beta\{F(\beta,N_1)+F(\beta,N_2)-F_{\rm tot}(\beta)\}],
\label{e:pN12der}
\end{eqnarray}
which is nothing but the desired formula \eqref{e:eqpN12}.

\subsection{Markov processes}
\label{s:sp}
Although statistical mechanics reproduces static aspects of thermodynamics,
it does not deal with dynamic properties such as the approach to equilibrium and
the second law.
To investigate these points from microscopic (deterministic)
dynamics is indeed a very difficult problem, whose understanding
is still poor.
For some of the known results, see, for example,
 \cite{PuszWoronowicz78,Lenard78,Tasaki98}.
If we become less ambitious and start from effective stochastic models,
then we have rather satisfactory understanding of these points.

\subsubsection{Definition of a general Markov process}
\label{s:sp1}
Again let $\calS$ be the set of all  microscopic states in a physical system.
A Markov process on $\calS$ is defined by specifying transition rates
$c(s\to s')\ge0$ for all $s\ne s'\in\calS$.
The transition rate $c(s\to s')$ is the rate (i.e., the probability divided by 
the time span) that the system changes its state from $s$ to $s'$.

Let $p_t(s)$ be the probability distribution at time $t$.
Then its time evolution is governed by the {\em master equation\/},
\begin{equation}
\frac{d}{dt}p_t(s)=\sumtwo{s'\in\calS}{(s'\ne s)}
\{-c(s\to s')\,p_t(s)+c(s'\to s)\,p_t(s')\},
\label{e:maeq}
\end{equation}
for any $s\in\calS$.

A Markov process is said to be {\em ergodic\/} if all the states are
``connected'' by nonvanishing transition rates\footnote{
More precisely, for any $s,s'\in\calS$, one can find a finite sequence
$s_1, s_2,\ldots,s_n\in\calS$ such that 
$s_1=s$, $s_n=s'$, and $c(s_{j}\to s_{j+1})\ne0$
for any $j=1,2,\ldots,n-1$.
}.
In an ergodic Markov process, it is known that the probability
distribution $p_t(s)$ with an arbitrary initial condition converges to a 
unique stationary distribution $p_\infty(s)>0$.
Then from the master equation \eqref{e:maeq}, one finds that the
stationary distribution is characterized by the equation
\begin{equation}
\sumtwo{s'\in\calS}{(s'\ne s)}
\{-c(s\to s')\,p_\infty(s)+c(s'\to s)\,p_\infty(s')\}=0,
\label{e:std}
\end{equation}
for any $s\in\calS$.
See, for example, \cite{Feller68}.

\subsubsection{Detailed balance condition}
\label{s:sp2}
The convergence to a unique stationary distribution suggests that,
if one wishes to model a dynamics around equilibrium, 
one should build a model so that the stationary distribution 
$p_\infty(s)$ coincides with the canonical distribution
$p_{\rm eq}(s)$ of \eqref{e:can}.
A sufficient (but far from being necessary) condition 
for this to be the case is that the
transition rates satisfy
\begin{equation}
c(s\to s')\,p_{\rm eq}(s)=c(s'\to s)\,p_{\rm eq}(s')
\label{e:db}
\end{equation}
for an arbitrary pair $s\ne s'\in\calS$.
By substituting \eqref{e:db} into \eqref{e:std},
one finds that each summand vanishes and 
\eqref{e:std} is indeed satisfied
with $p_\infty(s)=p_{\rm eq}(s)$.
The equality \eqref{e:db} is called the {\em detailed balance condition\/}
with respect to the distribution $p_{\rm eq}(s)$.

Today one always assumes the detailed balance condition
\eqref{e:db} when studying dynamics around equilibrium 
using a Markov process.
This convention is based on a deep reason, which was originally
pointed out by Onsager \cite{Onsager31a,Onsager31b}, 
that such models automatically satisfy macroscopic symmetry
known as ``reciprocity.''
See section~\ref{s:Ons}.

By substituting the formula \eqref{e:can} of the canonical distribution 
into the condition \eqref{e:db}, it is rewritten as
\begin{equation}
\frac{c(s\to s')}{c(s'\to s)}=\exp[\beta\{H(s)-H(s')\}],
\label{e:db2}
\end{equation}
for any $s\ne s'$ such that $c(s\to s')\ne 0$.
Usually the condition \eqref{e:db2} is also called the detailed
balance condition.
A standard example of transition rates satisfying \eqref{e:db2} is
\begin{equation}
c(s\to s')=a(s,s')\,\phi(\beta\{H(s')-H(s)\}),
\label{e:cae}
\end{equation}
where $a(s,s')=a(s',s)\ge0$ are arbitrary weights which ensure
the ergodicity\footnote{
A simple choice is to set $a(s,s')=1$ if $s'$ can be ``directly reached'' from $s$,
and $a(s,s')=0$ otherwise.
}, and $\phi(h)$ is a function which satisfies
\begin{equation}
\phi(h)=e^{-h}\,\phi(-h),
\label{e:phicond}
\end{equation}
for any $h$.
The standard choices of $\phi(h)$ are 
i)~the {\em exponential rule}\/ with $\phi(h)=e^{-h/2}$, ii)~the {\em heat bath (or Kawasaki) rule}\/ with $\phi(h)=(1+e^h)^{-1}$, and iii)~the {\em Metropolis rule}\/ with $\phi(h)=1$ if $h\le0$ and $\phi(h)=e^{-h}$ if $h\ge0$.
In equilibrium dynamics, these (and other) rules can be used rather arbitrarily depending on one's taste.
But it has been realized these days \cite{Tasaki04a,LefevereTasaki04} that the choice of rule crucially modifies the nature of the stationary state if one considers  nonequilibrium dynamics.
Indeed we will see a drastic example in section~\ref{s:pert} in the Appendix.

\subsubsection{The second law}
\label{s:sp3}
To study the minimum work principle \eqref{e:eqmwp}, we must 
theoretically formulate
mechanical operations by an outside agent.
When the agent moves a wall of the container, she is essentially changing the
potential energy profile for the fluid molecules.
We therefore consider a Hamiltonian $H^{(\alpha)}(s)$ with an additional control
parameter $\alpha$, and let $c^{(\alpha)}(s\to s')$ be transition rates whose
stationary distribution is the canonical distribution\footnote{
As an example, one replaces $H(s)$ in \eqref{e:cae} with  $H^{(\alpha)}(s)$.
}
\begin{equation}
p^{(\alpha)}_{\rm eq}(s)=\frac{1}{Z^{(\alpha)}(\beta)}
\exp[-\beta\,H^{(\alpha)}(s)].
\label{e:peqa}
\end{equation}

Suppose that the agent changes
this parameter according to a prefixed (arbitrary) function\footnote{
Here we are not including any feedback from the system to the agent.
(The agent does what she had decided to do, whatever the reaction 
of the system is.)
To include the effects of feedback seems to be a highly nontrivial
problem.
}
$\alpha(t)$ with $0\le t\le t_{\rm f}$.
($t_{\rm f}$ is the time at which the operation ends.)
Since the Hamiltonian $H^{(\alpha(t))}(s)$ is now time-dependent, 
the transition rates $c^{(\alpha(t))}(s\to s')$ also 
become time-dependent.

To mimic the situation in thermodynamics, we assume that,
at time $t=0$, the probability distribution coincides with the equilibrium
state for $\alpha=\alpha(0)$, i.e., we set $p_0(s)=p^{(\alpha(0))}_{\rm eq}(s)$.
The probability distribution $p_t(s)$ for $0\le t\le t_{\rm f}$
is the solution of the time-dependent master equation
\begin{equation}
\frac{d}{dt}p_t(s)=\sumtwo{s'\in\calS}{(s'\ne s)}
\{-c^{(\alpha(t))}(s\to s')\,p_t(s)+c^{(\alpha(t))}(s'\to s)\,p_t(s')\},
\label{e:tdme}
\end{equation}
which is simply obtained by substituting the time-dependent
transition rates into the master equation \eqref{e:maeq}.
Let us denote the average over the distribution $p_t(s)$ as
\begin{equation}
\bkt{g(s)}_t=\sum_{s\in\calS}g(s)\,p_t(s),
\label{e:gavt}
\end{equation}
where $g(s)$ is an arbitrary function on $\calS$.

Now, in a general time-dependent Markov process,
a theorem sometimes called the ``second law'' is known\footnote{
According to \cite{Kubo81}, such a theorem was first proved by Yosida.
See XIII-3 of \cite{Yosida68}.
}.
We shall describe it carefully in the Appendix~\ref{s:M2nd}.
The theorem readily applies to the present situation,
where the key quantity defined in \eqref{e:phia} becomes
\begin{equation}
\varphi^{(\alpha)}(s)=
-\log p^{(\alpha)}_{\rm eq}(s)=
\beta\{H^{(\alpha)}(s)-F(\beta,\alpha)\},
\label{e:phieq}
\end{equation}
with
\begin{equation}
F(\beta,\alpha)=-\frac{1}{\beta}\log\sum_{s\in\calS}
\exp[-\beta\,H^{(\alpha)}(s)]
\label{e:Fba}
\end{equation}
being the free energy with the parameter $\alpha$.
Then the basic inequality \eqref{e:M2nd} implies that,
for any differentiable function $\alpha(t)$, one has
\begin{equation}
\int_0^{t_{\rm f}}dt\,
\frac{d\alpha(t)}{dt}
\bkt{\left.
\frac{d}{d\alpha}H_\alpha(s)
\right|_{\alpha=\alpha(t)}
}_t
\ge
F(\beta,\alpha(t_{\rm f}))-F(\beta,\alpha(0)).
\label{e:spmwp}
\end{equation}

Let us claim that the left-hand side of \eqref{e:spmwp} is
precisely the total mechanical work done by the external agent.
Consider a change from time $t$ to $t+\Delta t$,
where $\Delta t$ is small.
The change of the Hamiltonian is 
$\Delta H(s)=H^{\alpha(t+\Delta t)}(s)-H^{\alpha(t)}(s)$.
Since the agent directly modifies the Hamiltonian,
the work done by the agent between $t$ and $t+\Delta t$  is
equal to\footnote{
Note that this is different from 
$\bkt{H^{\alpha(t+\Delta t)}(s)}_{t+\Delta t}-\bkt{H^{\alpha(t)}(s)}_t$.
The difference is nothing but the energy exchanged as ``heat.''
}
$\bkt{\Delta H(s)}_t+O((\Delta t)^2)$.
By summing this up (and letting $\Delta t\to0$),
we get the left-hand side of \eqref{e:spmwp}.
With this interpretation, the general inequality \eqref{e:spmwp} is nothing
but the minimum work principle \eqref{e:eqmwp}.

\section{Nonequilibrium steady states and local steady states}
\label{s:ss}
Equilibrium thermodynamics, equilibrium statistical mechanics, 
and Markov process description
of equilibrium dynamics, which we reviewed briefly in section~\ref{s:eq}
are {\em universal\/} theoretical frameworks that apply to
equilibrium states of arbitrary
macroscopic physical systems.
As we have discussed in section~\ref{s:intro},
our goal in the present paper is to construct such a universal thermodynamics
that applies to nonequilibrium steady states.

In the present section, we shall make clear the class of systems
that we study, and describe their nonequilibrium steady states (section~\ref{s:nss}).
We then discuss the important notion of {\em local steady state\/} (section~\ref{s:lss}).

\subsection{Nonequilibrium steady states}
\label{s:nss}
A macroscopic physical system is in a {\em nonequilibrium steady state\/}
if it shows no macroscopically observable changes while constantly
exchanging energy with the environment.

Although our aim is to construct a universally applicable theory,
it is useful (or even necessary) to work in concrete settings.
Let us describe typical examples that we shall study in the present paper.

\subsubsection{Heat conduction}
\label{s:hc}
\begin{figure}
\centerline{\epsfig{file=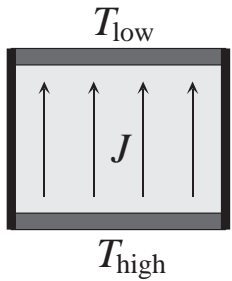,width=3cm}}
\caption[dummy]{
The upper and the lower walls have temperatures 
$T_{\rm low}$ and $T_{\rm high}$, respectively.
In the nonequilibrium steady state, the fluid in the container carries 
a steady heat current in the vertical direction.
We assume that there is no convection.
}
\label{f:hc}
\end{figure}
The first example is heat conduction in a fluid.
Suppose that a fluid consisting of a single substance is contained in a cylindrical
container as in Fig.~\ref{f:hc}.
The upper and the lower walls of the cylinder are kept at constant 
temperatures $T_{\rm low}$ and $T_{\rm high}$, respectively,
with the aid of external heat baths.
The side walls of the container are perfectly adiabatic.

If the system is kept in this setting for a sufficiently long time,
it will finally reach a steady state without any macroscopically 
observable changes.
We assume that convection does not take place,
so there is no net flow in the fluid.
But there is a constant heat current from the lower wall to the upper wall,
which constantly carries energy from one heat bath to the other.
This is a typical nonequilibrium steady state.

\subsubsection{Shear flow}
\label{s:sf}
\begin{figure}
\centerline{\epsfig{file=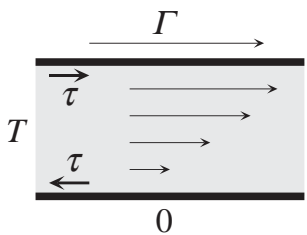,width=4cm}}
\caption[dummy]{
There is a fluid  between two ``sticky'' horizontal walls.
The upper wall moves with a constant speed $\Gamma$
while the lower wall is at rest.
In the nonequilibrium steady state, the fluid develops a velocity
gradient.
The forces that the upper and the lower walls exert on fluid are 
exactly opposite with each other.
}
\label{f:sf}
\end{figure}
The second example is a fluid under shear.
Consider a fluid in a box shaped container whose upper and lower walls are
made of a ``sticky'' material.
The upper wall moves with a constant speed $\Gamma$ while the lower wall is at rest\footnote{
One should device a proper geometry (periodic boundary conditions) to make it possible for the upper wall to keep on moving.
}.
We assume that the fluid is in touch with a heat bath at constant temperature $T$.
Since the moving wall does a positive work on the fluid,
the fluid must constantly throw energy away to the bath in order not to heat up.

If we keep the system in this setting for a sufficiently long time, it finally
reaches a steady state in which the fluid moves horizontally with varying speeds
as in Fig.~\ref{f:sf}.
The wall constantly injects energy into the system as a mechanical work while
the fluid releases energy to the heat bath.
This is another typical nonequilibrium steady state.

Since the fluid gets no acceleration in a steady state, the total 
force exerted on the whole fluid must be vanishing.
This means that the forces that the upper and the lower walls exert
on the fluid are exactly opposite with each other.
The same argument, when applied to an arbitrary region in the fluid,
leads to the well-known fact that the shear stress, defined as 
the flux of horizontal momentum in the vertical direction,  is constant 
everywhere in the sheared fluid.

\subsubsection{Electrical conduction in a fluid}
\label{s:ec}
\begin{figure}
\centerline{\epsfig{file=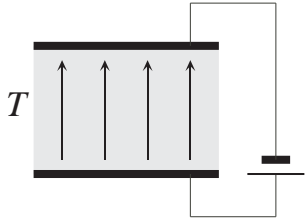,width=4cm}}
\caption[dummy]{
An electrically conducting fluid attached to a heat bath
is put in a uniform electric field.
In a steady state one has a constant electric current.
Joule heat generated in the fluid is absorbed by the heat bath.
}
\label{f:ec}
\end{figure}
The third example is electrical conduction in a fluid as in Fig.~\ref{f:ec}.
When a constant electric field is applied to a conducting fluid which is in touch with
a heat bath at a constant temperature, 
there appears a steady electric current.
It should be noted that the electric field does not generate particle flow in the fluid, but only generates a flow of electric carriers\footnote{
There must be a mechanism to move the carrier from one plate to the other so that
to maintain a steady current.
When the carrier is electron, this is simply done by using a battery as in Fig.~\ref{f:ec}.
}.
Since a normal conductor always generates Joule heat, 
there is a constant flow of energy to the heat bath.
This is also a typical nonequilibrium steady state.

\subsection{Local steady state}
\label{s:lss}
Let us discuss the notion of {\em local steady state\/}
which is central to our study.

To be concrete let us concentrate on the case of heat conduction
in a fluid (section~\ref{s:hc}).
In general the local temperature and the local density of the fluid vary
continuously as functions of the position\footnote{
Although the density is defined unambiguously in any situation,
the definition of temperature is much more delicate.
Here we simply assume that the local temperature can be measured by 
a small thermometer.
We will discuss more about the definition of temperature in 
section~\ref{s:faqt}.
}.
If one looks at a sufficiently small portion of the fluid, however,
both the temperature $T$ and the density $\rho$ are essentially constants.

In the standard treatment of weakly nonequilibrium systems
(see section~\ref{s:NETD}), one assumes that
the state within the small portion can be regarded as the equilibrium state
with the same $T$ and $\rho$.
Then the whole nonequilibrium state with varying temperature and density
is constructed by ``patching'' together these {\em local steady states\/}.

In general situations where the system is not necessarily close to equilibrium,
however, this treatment is not sufficient.
No matter how small the portion may be, there always exists a finite heat flux
going through it.
Therefore the local state in this small portion cannot be isotropic.
Since equilibrium states are always isotropic in a fluid, 
this means that the local state
cannot be treated as a local equilibrium state.
It should be treated rather as a {\em local steady state\/}.

A local steady state is in general anisotropic.
It is characterized by the temperature $T$, the density $\rho$, and (at least)
one additional parameter (which we do not yet specify) which measures the
``degree of nonequilibrium.''
Macroscopic quantities of the heat conducting fluid, such as the pressure,
viscosity, and heat conductivity, should in principle depend not only on 
$T$ and $\rho$ but also on the additional nonequilibrium parameter.
The main goal of our work is to present a thermodynamics
that applies to local steady states\footnote{
As a next step, one wishes to see how these local steady states can be
``patched'' together to form a global nonequilibrium steady state.
We hope this will be a topic of our future works.
}.


\subsection{Realization of local steady states}
\label{s:rlss}
As a next step we discuss how one can realize local steady states in 
each of the concrete examples.

\subsubsection{Heat conduction}
\label{s:rhc}
\begin{figure}
\centerline{\epsfig{file=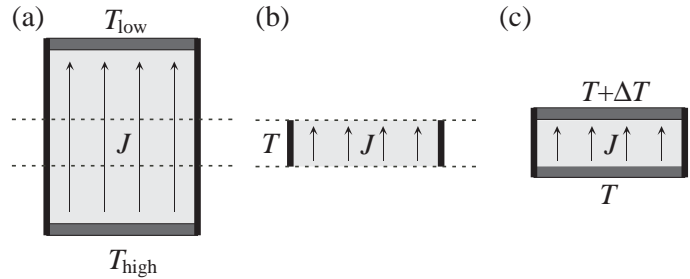,width=9cm}}
\caption[dummy]{
Local steady state for heat conduction.
(a)~There is a temperature gradient and heat flux in the vertical direction.
(b)~If one concentrates on a thin region of the system, the temperature $T$
and the density are essentially constant.
This is a local steady state.
(c)~We further assume that the same local steady state can be realized
in a thin system by adjusting the temperature of the upper and the lower
walls.
}
\label{f:hcl}
\end{figure}
Consider again heat conduction in a fluid.
Suppose that the system has a nice symmetry and we get a steady state
which is transitionally invariant in the horizontal directions.
The heat flux flows in the vertical direction as in Fig.~\ref{f:hcl}~(a).
We let the heat flux $J$ be the total amount of heat that passes through
an arbitrary horizontal plane in the fluid within a unit time.
Note that the heat flux $J$ is independent of the choice of the plane
because of the energy conservation.

Take a region in the fluid in between two (fictitious) horizontal
planes.
If the width of the region is sufficiently small, the temperature 
and the density in the region may
be regarded as constant.
This thin system realizes a local steady state for heat conduction
as in Fig.~\ref{f:hcl}~(b).

Suppose that one inserts into the fluid a horizontal
wall with very efficient thermal
conductivity and negligible thickness.
Since there is no macroscopic flow of fluid to begin with, and
the temperature is constant on any horizontal plane,
it is expected that the insertion of the wall does not cause
any macroscopically observable changes\footnote{
This statement is not as obvious as it first seems.
In reality there often appears a seemingly discontinuous temperature jump
between a fluid and a wall.
Our assumption relies on an expectation that this jump can be made
negligibly small by using a wall made of a suitable material
with a suitable surface condition.
}.
Then one can replace the two (fictitious) planes that determine
the thin region with two conducting horizontal walls 
without making any macroscopic changes.
Moreover, by connecting the two walls to heat baths with precise
temperatures, one can ``cut out'' the thin region from the rest
of the system as in Fig.~\ref{f:hcl}~(c). 
In this manner we can realize a local steady state
in an isolated form.

\subsubsection{Shear flow}
\label{s:rsf}
In the case of sheared fluid (section~\ref{s:sf}) identification of
a local steady state is (at least conceptually) much easier.
If the contact with the heat bath is efficient enough,
one may regard that the whole system has a uniform temperature.
If this is the case, the state of the whole system is itself a local steady state.

When the temperature difference within the fluid is not negligible,
one may again focus on a thin region to get a local steady state.
The technique of inserting thin walls can be used in this situation as well.
We here use a sticky wall with a negligible width and insert it horizontally
in such a way that it has precisely the same velocity as the fluid around it.
We can then isolate a local steady state\footnote{
When there is a temperature gradient, one gets a local steady state with a heat current as well as a shear.
We here assume that the latter has a dominant effect.
}.

\subsubsection{Electrical conduction in a fluid}
\label{s:rec}
The case of electrical conduction in a fluid (section~\ref{s:ec})
can be treated in a similar manner as the previous two examples.
If there are variations in the temperature or the density, we again 
restrict ourselves to a thin horizontal region to get a local steady state.
When electrons carry current,
an electrically conducting thin wall with a precisely fixed electric potential
may be inserted to the fluid without changing macroscopic behavior.

\section{Basic framework of steady state thermodynamics}
\label{s:bf}
As a first step of the construction of steady state thermodynamics (SST),
we carefully examine basic operations to local steady states (section~\ref{s:bo}).
Then we discuss how we should choose nonequilibrium thermodynamic
variables (section~\ref{s:var}).
To make the discussions concrete, we first restrict ourselves to the case
of heat conduction.
Other cases are treated separately (sections~\ref{s:bsf} and \ref{s:bec}).

\subsection{Operations to local steady states}
\label{s:bo}
As we saw in section~\ref{s:td}, various operations
(i.e., decomposition, combination, and scaling) on equilibrium states
are essential building blocks of equilibrium thermodynamics.
We shall now examine how these operations should be generalized to
nonequilibrium steady states.
This is not at all a trivial task since 
nonequilibrium steady states are inevitably anisotropic,
and there is a steady flow of energy going through it.

\begin{figure}
\centerline{\epsfig{file=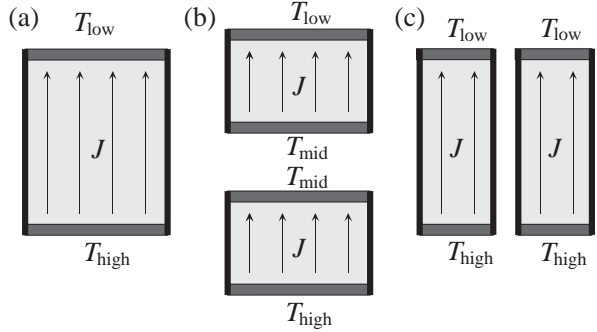,width=8cm}}
\caption[dummy]{
Two possible ways (b), (c) to decompose a heat conducting steady state (a).
Both the ways are theoretically sensible.
}
\label{f:hdec}
\end{figure}

We examine the case of steady heat conduction in a fluid.
In order to find general structures of steady states, 
we examine a heat conducting state between the temperatures $T_{\rm low}$
and $T_{\rm high}$ (Fig.~\ref{f:hdec}~(a)).
We still do not take the limit of local steady states.

There are two natural (and theoretical sensible) ways of decomposing the 
steady state.
In the first way, one inserts a thin horizontal wall with efficient heat conduction
as in section~\ref{s:rhc}.
Then one measures the temperature of the wall (which we call $T_{\rm mid}$)
and attach the wall to a heat bath with the same temperature $T_{\rm mid}$.
We expect that these procedures do not cause any macroscopically observable
changes.
Finally one splits the middle wall into two, and gets
the situation in Fig.~\ref{f:hdec}~(b), where one has two steady states.
In the second way, which is much more straightforward, one simply inserts
a thin adiabatic wall vertically to split the system into two as in Fig.~\ref{f:hdec}~(c).
One can of course revert these procedures, and combine the two states to 
get the original one.

\begin{figure}
\centerline{\epsfig{file=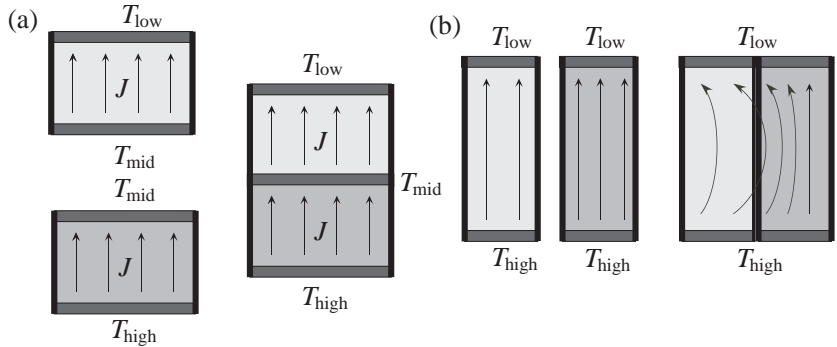,width=11cm}}
\caption[dummy]{
Two possible ways of combining two heat conducting states which
have different densities.
(a)~Contact through a horizontal wall does not change the two states as long
as the temperatures at the attached walls (denoted as $T_{\rm mid}$) coincide and the heat flux $J$ in the two states are identical.
(b)~Contact through a conducting vertical wall may inevitably lead to 
a modification of the heat flow pattern.
We are therefore led to consider only the vertical combination/decomposition
scheme (a).
}
\label{f:hcom}
\end{figure}

We next examine how one should combine two heat conducting states
which have different densities (or which contain different kinds of fluids).
One natural way is a combination in the vertical direction.
We prepare two heat conducting states 
between $T_{\rm low}$ and $T_{\rm mid}$ and between
$T_{\rm mid}$ and $T_{\rm high}$.
The two systems have the same horizontal cross sections.
We then attach the two walls with the temperature  $T_{\rm mid}$ 
together as in Fig.~\ref{f:hcom}~(a).
If the two states have exactly the same heat flux $J$,
there is no heat flow between the middle wall and the
heat bath with $T_{\rm mid}$.
This means that we can simply disconnect this heat bath without
making any changes to the combined steady states.
This way of combining two steady states always works provided that
the temperatures at the attached walls are the same (which is $T_{\rm mid}$) 
and the heat flux $J$ in the two states are identical with each other.
We can regard this as a natural extension of the combination of two
states frequently used in equilibrium thermodynamics.

As in the decomposition scheme, we can also think about combinations
in the horizontal direction.
We can put two heat conducting states together along a vertical
heat conducting wall as in Fig.~\ref{f:hcom}~(b).
This combination scheme too may look reasonable at first glance.
But note that the contact always modifies the heat flux pattern
unless the vertical temperature profiles of the two states before
the contact are exactly identical.
Since two different fluids (or fluid in two different densities) 
generally develop different (nonlinear) temperature profiles,
we must conclude that in general this horizontal contact 
modifies the two states.
It therefore cannot be used as a combination scheme in
thermodynamics\footnote{
If the temperature profiles are always linear, then one can say that the 
two profiles with the same terminal temperatures are identical.
One might think this is always the case in local steady states realized
in very thin systems.
But we point out that, no matter how thin a system may be, there can
be a phase coexistence in it, which leads to a nonlinear temperature 
profile.
Therefore the horizontal combination scheme in heat conduction may 
be useful only when one (i)~restricts oneself to local steady states,
and (ii)~rules out the possibility of phase coexistence.
We still do not know if we can construct a meaningful thermodynamics
starting from this observation.
}.

In conclusion, the decomposition/combination in the vertical direction
(in which one separates a system, or puts two systems together 
along a horizontal plane)
works in any situation, while that in the horizontal direction
is less robust.
The advantage of the former scheme is that it relies only on a
conservation law that is independent of thermodynamics.
More precisely the constancy of the heat flux $J$ is guaranteed by the energy conservation law and the steadiness of the states.
We are therefore led to a conclusion that, in nonequilibrium steady states
for heat conduction, the decomposition and combination of states should
be done in the vertical direction using horizontal planes,
keeping the horizontal cross section constant.
See, again, Figs.~\ref{f:hdec}~(b) and \ref{f:hcom}~(a).

For a local steady state, which is defined on a sufficiently thin system,
one can define a scaling operation.
Following the decomposition/combination scheme,
we shall fix the cross section of the system and scale only in the vertical 
direction\footnote{
The scaling factor $\la$ should not bee too large to keep the state 
a local steady state.
}.
When doing this we must carefully chose the (small)
temperature difference so that the heat flux $J$ is kept constant.
See Fig.~\ref{f:sc}.

\subsection{Choice of nonequilibrium variables}
\label{s:var}
\begin{figure}
\centerline{\epsfig{file=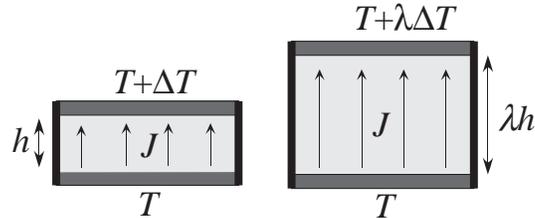,width=7cm}}
\caption[dummy]{
Scaling in a local steady state of a heat conducting fluid.
We fix the horizontal cross section and scale only in the vertical
direction by a factor $\la>0$.
The height $ h$ (and hence the volume), the amount of substance $N$,
and the (small) temperature difference $\deltaT$ are scaled by $\la$,
while the temperature $T$ and the heat flux $J$ are unchanged.
}
\label{f:sc}
\end{figure}
We now turn to the problem of describing local steady states
in a qualitative manner.
The main issue here is how one should choose a new thermodynamic variable
representing the ``degree of nonequilibrium.''
Rather surprisingly, we will see that,
by assuming that a reasonable thermodynamics exists,
we can determine the nonequilibrium variable almost uniquely.

Let us again use the heat conduction as an example, and take a sufficiently
thin system to realize a local steady state.
To characterize the local steady state, we definitely need the temperature
$T$, the volume $V$, and the amount of substance $N$.
Note that we only need a single temperature $T$ since a local steady state
has an essentially constant temperature.
In addition to these three variables, we need a
``nonequilibrium variable'' as we discussed in section~\ref{s:lss}.

When choosing the nonequilibrium variable,  we first postulate that the variable
should correspond to a physically ``natural'' quantity.
Then, in a local steady state for heat conduction, there are essentially
two candidates.
One is the heat flux $J$, which is the total energy that passes through any
horizontal plane within a unit time.
The other is the temperature difference $\deltaT$ between the upper and
the lower walls\footnote{
We have assumed that the temperature in a local steady state is essentially
constant.
But there must be a nonvanishing temperature difference $\deltaT$ to maintain the heat
conduction.
Of course we have $\deltaT\ll T$.
}.

To see the characters of these nonequilibrium variables,
we examine the scaling transformation of the local steady state
(Fig.~\ref{f:sc}).
When the system is scaled by a factor $\la>0$ in the vertical direction,
the extensive variables $V$ and $N$ are scaled to become
$\la V$ and $\la N$, respectively,
while the intensive variable $T$ is unchanged.
The heat flux $J$ is is unchanged because we want to keep the
local state unchanged.
(More formally speaking,  it is our convention,
which followed almost inevitably from the considerations
in section~\ref{s:bo},
to keep $J$
constant when extending the system in the vertical direction.)
The temperature difference $\deltaT$, on the other hand, must be scaled
to $\la \deltaT$ in order to maintain the same heat flux.
Therefore, within our convention of scaling, 
the nonequilibrium variable $J$ acts as an intensive variable
while $\deltaT$ acts as an extensive variable.

Thus our parameterization of a local steady state can either be\footnote{
Recall the notation to separate intensive and extensive variables by
a semicolon.
See section~\ref{s:td1}.
}
$(T,J;V,N)$ or $(T;V,N,\deltaT)$.
We wish to examine whether we can get consistent thermodynamics
by using these parameterizations.

\begin{figure}
\centerline{\epsfig{file=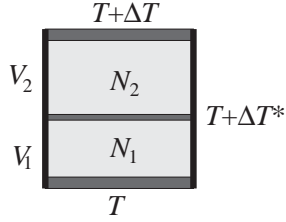,width=4cm}}
\caption[dummy]{
A local steady state with heat conduction is separated into 
lower and upper parts (with the volumes and the 
amounts of fluid equal to $V_1$, $V_2$, and 
$N_1$, $N_2$, respectively)
by a thermally conducting 
horizontal wall.
The temperatures $T$ and $T+\deltaT$ of the lower and the upper walls,
respectively, are fixed.
We want to determine
the temperature $T+\deltaT^*$ at the separating wall.
}
\label{f:tau}
\end{figure}

We first argue that a thermodynamics with the
parameterization $(T;V,N,\deltaT)$ is inconsistent, or, to say the least,
not useful.
To demonstrate this we consider the situation in 
Fig.~\ref{f:tau}, where a single fluid of volume $V$ is separated into 
two parts by a horizontal wall.
The volumes and the 
amounts of fluid in the lower and upper parts are 
fixed to
 $V_1$, $V_2$, and 
$N_1$, $N_2$, respectively.
The separating wall is thermally conducting.
The upper and the lower walls are fixed, and have fixed temperatures $T+\deltaT$
and $T$, respectively, where $\deltaT\ll T$.
The temperature $T+\deltaT^*$ at the separating wall is not fixed, but should
be uniquely determined in the steady state.

Let us note that the situation here is completely analogous to
those we have  seen in section~\ref{s:td3}.
(See, in particular, the discussion at the end of the section.)
Here the temperature difference in the lower part
(i.e., the difference between the temperatures of the lower wall
and the separating wall)
is $\deltaT_1=\deltaT^*$, while that in the upper part is
$\deltaT_2=\deltaT-\deltaT^*$.
Since $\deltaT$ is fixed, their sum $\deltaT_1+\deltaT_2$ is fixed.
Let us assume that the general structure of thermodynamics
is maintained here.
Then there should be an intensive quantity
$\nu(T;V,N,\deltaT)$ which is conjugate to $\deltaT$, and 
$\deltaT^*$ should be determined by the balance condition
\begin{equation}
\nu(T;V_1,N_1,\deltaT^*)=\nu(T;V_2,N_2,\deltaT-\deltaT^*).
\label{e:taununu}
\end{equation}

We know, on the other hand, that the balance of heat flux
between two subsystems can be universally
expressed by the equality
\begin{equation}
J(T;V_1,N_1,\deltaT^*)=J(T;V_2,N_2,\deltaT-\deltaT^*),
\label{e:tauJJ}
\end{equation}
where $J(T;V,N,\deltaT)$ is the heat flux written as a function of
$T$, $V$, $N$, and $\deltaT$.
Since the two conditions \eqref{e:taununu} and \eqref{e:tauJJ}
should be equivalent for each $T$,
there must be a function $g(J,T)$ such that
\begin{equation}
\nu(T;V,N,\deltaT)=g(J(T;V,N,\deltaT),T),
\label{e:nug}
\end{equation}
for any $T$, $V$, $N$, and $\deltaT$.
But note here that the quantity $\nu$
(which is conjugate to $\deltaT$) is dimensionless\footnote{
Our convention is that temperature has the dimension of energy.
}
while the heat flux $J$ has the dimension of energy divided by time.
This means that the function $g(J,T)$ must include (at least) one universal
constant which has the dimension of time\footnote{
For example, $g(J,T)=t_0J/T$ is dimensionless when $t_0$ is a constant
with the dimension of time.
}.
But it is quite unlikely (if not impossible) that a theory requires such a new
universal constant.
We must conclude that a thermodynamics with the parameterization
$(T;V,N,\deltaT)$ of local steady states 
(even if it exists) is highly unnatural.

On the other hand, a thermodynamics with the parameterization
$(T,J;V,N)$ does not suffer from such a pathology.
Here the ``degree of nonequilibrium'' is directly taken into account
through the intensive variable $J$, which directly accounts for the
energy conservation law.
Since the role of an intensive variable in thermodynamics
is to set the environment for the system, there is no room for 
internal inconsistencies to appear.

We are therefore led to the conclusion that {\em 
a heat conducting local steady state 
should be parameterized as $(T,J;V,N)$,
where the nonequilibrium variable $J$ is intensive}\/\footnote{
There is a possibility that a natural thermodynamic variable is a function
$\psi(J)$ rather than $J$ itself.
See footnote~\ref{fn:psi} in section~\ref{s:ftn}.
}.

The general lesson is as follows.
To get a sensible thermodynamics, {\em the ``degree of nonequilibrium''
should be taken into account through an intensive variable which
manifestly represents a conservation law that is imposed by physical
laws out of thermodynamics\/}.
An extensive nonequilibrium variable which looks natural from a physical point of view
may not be natural for a thermodynamic theory.
This is because the corresponding variational principle may conflict 
(or may become redundant) with the already existing conservation law.

\subsection{Shear flow}
\label{s:bsf}
Let us examine how the discussions in sections~\ref{s:bo} and \ref{s:var}
should be extended to local steady states of a sheared fluid.

Let us start from the decomposition/combination scheme.
If the fluid is two-dimensional (as in Fig.~\ref{f:sf}),
then it is obvious that decomposition and combination must be done
in the vertical direction along a horizontal line.
Decomposition of combination in other directions are simply 
impossible because of the horizontal flow.
In a three dimensional fluid, one might first imagine that 
decomposition in other directions\footnote{
For example one can think about cutting
the system into two along a plane
parallel to the page.
} can be used, but it turns out that it is not possible when there
is a phase separation within the system.
Anyway it is most natural to stick on a scheme which does not
depend on the dimensionality.
We shall always decompose or combine systems in
the vertical direction, along horizontal planes.

\begin{figure}
\centerline{\epsfig{file=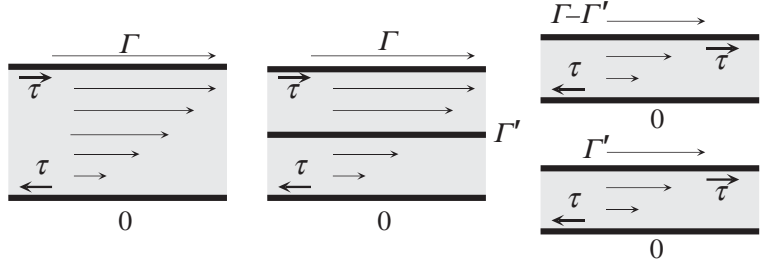,width=10cm}}
\caption[dummy]{
Decomposition of a steady state of sheared fluid.
One inserts a thin horizontal wall with the precise velocity 
$\Gamma'$ as in (b).
By splitting the two parts and Galilei transforming the upper system,
one gets the situation in (c).
Note that the shear force $\tau$ is preserved in this decomposition.
When combining two states, one starts from (c) and goes back to (b).
}
\label{f:sfd}
\end{figure}

To decompose a steady state, one first inserts a thin sticky wall horizontally
in such a way that the velocity of the wall is identical to that of fluid around it.
We expect that this insertion does not modify the state macroscopically.
Then one decomposes the inserted wall into two, and splits the system into two parts.
Note that the shear force $\tau$ is preserved in this process.
If one wishes to bring both the states into the standard form (where the lower
wall is at rest), one performs a Galilei transformation to the upper system 
as in Fig.~\ref{f:sfd}.

When one combines two states into one, one first prepares two steady states
(as in Fig.~\ref{f:sfd}~(c)) with the identical shear force $\tau$,
and then combine them (after a Galilei transformation)
as in Fig.~\ref{f:sfd}~(b).

It is then obvious that the shear velocity $\Gamma$ (which is the difference
between the velocities of the upper and the lower walls) is the extensive
variable, and the shear force $\tau$ is the intensive variable.
From an argument parallel to that in section~\ref{s:var}, we find that a thermodynamics
with the parameterization $(T;V,N,\Gamma)$ is inappropriate.
We shall parameterize local steady states as $(T,\tau;V,N)$,
and look for a useful thermodynamics.

\subsection{Electrical conduction in a fluid}
\label{s:bec}
Treatment of steady states in an electrically conducting fluid is somewhat more
delicate than the other two examples.
It seems that, in some situations, there are {\em two\/} completely different
formulations of 
thermodynamics, whose relations are far from obvious (and not yet clear to us).

\begin{figure}
\centerline{\epsfig{file=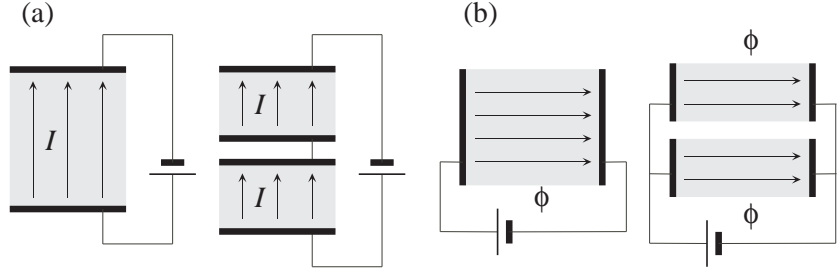,width=11cm}}
\caption[dummy]{
Two possible schemes of decomposition/combination in a 
steady state of fluid with electric current.
(a)~The first scheme is the same as that for heat conduction if we 
replace the heat flux $J$ with the total electric current $I$.
The charge conservation law ensures that the total current $I$ is a 
``good'' intensive nonequilibrium variable.
(b)~The second scheme makes use of the fact that the electric potential
difference $\phi$ is constant when one splits or combines systems along 
a plane parallel to the current.
Then $\phi$ becomes the intensive nonequilibrium variable.
}
\label{f:ecd}
\end{figure}

Obviously it is possible to develop a scheme completely parallel to that for
heat conduction.
For this, one simply identifies the total electric current $I$ with the heat flux $J$.
Then one decomposes, combines, and scales local steady states
in the direction of the current as in Fig.~\ref{f:ecd}~(a).
In the corresponding thermodynamics, the total electric current $I$
becomes the intensive nonequilibrium variable.
Local steady states are parameterized as $(T,I;V,N)$.

But this is not the only possible thermodynamics.
If one can neglect the effect of charge screening at the walls, then the
difference $\phi$ in the electric potential between the two walls 
is constant.
The potential $\phi$ is unchanged if one decomposes the system along a plane
parallel to the electric current as in Fig.~\ref{f:ecd}~(b).
We can use the corresponding combination and scaling convention to construct
a thermodynamics, where the electric potential difference $\phi$ becomes the
nonequilibrium intensive variable.
Local steady states are parameterized as $(T,\phi;V,N)$.

It is quite interesting that  there are two different formulations
of thermodynamics for a single physical system.
The two are truly different theories since they are based on different
schemes of scaling.
One must note that the two possible theories can never be related with each other
simply, for example, by a Legendre transformation.

It is an intriguing problem to find the true meaning of the (apparent) possibility of 
the two formulations of thermodynamics.
In particular it is exciting to find an ``unifying'' theory which includes the physics
of both the theories, provided that both the theories are found to be physically
meaningful\footnote{
A less interesting possibility is that only one of the two is physically meaningful.
(And much less interesting possibility is that none of them are.)
}.

\section{Operational determination of thermodynamic quantities}
\label{s:op}
In section~\ref{s:bf}, we have determined the basic structure of steady state
thermodynamics (SST), assuming that a sensible thermodynamics for
local steady states does exist.
We now turn to the task of determining thermodynamic quantities.
In doing so, we insist on defining everything through operational procedures 
which are (at least in principle) experimentally realizable.

In the present section, we discuss how one can determine the pressure
(section~\ref{s:pr})
and the chemical potential (section~\ref{s:cp}) of a local steady state $(T,\nu;V,N)$,
and further show that these quantities satisfy the Maxwell relation (section~\ref{s:mr}).
Here the variable $\nu$ represents a general nonequilibrium intensive
variable, and should be read as $J$, $\tau$, $I$, or $\phi$ depending
on the model that one has in mind.
Although the pressure $p(T,\nu;V,N)$ is fully determined,
we here determine only the $V$, $N$ dependence of the chemical
potential $\mu(T,\nu;V,N)$.
Determination of $T$, $\nu$ dependence will be discussed later in
section~\ref{s:cpf}.

The construction in the present section is fairly general and does not
depend on specific systems.
The convention is that when we say ``vertical'' direction, it means the
direction to which we scale our systems.
The reader should simply have in mind, for examples, 
Figs.~\ref{f:sc}, \ref{f:sfd}, and \ref{f:ecd}, and interpret the
word  ``vertical'' in the ordinary sense.
Recall that we can think about two different formulations of 
thermodynamics for electric
conduction in a fluid (section~\ref{s:bec}).

\subsection{Pressure --- a mechanical definition}
\label{s:pr}
\begin{figure}
\centerline{\epsfig{file=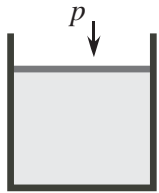,width=2cm}}
\caption[dummy]{
To be consistent with our convention of scaling,
we measure the (mechanical) pressure in the vertical
direction and identify it as the SST pressure
$p(T,\nu;V,N)$.
}
\label{f:pr}
\end{figure}
Pressure is a special thermodynamic quantity, which also has a purely mechanical
characterization\footnote{
We can say that equilibrium thermodynamics establishes 
quantitative connection with mechanics through the pressure
and the internal energy.
In our construction of SST, we make use of the pressure.
But it is extremely
difficult (if not impossible) to introduce the notion of internal energy
to nonequilibrium steady states since there always is a constant
flow of energy.
Oono and Paniconi \cite{OonoPaniconi98} indeed tried to define 
internal energy of SST by extending the notion of adiabaticy.
In our construction, we do not make use of adiabaticy.
}.

As usual the pressure of a local steady state is defined as the mechanical
pressure or through the work needed to make a small change of volume.
The only point we have to be careful about is that a nonequilibrium steady state
is anisotropic.
Since it is our convention to perform decomposition, combination,
and scaling only in the vertical direction, we shall only speak about
the pressure in the vertical direction.
It is obtained as the mechanical pressure exerted on the horizontal walls, 
or through the relation $\Delta W=p\,\Delta V$ where $\Delta W$
is the mechanical work needed to make a small volume change $\Delta V$
by moving a horizontal wall vertically.
See Fig.~\ref{f:pr}.

We denote the pressure thus determined as $p(T,\nu;V,N)$.
Since the vertical force should not change when one scales the system 
in the vertical direction, we have the intensivity
\begin{equation}
p(T,\nu;\la V,\la N)=p(T,\nu;V,N),
\label{e:pint}
\end{equation}
for any $\la>0$ (which is small enough to maintain local steadiness).
Note that in the case of electrical conduction in a fluid,
two pressures $p(T,I;V,N)$ and $p(T,\phi;V,N)$ corresponding
to the two different formulations of 
thermodynamics are in general different.

As in equilibrium,
 we expect the pressure $p(T,\nu;V,N)$ to be nonincreasing in
$V$.
Otherwise a small fluctuation in the volume may be magnified, leading to
a catastrophic volume change.

\subsection{Potential variation method and chemical potential}
\label{s:cp}
\begin{figure}
\centerline{\epsfig{file=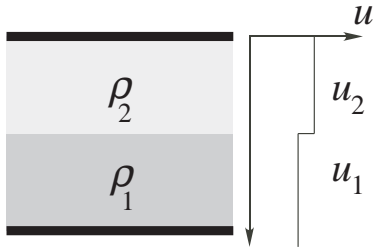,width=5cm}}
\caption[dummy]{
The method of potential variation which determines the difference
in chemical potentials.
An external potential which varies only in the vertical direction is
applied.
The densities of the fluid is $\rho_1$ and $\rho_2$ in the lower and
the upper regions where the potential takes constant values
$u_1$ and $u_2$, respectively.
Then the chemical potential difference is determined by
$\mu(\rho_1)+u_1=\mu(\rho_2)+u_2$.
}
\label{f:uv}
\end{figure}
In contrast to the pressure,
the chemical potential is a purely thermodynamic
quantity.
Therefore to define it for local steady states is a highly nontrivial
problem.
In the present section, we introduce the 
{\em method of potential variation\/},
which enables one to determine the difference in the chemical 
potential unambiguously.
By using this method, we determine $V$, $N$ dependence 
of the chemical potential in a purely operational manner.
To our knowledge this method of determining chemical potential
was first pointed out and used
by Hayashi and Sasa \cite{HayashiSasa03}.

Fix the temperature $T$ and the nonequilibrium variable $\nu$.
We denote by $\mu(\rho)=\mu(T,\nu;V,N)$ the chemical potential as a function of the density $\rho=N/V$.
We have assumed the intensivity
\begin{equation}
\mu(T,\nu;\la V,\la N)=\mu(T,\nu;V,N),
\label{e:muint}
\end{equation}
for any $\la>0$.

Now we apply an external potential which 
generates a force\footnote{
In the case of electrically conducting fluid, the present force is {\em not}\/ an electric one.
We have assumed that an electric field only affects the electric carriers, while the present potential couples with the whole fluid.
} acting on fluid particles.
The potential is equal to 
a constant $u_1$ in the lower half of the system, and is equal to
$u_2$ in the upper half.
We assume that the potential varies smoothly in the vertical direction between the two regions.

Let the system relax to its steady state under the applied potential.
We denote by $\rho_1$ and $\rho_2$ the densities of the fluid
in the lower and the upper regions, respectively.
Since the two regions can freely exchange fluid molecules and are
in  steady balance with each other, the chemical potential
$\mu(\rho)$ (if exists) should satisfy the balance relation
\begin{equation}
\mu(\rho_1)+u_1=\mu(\rho_2)+u_2,
\label{e:mumu}
\end{equation}
which is nothing but the nonequilibrium counterpart of the 
relation \eqref{e:eqmumu} in equilibrium thermodynamics.
Note that we are again speaking only about contacts of two regions
in the vertical direction.
As in the definition of the pressure, all the notions should be
consistent with our convention of scaling.

To be more logical, we are here defining the chemical potential
$\mu(\rho)$ as a quantity that satisfies the condition
\eqref{e:mumu}, under the assumption that \eqref{e:mumu} for various
$u_1$ and $u_2$ lead consistently to a  function $\mu(\rho)$.

It seems obvious that $\rho_1\ge\rho_2$ when $u_1\le u_2$, 
since the fluid feels a
downward force in the boundary of the two regions.
Then \eqref{e:mumu} implies that 
the chemical potential $\mu(\rho)$ is a nondecreasing function
of the density $\rho$.

\subsection{Maxwell relation}
\label{s:mr}
A crucial point in our operational definitions of the pressure and the
chemical potential is that they automatically imply the Maxwell relation
\begin{equation}
\partialf{p(\rho)}{\rho}=\rho\partialf{\mu(\rho)}{\rho},
\label{e:Max}
\end{equation}
which is exactly the same as its equilibrium counterpart \eqref{e:eqMax2}.
Here we still fix $T$, $\nu$, and write $p(\rho)=p(T,\nu;V,N)$
with $\rho=N/V$.

To show the Maxwell relation \eqref{e:Max}, we again consider the 
situation in Fig.~\ref{f:uv}, where a varying potential is applied to
a single fluid.
Suppose that $\Delta u=u_2-u_1>0$ is small.
Let us write densities as $\rho_1=\rho$ and $\rho_2=\rho-\Delta\rho$.
By writing the density $\rho(\rb)$ and the potential $u(\rb)$ as
a function of the position $\bf r$, we can evaluate the total
force exerted on the fluid from the potential as
\begin{equation}
F_u=-\int d^3\rb\,\rho(\rb)\,{\rm grad}\,u(\rb)
=-\{\rho+O(\Delta\rho)\}\int d^3\rb\,{\rm grad}\,u(\rb)
=-\{\rho+O(\Delta\rho)\}A\,\Delta u,
\label{e:Fu}
\end{equation}
where $A$ is the cross section area of the container, and 
we used the fact that $\rho(\rb)=\rho+O(\Delta\rho)$
everywhere in the fluid.
On the other hand, the forces exerted on the fluid from the 
upper and lower walls (as pressures) add up to
\begin{equation}
F_p=A\{p(\rho)-p(\rho-\Delta\rho)\}.
\label{e:Fp}
\end{equation}
Since the fluid is in a steady state, we must have
$F_u+F_p=0$, which leads to
\begin{equation}
p(\rho)-p(\rho-\Delta\rho)=\rho\,\Delta u+O((\Delta\rho)^2).
\label{e:ppr}
\end{equation}
From \eqref{e:mumu}, on the other hand, we have
\begin{equation}
\mu(\rho)-\mu(\rho-\Delta\rho)=\Delta u.
\label{e:mudu}
\end{equation}
From \eqref{e:ppr} and \eqref{e:mudu}, one readily gets
the desired Maxwell relation \eqref{e:Max}
by letting $\Delta u\to0$.

\section{SST free energy and its possible roles}
\label{s:f}
Since we have discussed the operational definitions of the pressure
$p(T,\nu;V,N)$ and the chemical potential $\mu(T,\nu;V,N)$,
we can now move on to the definition of the SST (Helmholtz)
free energy $F(T,\nu;V,N)$.
After noting the basic properties of the free energy (section~\ref{s:fd}), we discuss two
conjectures about its physical roles, namely, the extension of Einstein's formula for the density fluctuation (section~\ref{s:df}) and the minimum work principle (section~\ref{s:mw}).

As in section~\ref{s:op}, discussions in the present section are
fairly general, and apply to any of our examples.

\subsection{Definition of free energy}
\label{s:fd}
Let $(T,\nu;V,N)$ be a local steady state.
Since $T$ and $\nu$ are intensive, and $V$ and $N$ are extensive,
the extensivity of the free energy (to be defined) should read
\begin{equation}
F(T,\nu;\la\,V,\la\,N)=\la\,F(T,\nu;V,N),
\label{e:Fex}
\end{equation}
for any $\la>0$ (which is small enough to maintain local steadiness).
Since this is identical to its equilibrium counterpart \eqref{e:eqFex},
we expect our SST free energy to satisfy the same Euler equation
\eqref{e:eqEu} as the equilibrium free energy.
We therefore {\em require}
\begin{equation}
F(T,\nu;V,N)=-V\,p(T,\nu;V,N)+N\,\mu(T,\nu;V,N).
\label{e:Eu}
\end{equation}
Note that this is consistent with the
extensivity \eqref{e:Fex} of free energy and the
intensivity \eqref{e:pint} and \eqref{e:muint} of the pressure and the
chemical potential.

Since we have already determined $p(T,\nu;V,N)$ and (the $V$, $N$ dependence of)
$\mu(T,\nu;V,N)$, we can regard  the Euler equation \eqref{e:Eu} as our definition
of the free energy.
This determines the $V$, $N$ dependence of $F(T,\nu;V,N)$ completely
for each fixed $(T,\nu)$.
Dependence of $F(T,\nu;V,N)$ on $T$ and $\nu$ will be discussed later
in section~\ref{s:ftn}.

From the Maxwell relation \eqref{e:Max} (which should better be rewritten
in the form of \eqref{e:eqMax1}), we find that the free energy
defined by \eqref{e:Eu} satisfies
\begin{equation}
p(T,\nu;V,N)=-\partialf{F(T,\nu;V,N)}{V},
\label{e:pF}
\end{equation}
and
\begin{equation}
\mu(T,\nu;V,N)=\partialf{F(T,\nu;V,N)}{N},
\label{e:muF}
\end{equation}
as in the equilibrium thermodynamics (see \eqref{e:eqp} and \eqref{e:eqm}).

We can also show that $F(T,\nu;V,N)$ is jointly convex\footnote{
See footnote~\ref{fn:conv} in section~\ref{s:td2} for the definition.
The following discussion is quite standard.
} in the extensive variables $V$ and $N$.
To see this, fix $T$ and $\nu$ and define the specific free energy by
\begin{equation}
f(\rho)=F(T,\nu;1,\frac{N}{V})=\frac{F(T,\nu;V,N)}{V},
\label{e:fdef}
\end{equation}
with $\rho=N/V$.
Then the Euler equation \eqref{e:Eu} becomes
\begin{equation}
f(\rho)=-p(\rho)+\rho\,\mu(\rho).
\label{e:fEu}
\end{equation}
From this and the Maxwell relation \eqref{e:Max}, we find
$f''(\rho)=\mu'(\rho)$.
Since we have $\mu'(\rho)\ge0$ as argued in section~\ref{s:cp},
$f(\rho)$ is convex in $\rho$, and hence
\begin{equation}
f(\kappa\frac{N_1}{V_1}+(1-\kappa)\frac{N_2}{V_2})
\le
\kappa\,f(\frac{N_1}{V_1})+(1-\kappa)\,f(\frac{N_2}{V_2}),
\label{e:fk}
\end{equation}
for any $V_1$, $V_2$, $N_1$, $N_2$, and $0\le\kappa\le1$.
By setting $\kappa=\la\,V_1/\{\la\,V_1+(1-\la)\,V_2\}$
with $0\le\la\le1$,
\eqref{e:fk} becomes
\begin{equation}
F(T,\nu;\la\,V_1+(1-\la)\,V_2,\la\,N_1+(1-\la)\,N_2)
\le
\la\,F(T,\nu;V_1,N_1)+(1-\la)\,F(T,\nu;V_2,N_2),
\label{e:FVN}
\end{equation}
which shows the desired convexity.

\subsection{Density fluctuation}
\label{s:df}
Although we still do not know $T$, $\nu$ dependence of the
free energy $F(T,\nu;V,N)$, its dependence on $V$ and $N$
may give interesting physical information.
In the present and the next subsections, we discuss two of such conjectures.

\begin{figure}
\centerline{\epsfig{file=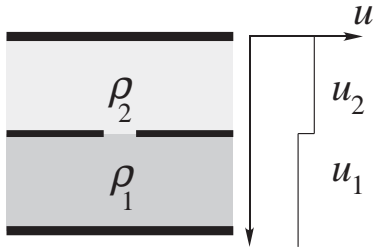,width=5cm}}
\caption[dummy]{
The setting for discussion the density fluctuation formula \eqref{e:pN12} for nonequilibrium steady states.
The situation is almost the same as that of Fig.~\ref{f:uv}, but the two parts are now separated by a horizontal wall with a small window in it.
}
\label{f:dfw}
\end{figure}

Consider the situation in Fig.~\ref{f:dfw}, where the potential which is equal to $u_1$  in the lower half and equal to $u_2$ in the upper half is applied.
This is almost the same as the situation in Fig.~\ref{f:uv}, but we separate the two parts by a horizontal wall with a small (but much larger from the molecular scale) window in it.
The wall is prepared so that the two parts maintain the same ``degree of nonequilibrium.''
Since the two parts will reach mechanical and thermodynamic balance via the window, the densities  $\rho_1$ and $\rho_2$ in the steady state is the same as that attained in the situation of Fig.~\ref{f:uv}.

In the average, the amounts of fluid in the two parts are equal to $\rho_1 V$ and $\rho_2 V$, respectively, where $V$ is the volume of each part.
As long as the volume is finite, however, there exists a fluctuation in the amounts of fluid or in the densities.
We conjecture that {\em this density fluctuation can be described by
the SST free energy}\/, just as in the equilibrium case \eqref{e:eqpN12}.
(See also \eqref{e:pN12der}.)
To our knowledge, such a fluctuation relation in steady states was first 
proposed by Hayashi and Sasa \cite{HayashiSasa03}.

To be precise, denote the
amounts of substance in the lower and the upper parts as
$N_1$ and $N_2$, respectively.
Since the fluid can move through the porous wall, $N_1$ and $N_2$
may vary while $N_1+N_2$ is exactly conserved.
We denote by $\tilde{p}(N_1,N_2)$ be the probability
density for the partition $(N_1,N_2)$ of the amounts of fluid.
Then, for each steady state with fixed $T$, $\nu$ $u_1$, and $u_2$, it is expected
that the Einstein's formula
\begin{equation}
\tilde{p}(N_1,N_2)
\simeq
{\rm const.}\,\exp[-\beta\{F(T,\nu;V,N_1)+F(T,\nu;V,N_2)+u_1\,N_1+u_2\,N_2\}],
\label{e:pN12}
\end{equation}
is valid, where $\beta=1/(k_{\rm B}T)$ is the inverse temperature.

We shall later show that the fluctuation formula \eqref{e:pN12}
holds exactly in the sheared fluid with weak contact (Appendix~\ref{s:WCSF}), and the driven lattice gas (Appendix~\ref{s:dfw}).

It is quite important to note that the conjectured relation
\eqref{e:pN12} 
can be checked by experiments which are independent of 
those necessary to determine the free energy $F(T,\nu;V,N)$.
Therefore we are proposing a highly nontrivial statement for 
general (not necessarily weak) nonequilibrium systems that can be verified (or falsified)
purely by experiments.
See section~\ref{s:exfl} for further discussions.

\bigskip\noindent
{\em Remark 1:}
When the density fluctuation is governed by the Einstein's formula \eqref{e:pN12}, it is likely that linear response relations for transport or relaxation phenomena in the {\em vertical}\/ direction are also valid.
For example, if one starts from the steady state with $u_1=u_2=0$ and suddenly applies at time $t=0$ a weak potential difference $\Delta u=u_2-u_1$, we expect the fluctuation response relation
\begin{equation}
\sbkt{\hat{N}_1(t)}_{\Delta u}=\sbkt{\hat{N}_1}_0+
\Delta u\,\beta\,\sbkt{\hat{N}_1\,\{\hat{N}_1(t)-\hat{N}_1(0)\}}_0
+O((\Delta u)^2),
\label{e:FRRgen}
\end{equation}
for $t\ge0$ to be valid.
Here $\sbkt{\cdots}_0$ is the expectation in the steady state with $u_1=u_2=0$.
See section~\ref{s:WCSL} and section~\ref{s:DLGLR} for more details in concrete examples.
To our knowledge, such linear response relations in a highly nonequilibrium steady state was first discussed in \cite{Hayashi05}.

\bigskip\noindent
{\em Remark 2:}
One might naively expect the fluctuation relation   \eqref{e:pN12} to hold in the situation of Fig.~\ref{f:uv}, where the two parts are separated by a  a {\em fictitious}\/ horizontal plane.
Although the corresponding relation in equilibrium holds either when the wall is real or fictitious, the situation may not be that simple in nonequilibrium steady states.
In  nonequilibrium steady states, one generally finds spatial long range correlations, which lead to anomalous density fluctuations \cite{DorfmanKirkpatrickSengers94}.
(See section~\ref{s:faqlrc}.)
It is likely that a simple Einstein type relation for fluctuation 
does not hold because of  this anomaly.
The contact via a small window considered above is expected to diminish the long range correlation between the two subsystems, while maintaining macroscopic balance of the two systems.
This is clearly seen in our treatment of the driven lattice gas in Appendix~\ref{s:mpq}.
See also Appendix~\ref{s:WCS} for still ``safer'' design of the contact.

\subsection{Minimum work principle}
\label{s:mw}
We also conjecture that there exists a steady state version of 
the minimum work principle (see section~\ref{s:td2})
and that the SST free energy plays a central role in it.

Let us discuss only the simplest version in the present section.
Later in Appendix~\ref{s:mwp} we shall discuss more about the minimum work principle 
in the contexts of driven lattice gas.

We conjecture that a straightforward generalization of 
the inequality \eqref{e:eqmwp} in equilibrium holds in a local
steady state.
However there is a sever restriction on mechanical operations performed
to the system.
To be consistent with our basic framework to change the geometry of the system
only in the vertical direction, we allow the external agent to change the volume
of the system only by moving the upper (or the lower) wall vertically.
We require that the intensive variables $T$ and $\nu$ are kept constant
during the operation.
Then the conjectured minimum work principle for local steady states is
\begin{equation}
W\ge F(T,\nu;V',N)-F(T,\nu;V,N),
\label{e:mwpV}
\end{equation}
where $V$ and $V'$ are the initial and the final volumes, respectively,
and $W$ is the mechanical work that the outside agent has done to the system.
It is crucial here that $W$ is the ordinary mechanical work, not an exotic 
(and often ill-defined) quantity like ``nonequilibrium work
\footnote{
In the framework of SST proposed by Oono and Paniconi
\cite{OonoPaniconi98}, the conjectured minimum work principle
is expressed in terms of  ``excess work''  which is obtained by
subtracting a ``house-keeping heat'' from the total mechanical work. 
Such a decomposition of the work with a generalized second law
was demonstrated  in a Langevin model \cite{HatanoSasa01}. 
Quite recently, an identity leading to this generalized second law
was tested experimentally \cite{Trepagnieretal04}. 
 }.''
The outside agent need not care whether the system is in an equilibrium 
state or in a steady state.

In the limit where the operation is indefinitely slow,
the relation \eqref{e:pF} for the pressure implies that the
minimum work principle \eqref{e:mwpV} is satisfied as an equality.
(This is indeed almost the definition of the pressure.)
Then the conjectured inequality \eqref{e:mwpV} looks plausible
since we usually have to do extra work when an operation is not 
slow enough for the fluid to follow.

We stress that the restriction on allowed operations is essential.
Since some of the nonequilibrium steady states have a macroscopic flow,
the agent may make use of it to reduce her work (or even to get a positive
amount of energy from, say, a waterwheel) if  arbitrary operations are allowed.
This is in a stark contrast with the
minimum work principle in the equilibrium thermodynamics, where the agent 
can perform any mechanical operations allowed by physics laws.

To summarize, our message is that {\em the minimum work principle may be
extended to  general nonequilibrium steady states provided that
one carefully restricts allowed operations}\/.
We stress that this too is a highly nontrivial conjecture that can be checked
purely by experiments.

\section{Steady state thermodynamics (SST) in a complete form}
\label{s:fu}
In the present section, we discuss an operational method to determine 
the dependence of the chemical potential on $T$ and $\nu$.
The basic idea depends on a new postulate that there are walls
(that we call $\mu$-walls) which realize a natural contact between
an equilibrium state
and a steady state (sections~\ref{s:muw} and \ref{s:cpf}).
This determines the free energy $F(T,\nu;V,N)$ (section~\ref{s:ftn}), and completes our
construction of steady state thermodynamics (SST).

The complete SST leads us to predictions of two new phenomena, namely,
the flux-induced osmosis (FIO) (sections~\ref{s:FIO} and \ref{s:am}), and a shift of coexistence temperature (section~\ref{s:sc}).
We show that both the phenomena are described by the new
nonequilibrium extensive quantity $\Psi(T,\nu;V,N)$ that we
shall introduce.
These two are intrinsically nonequilibrium phenomena that can never be
described within the equilibrium or the local equilibrium treatments.

\subsection{Contact of a steady state and an equilibrium state}
\label{s:muw}
As in sections~\ref{s:op} and \ref{s:f}, we treat a general local steady state
parameterized as $(T,\nu;V,N)$.

\begin{figure}
\centerline{\epsfig{file=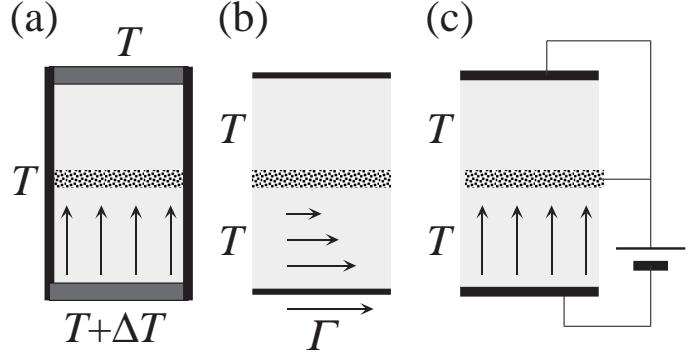,width=9cm}}
\caption[dummy]{
Combination of a nonequilibrium steady state (lower part)
and an equilibrium state (upper part) in
(a)~heat conducting fluid, (b)~sheared fluid, and 
(c)~electrical conduction in a fluid.
The separating walls are made of a porous material, and fluid can pass to
the other sides through narrow complicated paths.
This setting will be the basis for our determination of the chemical
potential and the free energy.
}
\label{f:cse}
\end{figure}

We shall now think about bringing a local steady state $(T,\nu;V,N)$
in contact with an equilibrium state $(T,0;V',N')$, allowing the two
states to slowly exchange the fluid.
To be consistent with the general scheme of SST, we put the two states together
in the vertical direction, separating them with a horizontal wall.

The separating wall should probably be made of a porous material which has many
narrow complicated paths through which the fluid can pass slowly.

In the case of a heat conducting fluid (section~\ref{s:hc}), we suppose that the
porous wall has very high heat conductivity, and is in touch with a heat bath
which has the same temperature $T$ as the upper most wall.
See Fig.~\ref{f:cse}~(a).
The lower most wall is kept at a different temperature $T+\deltaT$.
In this manner, we can realize a heat conducting (local) steady state in the
lower half of the system, and an equilibrium state with temperature $T$
in the upper half of the system.

In the case of a sheared fluid (section~\ref{s:sf}), we suppose that 
porous wall has ``sticky'' surfaces and is at rest.
See Fig.~\ref{f:cse}~(b).
The upper most wall is also at rest, while the lower most wall moves with
the speed $\Gamma$.
In this manner, we realize a (local) steady state with a constant shear
in the lower half, and an equilibrium state in the upper half.
We of course assume that the whole system is in an efficient contact 
with a heat bath at temperature $T$.

As for electrical conduction in a fluid
(section~\ref{s:ec}), we have to be careful.
When we employ the $(T,\phi;V,N)$ formalism of SST
(see Fig.~\ref{f:ecd}~(b)),
it is impossible to put an equilibrium state on top of a
steady state.
This is because a steady state in this setting has a uniform electric
field in the horizontal direction in it, while
an equilibrium state has no electric field.
Such a configuration is inhibited by the law of electrostatics, i.e.,
${\rm rot}\,{\bf E}=0$.
We therefore exclude the $(T,\phi;V,N)$ formalism from our considerations
in the present section.

As for the $(T,I;V,N)$ formalism of SST for electrical conduction in a fluid
(see Fig.~\ref{f:ecd}~(a)), the contact causes no apparent problems.
We suppose that the porous wall is electrically conducting, and let the 
porous wall and the upper most wall have the same electric potential.
See Fig.~\ref{f:cse}~(c).
By letting the the lower most wall have a different potential, we will get
a steady state with a constant electric current in the lower half,
and an equilibrium state in the upper half.
Again the whole system is assumed to be in touch
with a heat bath at temperature $T$.

\subsection{Complete determination of the chemical potential}
\label{s:cpf}
Our new (and fundamental) postulate is that, if a steady state
$(T,\nu;V,N)$ and an equilibrium state $(T,0;V',N')$ are in
contact with each other as in section~\ref{s:muw},
we have the equality
\begin{equation}
\mu(T,\nu;V,N)=\mu(T,0;V',N').
\label{e:msme}
\end{equation}
In equilibrium thermodynamics, two systems which 
exchange substance and are in balance with each other 
always have equal chemical potentials.
Our postulate \eqref{e:msme} is a straight generalization of this principle.

Since $\mu(T,0;V',N')$ in the right-hand side is the chemical potential
of an equilibrium state, it is fully determined within the equilibrium 
thermodynamics.
Therefore by preparing the contact between various steady states
and equilibrium states, one can fully determine the SST chemical potential
$\mu(T,\nu;V,N)$ from the proposed equality \eqref{e:msme}.

It must be noted, however, that \eqref{e:msme} is not a mere
definition.
In sections~\ref{s:cp}, we characterized the chemical potential
$\mu(T,\nu;V,N)$ using the method of potential variation, and
determined the $V$, $N$ dependence of $\mu(T,\nu;V,N)$
for each fixed $T$ and $\nu$.
The equality \eqref{e:msme} must reproduce the same  $V$, $N$ dependence.

Therefore what is essential in the postulate \eqref{e:msme} is the assumption
that {\em there exists a wall which (through \eqref{e:msme}) gives $\mu(T,\nu;V,N)$
consistent with the potential variation method\/}.
Let us call a wall with this property a {\em $\mu$-wall\/}.
The existence of a perfect $\mu$-wall is indeed far from obvious
(although it can be established for the driven lattice gas as we 
see in section~\ref{s:mmw}).
In fact we will see in section~\ref{s:am} that there are walls which are {\em not\/}
$\mu$-walls.
The validity of our postulate should ultimately be verified through series
of careful experiments.

We assume that the chemical potential thus defined satisfies the symmetry
\begin{equation}
\mu(T,\nu;V,N)=\mu(T,-\nu;V,N).
\label{e:musym}
\end{equation}
For a sheared fluid this is obvious from the original symmetry
(i.e., Galilei invariance) of the system.
For heat conduction and electrical conduction, we are assuming that the 
direction of the current does not affect the contact with equilibrium state.
This is not entirely obvious, but seems plausible.
Similarly we assume for the pressure that
\begin{equation}
p(T,\nu;V,N)=p(T,-\nu;V,N).
\label{e:psym}
\end{equation}

\subsection{Complete determination of the free energy}
\label{s:ftn}
Since we have completely determined the chemical potential
$\mu(T,\nu;V,N)$, and the pressure $p(T,\nu;V,N)$, 
we use the Euler equation \eqref{e:Eu}
to fully determine the free energy $F(T,\nu;V,N)$.

The free energy $F(T,\nu;V,N)$ thus defined satisfies the extensivity
\eqref{e:Fex}, is jointly convex 
in $V$ and $N$ (as we saw in section~\ref{s:fd}),
and has the symmetry
\begin{equation}
F(T,\nu;V,N)=F(T,-\nu;V,N),
\label{e:Fsym}
\end{equation}
because of \eqref{e:musym} and \eqref{e:psym}.
Trusting in robustness of the mathematical structure of thermodynamics,
we further assume that the free energy $F(T,\nu;V,N)$ is concave in the 
two intensive variables $T$ and $\nu$.
Although it is a general requirement in thermodynamics that the free energy
is concave in an intensive variable, we still do not know what this assumption
really means in the context of nonequilibrium steady states\footnote{
\label{fn:psi}
In equilibrium thermodynamics, the concavity is directly related to the
convexity and the variational principle for the conjugate extensive variable.
To find out whether we have similar structure in SST is one of the important
remaining issues.
There is a possibility that we should use a monotone function $\psi(\nu)$
of $\nu$ as the ``correct'' intensive nonequilibrium variable
to have a meaningful conjugate extensive variable.
}.

Assuming the differentiability of $F(T,\nu;V,N)$, we can define
(following equilibrium thermodynamics) the SST entropy
\begin{equation}
S(T,\nu;V,N)=-\partialf{F(T,\nu;V,N)}{T},
\label{e:S}
\end{equation}
and a new nonequilibrium extensive quantity
\begin{equation}
\Psi(T,\nu;V,N)=-\partialf{F(T,\nu;V,N)}{\nu}.
\label{e:Psi}
\end{equation}
Since the symmetry \eqref{e:Fsym} implies
$\Psi(T,\nu;V,N)=-\Psi(T,-\nu;V,N)$,
we find that $\Psi(T,0;V,N)=0$ provided that
$\Psi(T,\nu;V,N)$ is continuous in $\nu$.
Moreover the assumed concavity of $F(T,\nu;V,N)$ implies 
\begin{equation}
\partialf{\Psi(T,\nu;V,N)}{\nu}\ge0,
\label{e:Psii}
\end{equation}
and hence we have
\begin{equation}
\Psi(T,\nu;V,N)
\cases{
\ge0&if $\nu\ge0$;\cr
=0&if $\nu=0$;\cr
\le0&if $\nu\le0$.
}
\label{e:Psign}
\end{equation}
Since the inequalities \eqref{e:Psign} suggest that $\Psi(T,\nu;V,N)$ is a kind of
measure of the ``degree of nonequilibrium'', we call 
$\Psi(T,\nu;V,N)$ the {\em nonequilibrium
order parameter\/}\footnote{
As a simple minded analogy, imagine that $\nu$ is the external magnetic
field of a magnetic system.
Then $\Psi$ becomes the magnetization.
}.

\subsection{Flux induced osmosis (FIO)}
\label{s:FIO}
We continue to study the situation in Fig.~\ref{f:cse}, where a local
steady state $(T,\nu;V,N)$ and an equilibrium state $(T,0;V',N')$
are in contact with each other.
Thus the equality \eqref{e:msme} between the chemical potentials hold.

Let us think about changing
the nonequilibrium control parameter $\nu$ slightly 
while keeping the temperature $T$, the volume $V$, and the
equilibrium chemical potential $\mu_{\rm eq}=\mu(T,0;V',N')$ constant.
The last condition is met, for example, by making $V'$ and $N'$
much larger than $V$ and $N$, respectively.
Since the chemical potential $\mu(T,\nu;V,N)$ must be also constant 
because of \eqref{e:msme}, the amount of substance $N$ in the steady
state inevitably varies according to $\nu$.
In what follows we write $N(\nu)$ instead of $N$ to remind this fact.

By dividing the Euler equation \eqref{e:Eu} by $N$ and using the extensivity
\eqref{e:Fex} of $F$ and the intensivity \eqref{e:pint}, \eqref{e:muint} of
$p$, $\mu$, we get
\begin{equation}
F(T,\nu;v,1)=-v\,p(T,\nu;v,1)+\mu(T,\nu;v,1),
\label{e:mfp}
\end{equation}
where $v=V/N(\nu)$ is the specific volume.
When the nonequilibrium variable $\nu$ is varied, 
$T$ and $\mu(T,\nu;v,1)$  do not change while
$v$ may change.
Thus by differentiating \eqref{e:mfp} with respect to $\nu$, we get
\begin{equation}
-\Psi(T,\nu;v,1)-\partialf{v}{\nu}\,p(T,\nu;v,1)=
-\partialf{v}{\nu}\,p(T,\nu;v,1)-v\,\partialf{p(T,\nu;v,1)}{\nu},
\label{e:Ppp}
\end{equation}
where we used \eqref{e:Psi} and \eqref{e:pF}.
This implies
\begin{equation}
\partialf{p(T,\nu;V,N(\nu))}{\nu}=\frac{\Psi(T,\nu;V,N(\nu))}{V}.
\label{e:pPsi}
\end{equation}
Since the pressure $p(T,\nu;V,N(\nu))$ is equal to the equilibrium
pressure $p_{\rm ea}=p(T,0;V',N')$ when $\nu=0$,
and the sign of the right-hand side of \eqref{e:pPsi} is given by
\eqref{e:Psign},
we find that
\begin{equation}
p(T,\nu;V,N(\nu))\ge p_{\rm eq},
\label{e:pspe}
\end{equation}
in general.
For $\nu\ne0$, the inequality is expected to become strict
except in trivial systems where $\Psi(T,\nu;V,N)$ is vanishing.

We thus conclude that {\em there inevitably appears a difference in
the pressures of a nonequilibrium steady state and an
equilibrium state}\/ which are in contact with each other and
exchanging the fluid.
The assumed concavity of the free energy in the nonequilibrium intensive parameter
$\nu$ implies that the pressure of the steady state is always higher.
We call this pressure difference the {\em flux-induced osmosis\/} 
(FIO).

We stress that the FIO is an intrinsically nonequilibrium phenomenon, which 
can never be predicted within the standard local equilibrium approach
(see section~\ref{s:NETD}).
To confirm the existence of a FIO through careful experiments seems to be
a challenging task, which will shed completely new light on the physics of
nonequilibrium systems.
We note, however, that an actual design of experiment may be nontrivial.
See section~\ref{s:exFIO}.

The prediction of FIO in a heat conducting system may be rather
surprising since we assert that a transfer of heat leads
to a mechanical force acting on the porous wall.
Let us note, however, that the appearance of mechanical force 
may not be too radical at least
for dilute gases.
In a dilute gas with a nonuniform temperature profile,
it is known from the analysis of the Boltzmann equation that the
pressure tensor becomes anisotropic \cite{ChapmanCowling39}.
It is then possible that a nonequilibrium steady state and an
equilibrium state balance with each other to have different
vertical pressures.
In fact Kim and Hayakawa \cite{KimHayakawa03a},
in their careful reinvestigation of the Chapman-Enskog 
expansion, examined a naive contact between a 
nonequilibrium steady state and an equilibrium state,
and found that the pressure of the steady state is
indeed larger than that of the equilibrium state\footnote{
\label{fn:KH}
Although they found a FIO with the correct sign, their contact
is not a perfect $\mu$-wall since the equality \eqref{e:pprr}
is violated.
We still do not know how one should realize a perfect
$\mu$-wall in dilute gases.
}.
A recent calculation based on Enskog's equation leads to a different conclusion \cite{Ugawa05}.

\subsection{$\mu$-wall revisited}
\label{s:am}
Since the existence of a $\mu$-wall inevitably leads to the pressure
difference between the steady state and the equilibrium state,
we find that a $\mu$-wall must be able to support a pressure difference.
But clearly there exists a wall (which should be called a $p$-wall) which ensures
that the pressures on its two sides are identical.
For example, a ``wall'' made of a network of thin wires may be able to separate
a steady state and an equilibrium state, but cannot support a pressure difference.
In this case, we must conclude that there is a finite difference in the 
chemical potentials of the steady state and the equilibrium state, while
the pressures are the same.
This is in a sharp contrast with the situation in equilibrium thermodynamics,
where any contact which allows the exchange of fluid keeps the 
chemical potential on both sides equal.

We suspect that a perfect $\mu$-wall and a perfect $p$-wall represent
two different idealized limits of experimentally realizable walls.
A general wall separating a steady state and an equilibrium state
is expected to lie between these two limits.
For a non-ideal wall, we still expect to observe a FIO with the same sign, but
with a reduced magnitude.

Let us derive a useful relation which enables one to check if an
ideal $\mu$-wall
contact is realized or not.
Suppose that a steady state $(T,\nu;V,N)$ and an equilibrium state
$(T,0;V',N')$ are separated by a perfect $\mu$-wall and are in balance with 
each other.
We fix $T$ and $\nu$, and slightly change the density by varying 
$V$ or $V'$.
Since both the steady state pressure $p_{\rm ss}=p(T,\nu;V,N)$
and the equilibrium pressure $p_{\rm eq}=p(T,0;V',N')$ satisfy the 
same Maxwell relation \eqref{e:eqMax1} and \eqref{e:Max}, 
the changes of the pressures satisfy
$\Delta p_{\rm eq}=\rho_{\rm eq}\,\Delta\mu_{\rm eq}$
and 
$\Delta p_{\rm ss}=\rho_{\rm ss}\,\Delta\mu_{\rm ss}$,
where $\Delta\mu_{\rm eq}$ and $\Delta\mu_{\rm ss}$
are the changes in the chemical potentials.
But \eqref{e:msme} implies $\Delta\mu_{\rm eq}=\Delta\mu_{\rm ss}$,
and we get
\begin{equation}
\frac{\Delta p_{\rm eq}}{\Delta p_{\rm ss}}
=
\frac{\rho_{\rm eq}}{\rho_{\rm ss}}.
\label{e:pprr}
\end{equation}
The equality \eqref{e:pprr} may be checked experimentally to see
if an ideal $\mu$-wall is realized.

\subsection{Shift of coexistence temperature}
\label{s:sc}
\begin{figure}
\centerline{\epsfig{file=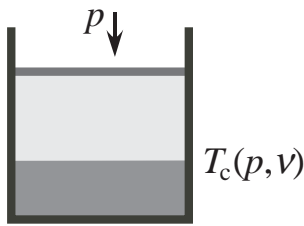,width=4cm}}
\caption[dummy]{
Two phases (say, liquid and vapor) separated by a horizontal plane coexist
within a nonequilibrium steady state.
We show that in general the coexistence temperature
$T_{\rm c}(p,\nu)$ shifts from its equilibrium value $T_{\rm c}(p,0)$.
}
\label{f:stc}
\end{figure}
Again we consider a single local steady state.
Suppose that there coexist two different phases, say, liquid and vapor,
within the local steady state.
We assume that the two phases are separated by a horizontal plane
as in Fig.~\ref{f:stc}.
By using standard techniques in thermodynamics, we derive a nonequilibrium
relation corresponding to the Clausius-Clapeyron relation.

By (fictitiously) splitting the system into two 
along the phase separation plane, we get
two local steady states in the low and the high temperature phases,
respectively.
We denote the local steady states in the
 low and the high temperature phases as
$(T,\nu;V_-,N_-)$ and $(T,\nu;V_+,N_+)$, respectively.

By dividing the Euler equation \eqref{e:Eu} by $N$ and using the extensivity
\eqref{e:Fex} of $F$ and the intensivity \eqref{e:pint}, \eqref{e:muint} of
$p$, $\mu$, we get for each phase
\begin{equation}
F(T,\nu;v_\pm,1)=-v_\pm\,p(T,\nu;v_\pm,1)+\mu(T,\nu;v_\pm,1),
\label{e:Fpm}
\end{equation}
where $v_\pm=V_\pm/N_\pm$ are the specific volumes.
It is convenient to fix the pressure $p=p(T,\nu;v_-,1)=p(T,\nu;v_+,1)$ constant.
We denote by $T_{\rm c}(p,\nu)$ the coexistence temperature at fixed
$p$ and $\nu$.
Since the two phases coexist, we have\footnote{
One may regard this balance condition as a special case of \eqref{e:mumu}.
} $\mu(T,\nu;v_-,1)=\mu(T,\nu;v_+,1)$, which, with \eqref{e:Fpm},
implies
\begin{equation}
F(T_{\rm c}(p,\nu),\nu;v_-,1)+v_-\,p=
F(T_{\rm c}(p,\nu),\nu;v_+,1)+v_+\,p.
\label{e:Fvp}
\end{equation}
We fix $p$ and differentiate \eqref{e:Fvp} with respect to $\nu$.
Noting that $v_\pm$ may depend on $\nu$ to keep the pressure constant,
we get
\begin{eqnarray}
&&-\partialf{T_{\rm c}(p,\nu)}{\nu}\,S(T_{\rm c}(p,\nu),\nu;v_-,1)
-\Psi(T_{\rm c}(p,\nu),\nu;v_-,1)
\ret
&&
=
-\partialf{T_{\rm c}(p,\nu)}{\nu}\,S(T_{\rm c}(p,\nu),\nu;v_+,1)
-\Psi(T_{\rm c}(p,\nu),\nu;v_+,1),
\label{e:TcS}
\end{eqnarray}
where we used the definitions \eqref{e:S} and \eqref{e:Psi} of
$S$ and $\Psi$, respectively.
We therefore find the following nonequilibrium relation analogous
to the Clausius-Clapeyron relation
\begin{equation}
\partialf{T_{\rm c}(p,\nu)}{\nu}=
-\frac{\psi_+-\psi_-}{s_+-s_-},
\label{e:CC}
\end{equation}
where
\begin{equation}
\psi_\pm=\frac{\Psi(T_{\rm c}(p,\nu),\nu;V_\pm,N_\pm)}{V_\pm},
\label{e:ppm}
\end{equation}
and
\begin{equation}
s_\pm=\frac{S(T_{\rm c}(p,\nu),\nu;V_\pm,N_\pm)}{V_\pm}
\label{e:spm}
\end{equation}
are the specific nonequilibrium order parameter and the
specific entropy of the two phases.

It is crucial to note that the nonequilibrium order parameter $\psi_\pm$
can be determined by experiments using
the FIO of section~\ref{s:FIO}.
Since the entropy $s_\pm$ may be approximated by their 
equilibrium values for small enough $\nu$, the right-hand side of
\eqref{e:CC} can be evaluated from experiments which do {\em not\/}
involve a phase coexistence.
This enables one to check (at least in principle)
for the {\em quantitative validity\/} of
the theory of SST in a purely experimental manner.

The concavity of the free energy implies that in general
the entropy $S(T,p,\nu;N)$ is nondecreasing\footnote{
To show this one Legendre transforms $F(T,\nu;V,N)$ to
$G(T,p,\nu;N)$ and use the relation $S=-\partial G/\partial T$
as in equilibrium thermodynamics.
} in $T$.
Thus we always have $s_+\ge s_-$, and expect
\begin{equation}
s_+-s_->0,
\label{e:spsm}
\end{equation}
in general.
Therefore the sign of the right-hand side of the Clausius-Clapeyron type
relation \eqref{e:CC} is determined by the sign of $\psi_+-\psi_-$.

Consider a sheared fluid and suppose that the low temperature phase
is a solid\footnote{
To consider a solid phase goes beyond our framework to treat only fluids.
But let us be optimistic.
}.
Since the solid phase is hardly affected by the shear, it is likely that 
$\psi_-$ is negligible.
Then the relation \eqref{e:CC} becomes
\begin{equation}
\partialf{T_{\rm c}(p,\tau)}{\tau}=
-\frac{\psi_+}{s_+-s_-}.
\label{e:CC2}
\end{equation}
The sign of the right-hand side of \eqref{e:CC2} is precisely known
by \eqref{e:Psign} and \eqref{e:spsm}.
We find that
\begin{equation}
T_{\rm c}(p,\tau)<T_{\rm c}(p,0),
\label{e:TcTc}
\end{equation}
in general.
Shear always induces melting of a solid.
This prediction is consistent with the recent numerical 
experiments \cite{ButlerHarrowell02} 
(where it was argued that the inequality \eqref{e:TcTc}
rules out (!) the possibility of a nonequilibrium thermodynamics).

It is especially interesting to investigate experimentally
the possible shift of coexistence temperature in a system with
 heat flow.
The right-hand side of \eqref{e:CC} may be positive or negative
depending on systems.
If $T_{\rm c}(p,J)>T_{\rm c}(p,0)$, one will find a remarkable
phenomenon of ``heat flux induced condensation'', i.e., one 
observes condensation of a low temperature phase near one of the
walls which have slightly {\em higher\/} temperature than the normal
transition temperature while the other wall has much higher temperature!

\section{Discussions}
\label{s:dis}

\subsection{Summary and perspective}
\label{s:sap}
In the present work, we have step by step developed a full fledged thermodynamics that is expected to apply to a wide class of nonequilibrium steady states.
We have tried hard to be as clear as possible in explaining our basic assumptions and  reasoning behind the construction.
We believe that our consideration about possible extensions of thermodynamics to nonequilibrium steady states is much more careful and rigorous than any other existing attempts.

We have developed our theory in the level of purely macroscopic phenomenology, and clarified what conclusions we get from only phenomenological considerations.
All thermodynamic quantities are defined through experimentally realizable operations.
We also made some nontrivial predictions that can be tested empirically.
We shall discuss about possible experimental tests in section~\ref{s:exp}.

The reader may have noticed that there are, roughly speaking,
three stages in our theory of SST.
The first stage, developed in sections~\ref{s:ss} and \ref{s:bf}, deals with very basic framework of SST.
We have examined basic symmetries of nonequilibrium steady states and fundamental structures of thermodynamics, and made a rather strong restriction on possible theories.
The second stage, developed in sections~\ref{s:op} and \ref{s:f}, deals with thermodynamics with fixed $T$ and $\nu$.
We have discussed important physics 
(the minimum work principle and the density fluctuation formula)
that can be read off from the $V$, $N$ dependence of the SST free energy.
The third stage, developed in section~\ref{s:fu}, deals with SST in a complete form.
We predicted the FIO and the shift of coexistence temperature.

We believe that the logic in the first and the second stage is reasonably firm and reliable.
Although a sound logic does not necessarily mean the validity of the theory, we are rather confident that our theory is realized in some nonequilibrium systems in nature.
The theory in the third stage is definitely most interesting, but its logic is not as firm as the previous ones.
First the notion of $\mu$-wall is not yet perfectly clear 
from an operational point of view, and must be further examined.
Secondly, our assumption about the concavity of the SST
free energy depends solely on the analogy with the conventional thermodynamics and has no operational foundation.
We nevertheless wish to encourage experimental works for testing our predictions related to $\mu$-wall contacts and clarifying the nature of contacts between equilibrium states and nonequilibrium steady states.
Either verification or falsification of our predictions will provide  hints for future development in nonequilibrium physics.

A crucial point about the potential significance of
SST is whether
it becomes a useful guide in the (future)
construction of statistical mechanics
for steady nonequilibrium states.
The fact that we have arrived at an
essentially unique theory
is rather encouraging.
We hope that, by trying to construct a statistical
theory that is consistent with the (unique)
nonequilibrium thermodynamics,
we are naturally led to a meaningful and
correct statistical mechanics
for steady nonequilibrium states.
We shall discuss related issues in 
the Appendix~\ref{s:spn}, 
where we look at SST from a microscopic point of view.

\subsection{Possibility of experimental tests}
\label{s:exp}
Let us briefly discuss possible experimental verification
of our predictions.
Our aim here is not to go into details of concrete experimental setups
but to make clear some essential points in our theory which need
to be tested empirically.

\subsubsection{Chemical potential and fluctuation}
\label{s:exfl}
Probably the most promising experiments of SST are those
designed to verify the formula for density fluctuation  in nonequilibrium steady
states (and the corresponding fluctuation-response relations) discussed in section~\ref{s:df}.

For this, one should prepare a system which is separated into two 
by a wall (which realizes a weak contact  of the two subsystems), 
and in which an 
external potential can be varied.
One then realizes a nonequilibrium steady state in the system.
See, for example, Figs.~\ref{f:dfw} and \ref{f:WCS}.
One should also be able to  measure the amounts of substance
in the two subsystems accurately\footnote{
A hopeful candidate is a system of charged plastic beads floating in water.
}.

In the first stage of experiments, one fixes $T$, $\nu$, and
measures the averaged amounts of substance for various
(fixed) values of the potential difference $u_2-u_1$.
Then by using these data and the definition \eqref{e:mumu}
of the chemical potential, one can experimentally determine
$\mu(\rho)$ (up to an arbitrary additive constant).
Of course nothing has been verified at this stage.

In the second stage of experiments (which may of course
be carried out at the same time as the first stage),
one measures the fluctuation of the amounts of 
substance $N_1$, $N_2$ in the two subsystems, again for
fixed $u_1$ and $u_2$.
Then our conjecture is that the probability to observe a partition
into $N_1$ and $N_2$ behaves as
\begin{equation}
\tilde{p}(N_1,N_2)\propto
\exp[-\beta\{F(V_1,N_1)+u_1N_1+F(V_2,N_2)+u_2N_2\}],
\label{e:ptilex}
\end{equation}
where we dropped $T$ and $\nu$.
We have extended \eqref{e:pN12} to treat the case where the two regions have
volumes $V_1$ and $V_2$.
This enables us to get as much information as possible from
this setting.

Now if the fluctuation is small (which seems to be always the case
in actual experiments for large systems),
one may expand the above formula \eqref{e:ptilex} to get
\begin{equation}
\tilde{p}(\Delta N)\propto
\exp\sqbk{-\frac{\beta}{2}\cbk{
\partialD{N_1}\mu(\frac{N_1^*}{V_1})+
\partialD{N_2}\mu(\frac{N_2^*}{V_2})}(\Delta N)^2},
\label{e:pdN}
\end{equation}
where $N_1^*$ and $N_2^*$ are most likely values of the amounts of
substance determined by
$\mu(N_1^*/V_1)+u_1=\mu(N_2^*/V_2)+u_2$,
and $\Delta N=N_1^*-N_1=N_2-N_2^*$ is the deviation.
Thus the variance of the deviation $\Delta N$ is given by
\begin{equation}
\bkt{(\Delta N)^2}=\frac{1}{\beta}
\cbk{\frac{\mu'(\rho_1^*)}{V_1}+\frac{\mu'(\rho_2^*)}{V_2}}^{-1},
\label{e:dN2}
\end{equation}
where $\rho_i^*=N_i^*/V_i$.

It must be noted that only $\mu'(\rho)$ (not $\mu(\rho)$  itself)
appears in the formula \eqref{e:dN2} of the variance.
Since one has determined $\mu'(\rho)$ unambiguously in the first stage,
one can now check for the quantitative validity of \eqref{e:dN2}.
Note also that one can increase the reliability of the conclusion by
carrying out experiments with various values of $u_2-u_1$,
and $V_1$, $V_2$.

As we shall discuss in section~\ref{s:faqt},
there is a possibility that the inverse temperature $\beta$
in \eqref{e:dN2} should be replaced with an ``effective inverse temperature.''
Even if this is the case, a series of experiments with varying $u_2-u_1$, $V_1$,
and $V_2$ is enough to check for the quantitative validity of the
conjectured formula \eqref{e:dN2}.

Moreover, one can determine the effective inverse temperature from a completely independent set of measurements.
One can measure relaxation process which takes place after suddenly changing the potential difference, and compare it with static temporal correlations through the fluctuation-response relations like \eqref{e:FRRgen} or those described in section~\ref{s:WCSL} and section~\ref{s:DLGLR}.
Since the parameter $\beta$ appearing in fluctuation-response relations should be the same as those in the density fluctuation formula, this makes our proposal of experiments completely nontrivial.
See also section~\ref{s:WCSO} for the similar discussion.

The minimum work principle \eqref{e:mwpV} may be also checked 
empirically, but we still do not know what can be conclusive
experiments.

\subsubsection{Flux induced osmosis}
\label{s:exFIO}
There is no doubt that the most exciting experimental verification of our SST
is to directly observe flux-induced osmosis (FIO),
especially in a heat conducting state,
and directly measure the nonequilibrium order parameter 
$\Psi$.
Unfortunately this project seems still not easy to carry out for
several reasons.
One essential difficulty is that we still do not know how one can realize
a perfect $\mu$-wall, which is necessary for a measurement of $\Psi$.
We imagine that a  search for sufficiently good $\mu$-walls 
should be done through a series of 
careful experiments using various
different walls.

Therefore the first step will be to confirm the existence of FIO
in a system with not necessarily perfect $\mu$-wall.
This alone, we believe, can be an essential step toward a better understanding
of truly nonequilibrium systems.
But even this may not be easy since we have almost no 
a priori estimate for the magnitude of the pressure difference.
The lack of quantitative estimates is characteristic to any
thermodynamic arguments.
Thermodynamics provides us with universal and exact relations,
but not with numerical estimates.

For dilute gases, however, we can make some rough estimates based on 
kinetic theory.
Consider a heat conducting state of a dilute gas.
We want to examine the dimensionless quantity
$\theta=(p_{\rm ss}-p_{\rm eq})/p_{\rm eq}$ which characterizes
the magnitude of the FIO.
Let $j$ be the heat flux per unit area.
Since $\theta$ should be an even function of $j$, it is expected 
that $\theta$
is proportional to $j^2$ when $j$  is small.
Then the dimensional analysis shows that the only dimensionless
combination (that includes $j^2$)  of basic quantities is
\begin{equation}
\theta\sim\frac{mj^2}{p^2kT},
\label{e:theta1}
\end{equation}
where $m$ is the mass of the gas molecule.
Of course $p$ (which may be either $p_{\rm eq}$ or $p_{\rm ss}$)
is the pressure and $T$ is the temperature.
This indeed is roughly equal to 
the magnitude of the anisotropy of the pressure tensor
obtained from the Chapman-Enskog expansion \cite{ChapmanCowling39}.
The FIO obtained in the dilute gas calculation
by Kim and Hayakawa \cite{KimHayakawa03a} also has the same
order of magnitude.

To first confirm that FIO is negligibly small in the ordinary environment,
let us examine the Ar gas at $T=273\ {\rm K}$  and $p=1\ {\rm atm}$.
The heat flux is given by $j=\kappa(\nabla T)$ with
the thermal conductivity 
$\kappa\simeq 2.1\times10^{-2}\ \rm J\, (s\,m\,K)^{-1}$.
As for the temperature gradient, let us set for the moment
$\nabla T=10^4\ \rm K/m$ (i.e., 
100 K difference within 1 cm, which is easily realizable in the kitchen).
By using $m\simeq 6.0\times 10^{-26}\ \rm kg$, we get
$\theta\sim5\times 10^{-11}$, which is miserably small 
as we anticipated.

To get more general estimate and see what we can do, we further use the
results from the gas kinetic theory to write
\begin{equation}
j\sim \sqrt{\frac{kT}{m}}\,\frac{k\,(\nabla T)}{d^2},
\label{e:jkin}
\end{equation}
where $d$ is the hard core diameter of the gas molecule.
Then we get
\begin{equation}
\theta\sim\frac{k^2\,(\nabla T)^2}{d^4p^2},
\label{e:theta2}
\end{equation}
which is independent of $T$.
The formula means that we must have low pressure as well as
large temperature gradient in order to observe
a large nonequilibrium effect.
Let us set $p=10\ \rm Pa$, for example, where
the mean free path at room temperature is still not too large.
Let us require $\theta\sim10^{-2}$ since a difference of
0.1 Pa may be detected by a diaphragm.
Then since $d\sim10^{-10}\ \rm m$, we need to have
$\nabla T \sim10^3\rm\ K/m$,
which may not be too unrealistic.
Perhaps a major challenge in the realization of this experiment is
to control convection and other non-thermodynamic effects.
It was clearly pointed out in  \cite{NishinoHayakawa05} that, in a naive setting of FIO, the pressure difference due to the boundary effect (namely, the temperature gap) is proportional to the heat flux $j$, and hence overwhelms the effect predicted by SST.
Therefore one should devise a special setting in which the temperature gap is eliminated to reveal the effect proportional to $j^2$.
(See also section~\ref{s:faqwall}.)

As for systems other than dilute gases, we have no estimates for FIO.
We may simply hope that the nonequilibrium effects become large
in some complex fluids.

The conjectured shift of coexistence temperature is also an interesting 
topic of experiments.
In this case, however, we have no a priori estimate at all.
It is quite interesting to know if experiments like the one described in \cite{MillsPhillips02,MillsPhillips03} 
have any relations with our predictions.

\subsection{Frequently asked questions}
\label{s:faq}
During these four years that we have spent developing the present theory,
we were repeatedly asked some important questions 
in conferences, seminars, personal discussions, and referee reports.
We shall summarize below our answers to some of these ``frequently asked
questions.''

Probably the most frequently asked question was about the possible
relation of SST to some of the existing attempts in nonequilibrium physics.
This is why we prepared section~\ref{s:EA}, where we listed some of the major
advances in nonequilibrium physics and discussed their relevance 
(or irrelevance) to SST.
We then concluded that there have been essentially nothing really similar to SST,
concerning either basic philosophy or concrete theories.

\subsubsection{What is ``temperature'' in nonequilibrium steady states?}
\label{s:faqt}
In most stages of our phenomenological construction of SST, the 
role of temperature has simply been  to keep the 
environment constant\footnote{
The only exception is the definition \eqref{e:S} of the SST entropy,
which involves a differentiation in $T$.
But we did not have much quantitative discussions about the 
SST entropy.
}.
Therefore we did not have to be so careful about precise parameterization 
of the temperature.
We simply used the same temperature scale as the heat baths, and
declared that the temperature may be measured by a thermometer.

In the future development of SST, however, it is quite likely that we must
examine the notion of temperature more carefully, and study the behavior
of the SST entropy in a quantitative manner.
We will face the problem to determine the temperature
which is intrinsic to a nonequilibrium steady state.
Let us discuss one possibility here.

We determined the $V$, $N$ dependence of the free energy
$F(T,\nu;V,N)$ in section~\ref{s:fd}, using the pressure and the
chemical potential.
Both the quantities were determined operationally in
section~\ref{s:op} {\em without references to the value
of the temperature\/}.
The only requirement was that the temperature was fixed.

In section~\ref{s:df}, on the other hand, we discussed the formula
\eqref{e:pN12} for density fluctuation.
In this formula, the free energy appears in the dimensionless
form $F/(k_{\rm B}T)$.
We have so far assumed simply that this $T$  is the same temperature
as the equilibrium temperature of the heat bath.
(This is indeed the
case in the driven lattice gas,
See section~\ref{s:dfw}.)
But there is a possibility that this $T$ should be replaced by an
``effective temperature'' intrinsic to the nonequilibrium steady
state, which may not necessarily be the same as the equilibrium temperature.

In other words,
the fluctuation formula \eqref{e:pN12} may be used 
to experimentally determine the intrinsic nonequilibrium temperature
since the free energy itself can be determined by other experiments
which do not involve precise values of temperatures.
Moreover, as we have discussed in section~\ref{s:exfl}, measurements related to linear response relations may be used to determine the nonequilibrium temperature directly.

From a theoretical point of view, the parameter $\beta$
that appears in the formula \eqref{e:wcan} for the
``weak canonicality'' may be used to determine the 
nonequilibrium temperature of a stochastic model
(which coincides with the usual temperature
in the models studied in Appendix~\ref{s:spn}).

\subsubsection{Don't long-range correlations destroy thermodynamics?}
\label{s:faqlrc}
It is well-known that nonequilibrium steady states in a system with a conservation law generically exhibits spatial long range correlations \cite{DorfmanKirkpatrickSengers94}.
More precisely correlation functions of some physical quantities decay 
slowly with a power law, even though the system is not at the critical point.

The mechanism of the long range correlation has been understood rather clearly from a phenomenological point of view, and the existence of long range correlations is confirmed in some experiments.
There are also  rigorous result \cite{Spohn83} and explicit perturbative calculations \cite{Tasaki04a,LefevereTasaki04,Sasa04} which show that 
there indeed are such long range correlations in concrete stochastic models.
No doubt the existence of 
long range spatial correlations is one of the intrinsic aspects of nonequilibrium steady states.

One might rather naively imagine that the existence of long range 
correlations is not consistent with the existence of thermodynamics 
since the correlations may destroy extensivity or locality.
But of course this is far from the case.
The best argument may be the fact (which will be rigorously established in Appendix~\ref{s:spn}) that we can construct perfectly consistent SST for the driven lattice gas, which is a typical model known to exhibit long range correlations.
We do not only obtain general formulae for the thermodynamic functions but also compute them in the limit of high temperature and low density.
We find nothing pathological that originates from the long range correlations.

The point is that the long range correlations show up (literally) in multi-point
correlation functions, not in local observables.
Thus they may lead to anomalous fluctuation of the sum of a certain quantity in a large region of the system.
A notable example is anomalous density fluctuation mentioned in section~\ref{s:df}.
If one looks at local quantities such as the pressure, the temperature or the density, 
on the other hand,
one observes nothing anomalous  originating from the long range correlations.
Thus all the thermodynamic quantities of SST, which are defined in terms of operations which involve local quantities, remain well-defined.

\subsubsection{Why do you consider walls and contacts?}
\label{s:faqwall}
The use of idealized walls and contacts may be the culture of thermodynamics.
In conventional thermodynamics, one freely uses such devices as infinitely thin walls or perfectly adiabatic walls, or sometimes more delicate ones like semipermeable membrane\footnote{
But see \cite{LiebYngvason99} for a sever criticism about the use of semipermeable membrane.
}.
Decompositions, combinations, or contacts of various states realized by these walls play fundamental roles in the construction of thermodynamics.
Of course boundaries are tricky objects even in equilibrium physics since properties of the system are inevitably modified near the boundaries.
Nevertheless one can (very successfully) characterize bulk properties (not boundary properties) of equilibrium states by using all these walls and contacts.
In many cases regions near boundaries are negligibly small compared with the bulk, and contacts are not modified by boundary effects.

Simply speaking, we wish to adopt the same strategy in our approach to nonequilibrium physics.
We do consider various walls and contacts, but our main concern is to reveal universal properties exhibited by bulk of nonequilibrium steady states.
Just as in equilibrium thermodynamics, walls and contacts are mere probes with which we investigate bulk properties.

Of course there is a possibility that one encounters various unexpected boundary effects caused by highly nonequilibrium nature of the states\footnote{
A typical example is the delicateness about contact between nonequilibrium states and equilibrium states as discussed in section~\ref{s:fu}.
}.
That is why we have carefully discussed the settings of the systems, the ways we insert and remove walls, the ways we put two systems into contact, and so on.
We have tried to minimize unwanted disturbance of the states by the walls or contacts.
Nevertheless there may still arise delicate material-dependent issues (like the temperature gap at boundaries) which complicate the analysis.
In such cases, one should carefully distinguish such non-universal effects specific to boundaries from universal properties of the bulk, and try to minimize the former either theoretically or by experimentally devising a suitable setting.
This may arouse nontrivial challenges for  experimental designs, but we believe it is always possible in principle.

\subsubsection{Is SST useful?}
\label{s:useful}
If one assumes that SST is a correct theory describing nature, one can still ask if the theory is useful.

To be honest, we admit that SST does not have much practical use at least in its present form.
As we have mentioned in section~\ref{s:PTSST}, for example, we cannot get any information about transport coefficients (linear or nonlinear) from the present form of SST.
The present SST only deals with the response of a system at a fixed ``degree of nonequilibrium.''
As for the SST entropy, we do not expect it to play a significant role like its equilibrium counterpart.
This is because we do not have (and do not expect to get) a natural 
operational definition of adiabatic operations for nonequilibrium systems.
(But see \cite{OonoPaniconi98} for an attempt.)

We believe that the significance of SST should be sought in more conceptual and theoretical aspects.
First of all, the mere fact that there exists a consistent thermodynamics for a large class of nonequilibrium steady states (if true) must be regarded as news of fundamental importance.
It shows that nonequilibrium states are not generated arbitrarily by time evolution, but are subject to a strict thermodynamic structure.
More importantly from a practical point of view, establishment of thermodynamics can be a first important step toward a construction of statistical mechanics for steady states.
We believe that such a statistical theory will not only of great theoretical interest but will have many practical applications.

\appendix

\section{Weak contact, density fluctuation, and linear response in a sheared fluid}
\label{s:WCS}
In the present appendix, we discuss a particular scheme for making two subsystems of sheared fluid into contact, and examine some of the conjectures of SST.
Analysis presented here is limited to the first and the second stages of SST (see Section~\ref{s:sap}), developed in Sections~\ref{s:ss} to \ref{s:f}, where we dealt with thermodynamics with a ``fixed degree of nonequilibrium.''
Based on some plausible assumptions, we can confirm our predictions about the density fluctuation stated in section~\ref{s:df}.
We can go further to discuss linear response theory and a determination of the ``nonequilibrium temperature.''

Compared with another ``existence proof of SST'' for the driven lattice gas presented in 
 the next Appendix~\ref{s:spn}, the present approach is more heuristic.
Nevertheless, the present setting has a clear advantage of being much closer to realistic systems.
We hope that the results in the present Appendix and the next Appendix~\ref{s:spn} play complementary roles in reinforcing the logic of SST developed in the main body of the paper.

\subsection{Weak contact scheme}
\label{s:WCSD}
\begin{figure}
\centerline{\epsfig{file=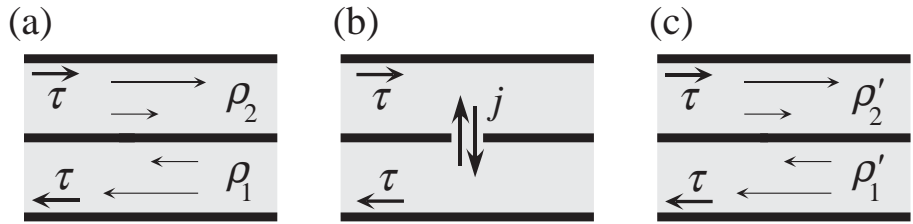,width=12cm}}
\caption[dummy]{
The scheme of weak contact for a sheared fluid.
The upper and the lower parts are separated by a horizontal wall which has a small window in it.
(a)~Both the lower and the upper parts are in their steady states.
(b)~The window opens for a finite interval of time, allowing some fluid to move.
We denote by $j$ the total amount of fluid that have moved from the lower to the upper part.
(c)~The window closes, and again both parts settle into their steady states with new densities.
We repeat this process many times and discuss the balance between the two parts,  density fluctuation, and linear response.
}
\label{f:WCS}
\end{figure}

Consider a system as in Fig.~\ref{f:WCS}~(a) in which a sheared fluid is separated into two parts by a horizontal wall.
Both the lower and the upper parts have volume $V$ and are characterized by the same ``degree of nonequilibrium'' (i.e., shear force) $\tau$.
Following the strategy of section~\ref{s:cp}, potential which is equal to $u_1$ and $u_2$ in the lower and the upper parts, respectively, is applied.
We assume that the lower and the upper parts are separately  in nonequilibrium steady states with densities $\rho_1$ and $\rho_2$, respectively.

The horizontal wall has a window on it, which opens for a finite interval of time.
We assume that the size of the window is much smaller than that of the wall, but is much larger than the molecular scale.
Likewise the interval during which the window is open is much shorter than the relaxation time of macroscopic quantities, but is much longer than the molecular time scale.
By $j$ we denote the amount of fluid that passes through the window from the lower part to the upper part during the interval that the window is open.

After the window closes, we keep the system as it is for a sufficiently long time so that the two parts reach their steady states with the new densities $\rho'_1=\rho_1-(j/V)$, and $\rho'_2=\rho_2+(j/V)$, respectively.

We repeat many times the above procedure of (short) opening of the window followed by a (long) relaxation period.
Then there takes place a very slow exchange of fluid between the lower and the upper parts of the system.
This process of exchange may be described as a discrete time Markov process where the state is characterized by the densities $(\rho_1,\rho_2)$ of the two parts.
By analyzing this Markov process, we can examine the construction and the predictions of SST.

The main motivation for devising this weak contact is that nonequilibrium steady states generically develop spatial long-range correlations (see section~\ref{s:faqlrc}), which lead to anomalous density fluctuation and transport.
Since our basic strategy in the present work (see section~\ref{s:faqwall}) is to characterize the steady state in each system by examining the contact, the long range correlation {\em between}\/ the two parts introduces unnecessary complication.
By separating the two parts by a wall and allowing the exchange of fluid only via the small window during a finite interval, we can  inhibit the system from developing long range correlations between the upper and the lower parts.
(Of course long range correlation within each part is developed.)

Now we shall make some plausible assumptions about the behavior of the quantity $j$, the amount of fluid that have moved in a single opening.
The most important assumption is that $j$ is a random quantity whose behavior is determined solely by local properties of the nonequilibrium steady states at the both sides of the window.
This is quite likely since the small window is open during the time interval which is much shorter than the relaxation time.
An important consequence of this assumption is that the behavior of $j$ essentially does not change when one increases the total volume $V$ of the system (fixing the size and the opening interval of the window).
This means that the change of the density in a single step, which is $j/V$ , is proportional to $V^{-1}$.
Thus $j/V$  can be made much smaller than the change of the average density in response to a change of potential, and than the typical magnitude of the density fluctuation in a steady state (since the former change is independent of $V$, and the latter fluctuation is proportional to $V^{-1/2}$).
We shall make use of this smallness of $j/V$ repeatedly in what follows.

To simplify the notation, we write the potential difference as $\Delta u=u_2-u_1$, and write the densities in the lower and the upper parts as $\rho_1=\rho_0-\theta$ and $\rho_2=\rho_0+\theta$, respectively.
Here $\rho_0$ is the total density, i.e., the total amount of fluid divided by the total volume.
From now on we shall always describe the densities in the two parts using the deviation $\theta$.

When $\Delta u$ and $\theta$ are given, we denote by $\psi_{\Delta u,\theta}(j)$ the probability density for the variable $j$.
We assume that  $\psi_{\Delta u,\theta}(j)$ is independent of the volume $V$.
We have omitted the dependence on $T$ and $\tau$ since these variables are always fixed.

When $\Delta u=0$ and $\theta=0$, the two parts of the system are completely identical.
Thus the corresponding probability density $\psi_{0,0}(j)$ is an even function of $j$ with a peak at $j=0$.
We assume that $\psi_{0,0}(j)$ decays rapidly for large $|j|$, and is normalized as $\int dj\,\psi_{0,0}(j)=1$.

When $\theta\ne0$, i.e., when the densities become uneven, the probability density $\psi$ becomes asymmetric.
In the lowest order, this effect may be written as
\begin{equation}
\psi_{0,\theta}(j)
\simeq \mathcal{N}(0,\theta)\,\psi_{0,0}(j)\,\exp[-\alpha_0\,\theta\,j],
\label{e:psiod}
\end{equation}
where $\alpha_0$ is a certain constant, and $\mathcal{N}(0,\theta)$ is the normalization constant.
When there is a difference $\Delta u=u_2-u_1$ in the potential, this is further modified as
\begin{equation}
\psi_{\Delta u,\theta}(j)
\simeq  \mathcal{N}(\Delta u,\theta)\,\psi_{0,0}(j)\,\exp[-\alpha_0\,\theta\,j-\frac{\beta}{2}\Delta u\,j],
\label{e:psiud}
\end{equation}
again in the lowest order.
Here $\beta$ is a certain constant, which can of course be identified as an inverse (effective) temperature.
Again the constant $\mathcal{N}(\Delta u,\theta)$ is chosen to ensure $\int dj\,\psi_{\Delta u,\theta}(j)=1$.

\subsection{Balance condition and the chemical potential}
\label{s:WCSB}
Suppose that we apply a small potential difference $\Delta u=u_2-u_1$ to the system, and repeat the process of opening the window sufficiently many times to have a steady balance between the lower and the upper parts.
Since the balance is attained when $\psi_{\Delta u,\theta}(j)$ is symmetric in $j$, we get from \eqref{e:psiud} that
\begin{equation}
\alpha_0\,\theta=-\frac{\beta}{2}\,\Delta u.
\label{e:adbu}
\end{equation}
Since the definition \eqref{e:mumu} of the chemical potential implies
\begin{equation}
\Delta u=u_2-u_1=\mu(\rho_0-\theta)-\mu(\rho_0+\theta)
\simeq-2\theta\,\mu'(\rho_0),
\label{e:u21mumu}
\end{equation}
we can represent the derivative of the chemical potential as
\begin{equation}
\mu'(\rho_0)=\frac{\alpha_0}{\beta}.
\label{e:mupab}
\end{equation}

We note that when the two parts are in balance with each other, mechanical balance as we discussed in section~\ref{s:mr} also holds.
This is because the window is sufficiently large and opens for a sufficiently long time for hydrodynamics to be efficient.
Consequently, the chemical potential expressed as \eqref{e:mupab} satisfies the Maxwell relation \eqref{e:Max} along with the pressure $p(\rho)$ defined from  mechanical forces.

\subsection{Density fluctuation}
\label{s:WCSF}
We shall analyze the density fluctuation in the steady balance condition that we have characterized, and derive the conjectured Einstein's formula \eqref{e:pN12}.

We fix the potential difference $\Delta u=u_2-u_1$, and describe the densities in the lower and the upper parts as $\rho_1=\rho_0-\theta$ and $\rho_2=\rho_0+\theta$, respectively.
From the balance condition \eqref{e:adbu}, we see that the most probable value of the density deviation $\theta$ is given by
\begin{equation}
\theta_0=-\frac{\beta\,\Delta u}{2\,\alpha_0}.
\label{e:theta0}
\end{equation}

To describe the fluctuation around $\theta_0$, we denote by $p(\theta)$ the stationary probability density that densities in the two parts are $\rho_0-\theta$ and $\rho_0+\theta$, respectively.
The stationary probability is the unique solution of 
\begin{equation}
p(\theta)=\int d\theta'\,p(\theta')\,c(\theta'\to\theta),
\label{e:dpcpc}
\end{equation}
for any $\theta$, where
\begin{equation}
c(\theta\to\theta+\frac{j}{V})=V\,\psi_{\Delta u,\theta}(j)
\label{e:cdpsi}
\end{equation}
is  the transition probability density.
It is normalized as 
\begin{equation}
\int d\theta'\,c(\theta\to\theta')=1.
\label{e:cnorm}
\end{equation}
The condition \eqref{e:dpcpc} is the continuous state-variable version  of \eqref{e:std}.

As we shall see below, the solution of \eqref{e:dpcpc} can be obtained by assuming the detailed balance condition (see section~\ref{s:sp2})
\begin{equation}
p(\theta)\,c(\theta\to\theta')=p(\theta')\,c(\theta'\to\theta),
\label{e:pcpcpzero}
\end{equation}
for any $\theta$ and $\theta'$.
Clearly \eqref{e:pcpcpzero} along with \eqref{e:cnorm} implies \eqref{e:dpcpc}, but not vice versa.

If we substitute $\theta'=\theta+(j/V)$, \eqref{e:cdpsi}, and \eqref{e:psiud}, the detailed balance condition \eqref{e:pcpcpzero} becomes
\begin{eqnarray}
&&p(\theta)\,\mathcal{N}(\Delta u,\theta)\,\exp[-\alpha_0\,\theta\,j-\frac{\beta}{2}\,\Delta u\,j]
\ret
&&=p(\theta+\frac{j}{V})\,\mathcal{N}(\Delta u,\theta+\frac{j}{V})\,\exp[\alpha_0\,(\theta+\frac{j}{V})\,j+\frac{\beta}{2}\,\Delta u\,j].
\label{e:pcpcpzero2}
\end{eqnarray}
We then have
\begin{equation}
\frac{p(\theta)}{p(\theta+(j/V))}\simeq\exp[2\alpha_o\,\theta\,j+\beta\,\Delta u\,j],
\label{e:ptptpetj}
\end{equation}
where we have used the smallness of $j/V$ to neglect $j^2/V$ in the exponential and set $\mathcal{N}(\Delta u,\theta+(j/V))/\mathcal{N}(\Delta u,\theta)\simeq1$.
By taking the logarithm and again noting that $j/V$ is small, we get
\begin{equation}
-\frac{j}{V}\,\frac{\partial}{\partial\theta}\log p(\theta)\simeq
2\alpha_o\,\theta\,j+\beta\,\Delta u\,j,
\label{e:dpdt2aV}
\end{equation}
which can readily be solved to give
\begin{equation}
p(\theta)\propto\exp[-\alpha_o\,V\,(\theta-\theta_0)^2]
=\exp[-\beta\,\mu'(\rho_0)\,\frac{(\Delta N)^2}{V}]
\label{e:WCEins}
\end{equation}
where $\Delta N=(\theta-\theta_0)\,V$ is the deviation measured by the amount of fluid.
This is nothing but the Einstein's formula \eqref{e:pN12}, expanded to the lowest order in $\Delta N$.

\subsection{Linear transport}
\label{s:WCSL}
We finally treat time-dependent linear transport through the weak contact, and prove the corresponding fluctuation-response relation.
This is especially useful since the resulting relations can be used to determine the parameter $\beta$ in \eqref{e:psiud} by using measurable quantities.
See Section~\ref{s:WCSO}.

It may appear surprising that one can deal with transport phenomena in a highly nonequilibrium system.
In this case, however, we are dealing with a very weak flow in the direction orthogonal to the (strong) shear flow.
This enables us to treat the transport by using only elementary techniques in Markov processes.

Suppose, for simplicity, that the two parts have been in a steady contact with vanishing potential difference $\Delta u=u_2-u_1=0$.
At an instant, we apply a nonvanishing but small potential difference\footnote{
It is also easy to treat time-dependent perturbation as we shall do in section~\ref{s:DLGLR}.
} $\Delta u$.
Then there takes place a relaxation from a nonequilibrium steady state with $\Delta u=0$ to another nonequilibrium steady state with nonvanishing $\Delta u$.

Let us define the single-step time evolution operator by
\begin{equation}
[\hat{T}_{\Delta u}\,p](\theta)=
\int d\theta'\,p(\theta')\,c_{\Delta u}(\theta'\to\theta),
\label{e:Tdef}
\end{equation}
 where we have explicitly labeled $c(\theta\to\theta')$ (defined in \eqref{e:cdpsi}) with the potential difference $\Delta u$.
 The probability density after $M$ openings of the window is given by
 \begin{equation}
p_M(\theta)=[(\hat{T}_{\Delta u})^M\,p_0](\theta),
\label{e:pMt1}
\end{equation}
where $p_0(\theta)$ is the stationary distribution for the case $\Delta u=0$.

We wish to express $\hat{T}_{\Delta u}$ as a perturbation to  $\hat{T}_{0}$.
As for the transition probability, we find from \eqref{e:cdpsi} and \eqref{e:psiud} that 
\begin{eqnarray}
c_{\Delta u}(\theta'\to\theta)
&=&\frac{\mathcal{N}(\Delta u,\theta')}{\mathcal{N}(0,\theta')}\,
\exp[-\frac{\beta}{2}\,\Delta u\,(\theta-\theta')\,V]\,c_{0}(\theta'\to\theta)
\ret
&=&c_{0}(\theta'\to\theta)
-\frac{\beta}{2}\,\Delta u\,(\theta-\theta')\,V\,c_{0}(\theta'\to\theta)
\ret
&&
+\frac{\beta}{2}\,\Delta u\int d\theta''\,(\theta''-\theta')\,V\,c_{0}(\theta'\to\theta'')
+O((\Delta u)^2)
\label{e:cuex1}
\end{eqnarray}
where the contribution from the normalization factor has been determined by noting that $\int d\theta\,c_{0}(\theta'\to\theta)=\int d\theta\,c_{\Delta u}(\theta'\to\theta)=1$.
To proceed we need to work on the third term in the right-hand side of \eqref{e:cuex1}.
By writing $\theta''=\theta'+(j/V)$, the relevant integral becomes
\begin{equation}
\int d\theta''\,(\theta''-\theta')\,V\,c_{0}(\theta'\to\theta'')
=
\frac{1}{V}\int dj\,j\,c_{0}(\theta'\to\theta'+\frac{j}{V})
\simeq
-(\alpha_0\,\theta'+\frac{\beta}{2}\,\Delta u)\,(j_0)^2,
\label{e:intVc0}
\end{equation}
where the final approximate expression is obtained from \eqref{e:cdpsi}, \eqref{e:psiud}, and a new (but plausible) assumption that $\psi_{0,0}(j)\propto\exp(-j^2/\{2(j_0)^2\})$.
Here $j_0>0$ is the typical value for the quantity $j$.
We now note that $\theta$ and $\theta'$ differ only by $O(j_0/V)$.
Therefore \eqref{e:intVc0} implies that one can  replace $\theta'$ by $\theta$ as
\begin{equation}
\int d\theta''\,(\theta''-\theta')\,V\,c_{0}(\theta'\to\theta'')
=
\int d\theta''\,(\theta''-\theta)\,V\,c_{0}(\theta\to\theta'')
+O(\frac{\alpha_0\,(j_0)^3}{V}),
\label{e:inttpt}
\end{equation}
where the difference is so small for large systems that we shall neglect it from now on.
By substituting the replacement \eqref{e:inttpt} into \eqref{e:cuex1}, we have
\begin{eqnarray}
c_{\Delta u}(\theta'\to\theta)&=&
c_{0}(\theta'\to\theta)
-\frac{\beta}{2}\,\Delta u\,(\theta-\theta')\,V\,c_{0}(\theta'\to\theta)
\ret
&&
+\frac{\beta}{2}\,\Delta u\int d\theta''\,(\theta''-\theta)\,V\,c_{0}(\theta\to\theta'')+O((\Delta u)^2).
\label{e:cuex}
\end{eqnarray}

By substituting \eqref{e:cuex} into \eqref{e:Tdef}, we get
\begin{equation}
[\hat{T}_{\Delta u}\,p](\theta)\simeq
[\hat{T}_{0}\,p](\theta)-\frac{\beta\,\Delta u}{2}
\int d\theta'\,(\theta-\theta')\,V\,\{p(\theta')\,c_{0}(\theta'\to\theta)
+p(\theta)\,c_0(\theta\to\theta')\}.
\label{e:Tp1}
\end{equation}
When applied to $p_0(\theta)$, the stationary condition $[\hat{T}_0\,p_0](\theta)=p_0(\theta)$ and the detailed balance condition \eqref{e:pcpcpzero} makes the expression \eqref{e:Tp1} much simpler as
\begin{eqnarray}
[\hat{T}_{\Delta u}\,p_0](\theta)&\simeq&
p_0(\theta)-
\beta\,\Delta u\,
\int d\theta'\,(\theta-\theta')\,V\,p_0(\theta')\,c_{0}(\theta'\to\theta)
\ret
&=&
p_0(\theta)-\beta\,\Delta u\,[\hat{j}\,p_0](\theta)
\label{e:Tuppj}
\end{eqnarray}
We have defined the operator $\hat{j}$ by
\begin{equation}
[\hat{j}\,p](\theta)=\int d\theta'\,(\theta-\theta')\,V\,p(\theta')\,
c_0(\theta'\to\theta),
\label{e:jdef}
\end{equation}
which counts the total amount of the fluid that have moved and at the same time generates a single-step time evolution.

By substituting \eqref{e:Tuppj} into \eqref{e:pMt1}, and expanding in $\Delta u$, we find
\begin{equation}
p_M(\theta)=p_0(\theta)-\beta\,\Delta u\sum_{m=0}^{M-1}
[(\hat{T}_0)^m\,\hat{j}\,p_0](\theta)+O((\Delta u)^2).
\label{e:pMp0bdu}
\end{equation}
This is our basic equation for the linear response theory.

Let $N_1(\theta)=(\rho_0-\theta)\,V$ be the total amount of fluid in the lower part.
We denote its average after $M$ openings of the window as $\sbkt{\hat{N}_1(M)}_{\Delta u}=\int d\theta\,N_1(\theta)\,p_M(\theta)$.
Here $\sbkt{\hat{N}_1(M)}_{\Delta u}$, as a function of $M$, describes the relaxation phenomenon after the potential difference $\Delta u$ was turned on.
By using \eqref{e:pMp0bdu}, we find that
\begin{eqnarray}
\sbkt{\hat{N}_1(M)}_{\Delta u}&=&
N_0-\beta\,\Delta u
\sum_{m=0}^{M-1}N_1(\theta)\,[(\hat{T}_0)^m\,\hat{j}\,p_0](\theta)
+O((\Delta u)^2)
\ret
&=&
N_0-\beta\,\Delta u
\sum_{m=0}^{M-1}\sbkt{\hat{N}_1(M)\,\hat{j}(M-m)}_0
+O((\Delta u)^2),
\label{e:WCSFRR}
\end{eqnarray}
where $N_0=V\,\rho_0=\int d\theta N_1(\theta)\,p_0(\theta)$ is the amount of fluid in the lower part in the steady state with $\Delta u=u_1-u_2=0$.
The definition of the temporal correlation function $\sbkt{\hat{N}_1(M)\,\hat{j}(M-m)}_0$ can be read off from \eqref{e:WCSFRR}.

Reflecting the fact that $\hat{j}$ measures the amount of fluid moved from the lower to the upper part, \eqref{e:WCSFRR} can be further rewritten in the form
\begin{equation}
\sbkt{\hat{N}_1(M)}_{\Delta u}=
N_0+\beta\,\Delta u\,
\bkt{\hat{N}_1(M)\,\{\hat{N}_1(M)-\hat{N}_1(0)\}}_0
+O((\Delta u)^2).
\label{e:WCSFRR2}
\end{equation}
Thus the relaxation phenomenon is fully described (in the lowest order in $\Delta u$) in terms of the temporal correlation function in the steady state without a potential difference.
This means that one can unambiguously determine the inverse temperature $\beta$ by measuring the behavior of the variable $N_1$.

\subsection{Discussion}
\label{s:WCSO}
In the present appendix, we have devised a weak contact realized by a small window in the setting of a sheared fluid.
Based on plausible assumptions about the exchange of fluid through the window, we were able to recover the conjecture of SST about the density fluctuation.
We were further able to show the fluctuation-response relation for the time-dependent relaxation phenomena.

It is crucial that we have discussed three different settings and the corresponding measurements, namely, to measure i)~the most probable densities when a potential difference is applied (section~\ref{s:WCSB}), ii)~the density fluctuation for a fixed potential difference (section~\ref{s:WCSF}), and iii)~the relaxation phenomena (section~\ref{s:WCSL}) when a potential difference is suddenly applied.

From i), we get information about $\mu'(\rho)$, and, from ii), we get $\beta\,\mu'(\rho)$.
These two already give nontrivial prediction to experimental results provided that the parameter $\beta$ can be identified as the inverse temperature.
This is what we expect most naively, but there is a possibility that $\beta$ should be regarded as the nonequilibrium inverse temperature which deviates slightly from its equilibrium counterpart (see section~\ref{s:faqt}).
In such a case, iii) provides a definite operational method for measuring the ``nonequilibrium inverse temperature'' $\beta$.
Thus the three measurements proposed in i), ii), and iii) together form a complete operational test about the validity of (a part of) SST.

It is quite interesting whether the present weak contact scheme can be applied to  systems other than the sheared fluid.
Strictly speaking the present argument works when there is a symmetry between the two parts.
Thus it can be applied directly only to systems with a symmetry such as the $(T,\phi;V,N)$ formalism of electrically conducting fluid.
It would be very useful if similar scheme can be developed in more general systems.
Needless to say, a major remaining challenge is to devise similar realization of the $\mu$-wall contact that we have discussed in section~\ref{s:fu}.
For the moment this problem seems quite delicate, and we have concrete results only in the driven lattice gas (see section~\ref{s:mmw}).

\section{SST in the driven lattice gas}
\label{s:spn}
In the main body of the paper, we have developed the framework of
steady state thermodynamics (SST) from a macroscopic
phenomenological point of view.
In the present Appendix, we will develop a microscopic point of view, and  demonstrate that 
SST can be  realized in 
nonequilibrium steady states of a standard 
Markov process called the driven lattice gas.
This provides a complementary analysis to the ``mesoscopic'' approach developed in the Appendix~\ref{s:WCS}.
The argument here is quite general and can be extended to
a much larger class of
Markov processes.

With this ``existence proof'' we can be sure that our 
framework of SST is theoretically consistent.
Moreover concrete mathematical results in the driven lattice gas
may give us hints for further development of our phenomenology.

Here we introduce the driven lattice gas, and realize a weak contact 
 to determine the chemical potential, the pressure, 
and the free energy.
We get general formulae for these thermodynamic quantities, which allow
us to compute them 
explicitly (by using, for example, a computer or a systematic expansion).
We also present results of the simplest calculation in the limit of
high temperature and low density.

We stress that we are {\em not\/} defining these
thermodynamic quantities based on
formal analogies with the equilibrium statistical mechanics, but rather
defining everything based on the operational procedures discussed
carefully in section~\ref{s:op}.
Therefore our free energy $F$ is (at least for the moment) not related to 
a statistical quantity through a relation like $F=-(1/\beta)\log Z$ or $S=k_{\rm B}\log W$.

We also show that the fluctuation relation (see section~\ref{s:df})
and the minimum work principle (see section~\ref{s:mw}) hold 
exactly in these models.
We argue that an ideal $\mu$-wall (see section~\ref{s:cpf}) may be constructed for the driven lattice gas.
We can go further to develop a linear response theory for time-dependent particle exchange process through weak coupling (see section~\ref{s:WCSL}).

As is clear from the summary in the above paragraphs, we can reproduce
most aspects of our phenomenological construction.

\Rem
If one applies the construction in the present Appendix to a Markov process
whose stationary distribution corresponds to an equilibrium state,
then we get thermodynamic quantities which exactly coincide
with those obtained from equilibrium statistical mechanics.
This fact alone is of some interest since our derivation involves
only local correlation functions, nothing analogous to the partition function.
One may regard our construction as a robust method of 
``getting thermodynamics without a partition function'',
which can be applied to nonequilibrium steady states
as well as equilibrium states.

\subsection{Definition}
\label{s:bs}
Let us define the driven lattice gas.
It belongs to a class of models
which are obtained by making minimum modifications to
Markov processes for equilibrium dynamics discussed in section~\ref{s:sp}.
As we have stressed in section~\ref{s:sp2}, the detailed balance condition 
\eqref{e:db} with
respect to the equilibrium state is a fundamental requirement for constructing
physically meaningful models for equilibrium dynamics.
Unfortunately we still do not know what are corresponding guiding principles 
in nonequilibrium stochastic dynamics.
This means that we must proceed carefully, not over-trusting results from model
studies, and always questioning robustness of conclusions.

\begin{figure}
\centerline{\epsfig{file=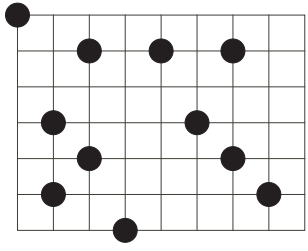,width=4cm}}
\caption[dummy]{
Basic setup of lattice gas models.
Particles live on sites of the $\ell\times h$ two dimensional
lattice.
Particles hop around the lattice according to 
given transition rates.
}
\label{f:lg}
\end{figure}

For simplicity we treat models in two dimensions, but extensions
to higher dimensions are automatic.
Define an $\ell\times h$ lattice $\La$ by
\begin{equation}
\La=\set{\xb=(x_1,x_2)}{x_1=0,\pm1,\ldots,\pm\frac{\ell-1}{2},\ x_2=0,\pm1,\ldots,\pm\frac{h-1}{2}},
\label{e:Lambda}
\end{equation}
where (as usual) $x_1$ and $x_2$ are the horizontal and the vertical coordinates,
respectively.
Lattice sites are denoted as $\xb,\yb,\ldots\in\La$.
We impose periodic boundary conditions 
(i.e., we identify $x_1=(\ell+1)/2$ with $x_1=-(\ell-1)/2$, and $x_2=(h+1)/2$ with $x_2=-(h-1)/2$).
For each site $\xb\in\La$, we associate an occupation variable 
$\etax$.
We set $\etax=1$ if the site $\xb$ is occupied by a particle, and 
$\etax=0$ if $\xb$ is empty.
We do not allow more than two particles to occupy a single site.
See Fig.~\ref{f:lg}.

We denote a collection of $\etax$ for all $\xiL$ as
\begin{equation}
\etab=(\etax)_{\xiL},
\label{e:etab}
\end{equation}
and call it a {\em configuration\/}.
A configuration $\etab$ corresponds to a microscopic state $s$
in sections~\ref{s:sm} and \ref{s:sp}.
For a given configuration $\etab$, we write
\begin{equation}
|\etab|=\sum_{\xiL}\etax,
\label{e:etat}
\end{equation}
which is the total number of particles in $\etab$.

The Hamiltonian $\HL(\cdot)$ gives the energy $\HL(\etab)$
for a configuration $\etab$ on the lattice $\La$.
Although our discussion does not depend on the specific choice of $\HL(\cdot)$  (except in section~\ref{s:pert}), let us take for concreteness  the Ising Hamiltonian
\begin{equation}
\HL(\etab)=-J\sum_{\bkt{\xb,\yb}}\etax\etay,
\label{e:Hgee}
\end{equation}
where $J$ is the coupling constant, and the sum is over all pairs of 
nearest neighbor sites (according to the periodic boundary conditions).
Although the symbol $J$ is used to denote the flux in the main body of the paper, we here follow the standard notation in the driven lattice gas.

For a site $\xiL$ and a configuration $\etab$, we denote by $\etabx$
the new configuration obtained by changing\footnote{
Since $\etax=0,1$, the change means $0\to1$ or $1\to0$.
} 
the value of $\etax$ in $\etab$.
More precisely, we set
\begin{equation}
(\etabx)_{\yb}=\cases{
1-\etax&if $\yb=\xb$;\cr
\etay&if $\yb\ne\xb$.
}
\label{e:etaxy}
\end{equation}
Similarly for $\xb,\yb\in\La$ and a configuration $\etab$,
we denote by $\etabxy$ the configuration obtained by 
changing both $\etax$ and $\etay$.

We can define a Markov process using the general discussion in section~\ref{s:sp1} once we specify the transition rates $\cTEL(\etab\to\etab')$.
For any configuration $\etab$ and any $\xb,\yb\in\La$ such that $|\xb-\yb|=1$, we set
\begin{eqnarray}
\cTEL(\etab\to\etabxy)
&=&\etax(1-\etay)\,\phi[\beta\{\HL(\etabxy)-\HL(\etab)+E(x_1-y_1)\}]
\ret
&&+(1-\etax)\etay\,\phi[\beta\{\HL(\etabxy)-\HL(\etab)+E(y_1-x_1)\}],
\label{e:CTE1}
\end{eqnarray}
where $\beta=1/(k_{\rm B}T)$ is the inverse temperature and $E$ is the ``electric field'' in the $x_1$ direction.
Here $\phi(h)$ is the function introduced in  section~\ref{s:sp1}, and again satisfies \eqref{e:phicond}.
The first term in the transition rate \eqref{e:CTE1} represents a hop of particle from $\xb$ to $\yb$ and the second term from $\yb$ to $\xb$.
We set $\cTEL(\etab\to\etab')=0$ for $\etab$, $\etab'$ not of the form \eqref{e:CTE1}.
Note that the particle number $|\etab|$ is conserved in the allowed transitions.

A Markov process with the transition rates \eqref{e:CTE1} is ergodic (see section~\ref{s:sp1})
in a state space with a fixed $|\etab|$.
Thus for each positive integer $N\le\ell h$, there is a unique stationary distribution
$p^{(T,E)}_{\La,N}(\etab)$ which is nonvanishing only
when $|\etab|=N$.
When $E\ne0$, this stationary distribution represents
the nonequilibrium steady state of the system.
We denote the average over the steady state distribution as
\begin{equation}
\bkt{g(\etab)}^{(T,E)}_{\La,N}=\sum_{\etab}g(\etab)\,
p^{(T,E)}_{\La,N}(\etab),
\label{e:gTn}
\end{equation}
where $g(\etab)$ is an arbitrary function of configurations.

In the steady state of the present driven lattice gas, there is a constant 
flow of particles toward the right.
In this sense the model has some resemblance with the $(T,\phi;V,N)$
formalism of electrical conduction.
See section~\ref{s:bec}, especially Fig.~\ref{f:ecd}~(b).
Note that we do not have electric plates here, which are effectively replaced by
the periodic boundary conditions.
Since the parameter $E$ corresponds to electric field, we can identify
the nonequilibrium intensive variable as
$\phi=E\ell$.
For convenience, however, we shall use the parameter $E$ (which may also be regarded as an intensive nonequilibrium variable) to characterize the steady states.

\subsection{Local detailed balance condition}
\label{s:LDB}
Take any $\xb,\yb\in\La$ such that $|\xb-\yb|=1$, and let $\etab$ be any configuration such that $\etax=1$ and $\etay=0$.
Then the definition \eqref{e:CTE1} and the condition \eqref{e:phicond} for $\phi(h)$ implies
\begin{equation}
\frac{\cTEL(\etab\to\etabxy)}{\cTEL(\etabxy\to\etab)}
=\exp[\beta\{\dHxy+E(y_1-x_1)\}],
\label{e:ldb}
\end{equation}
which should be compared with the detailed balance condition
\eqref{e:db2} in equilibrium dynamics.
It must be noted that there is no single function\footnote{
If there was one, we could write the steady state distribution
as $p_{\rm ss}(\etab)=\tilde{Z}^{-1}\exp[-\beta\tilde{H}(\etab)]$,
and the Markov process would satisfy the detailed balance condition
with respect to $p_{\rm ss}(\etab)$.
} $\tilde{H}(\cdot)$ which
enables us to express the right-hand side of  \eqref{e:ldb} simply as
$\exp[\{\tilde{H}(\etab)-\tilde{H}(\etabxy)\}]$.
Try, for example, 
$\tilde{H}(\etab)=\HL(\etab)-E\sum_{\xiL}x_1\etax$.
Then \eqref{e:ldb} is equal to
$\exp[\beta\{\tilde{H}(\etab)-\tilde{H}(\etabxy)\}]$ for most $\xb$ except for those
at boundaries.
For any $\xiL$ one can choose a  $\tilde{H}(\cdot)$ which covers sites
around $\xb$, but not the whole lattice.
In this sense, the relation\eqref{e:ldb} is sometimes called\footnote{
The term ``local detailed balance'' might be a bit confusing.
A detailed balance condition, as in \eqref{e:db}, is always stated
with respect to a particular stationary distribution.
Here one still does not know what the stationary distribution is.
A condition like  \eqref{e:ldb} may be better called
``local energy conservation.''
} the ``local detailed balance.''

\subsection{Determination of thermodynamic quantities}
\label{s:mpq}
We will see how we can realize the weak contact of two systems,
and evaluate thermodynamic quantities.

\subsubsection{Weak contact in lattice gases}
\label{s:wc}
\begin{figure}
\centerline{\epsfig{file=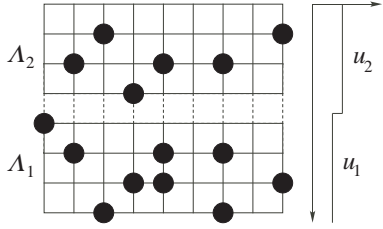,width=5cm}}
\caption[dummy]{
Weak contact of two steady states in a lattice gas model.
Potential which is equal to $u_1$ and $u_2$ on the lattices
$\La_1$ and $\La_2$, respectively, is applied to the system.
The potential only modifies the transition rates between the two lattices.
By using this setting we can define the chemical potential of a lattice gas model.
In fact the same setting can be used to realize a perfect $\mu$-wall
that connects a nonequilibrium steady state with an equilibrium state.
See section~\ref{s:mmw}.
}
\label{f:mwc}
\end{figure}
We shall now realize in the context of lattice gases the idea of 
weak contact.

Let $\La_1$ and $\La_2$ be $\ell\times h$ lattices
identical to $\La$ of \eqref{e:Lambda}.
(It is automatic to extend the present discussion to cases where $\La_1$ and $\La_2$ are not identical.)
We denote by $\etab$ and $\zetab$ lattice gas configurations on
$\La_1$ and $\La_2$, respectively.
The total Hamiltonian is 
\begin{equation}
H_{\rm tot}^{(u_1,u_2)}(\etab,\zetab)=
H_{\La_1}(\etab)+H_{\La_2}(\zetab)+u_1|\etab|+u_2|\zetab|,
\label{e:Htot}
\end{equation}
where $H_{\La_i}(\cdot)$ ($i=1,2$) are copies of the Hamiltonian $\HL(\cdot)$,
and $u_1$ and $u_2$ are uniform potentials applied to $\La_1$ and $\La_2$,
respectively.
Note that such a potential does not affect the dynamics within $\La_1$ or $\La_2$
at all, and only modifies hopping between the two lattices.

Our Markov process is defined by the transition rates
$c^{(T,E)}_{\La_1}(\etab\to\etabxy)$ and
$c^{(T,E)}_{\La_2}(\zetab\to\zetab^{\xb,\yb})$
(which are faithful copies of $\cTEL(\etab\to\etabxy)$
of \eqref{e:CTE1}) within each sublattice, and additional transition rates
for hops between the two sublattices $\La_1$ and $\La_2$.

\begin{figure}
\centerline{\epsfig{file=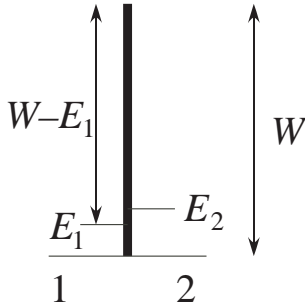,width=4cm}}
\caption[dummy]{
A system consisting of two states 1 and 2 which are separated by a large energy barrier $W$.
By examining equilibrium dynamics that satisfies the detailed balance condition, we can determine the transition rates $c(1\to2)$ and $c(2\to1)$.
}
\label{f:E12W}
\end{figure}

Let us pose here and examine how one should design transition rates for two subsystems that are coupled very weakly.
Here we assume  that the weak coupling is realized by the presence of a  very large energy barrier for transitions between the two subsystems.
For simplicity let us first consider a system with two states 1 and 2 with the energies $E_1$ and $E_2$, and assume that the two states are separated by an energy barrier $W\gg E_1,E_2$.
See Fig.~\ref{f:E12W}.
We shall construct a Markov process describing the equilibrium of this system.
Suppose that the system is in the state 1.
In order for the system to jump into the state 2, it must first reach the top of the energy barrier, raising the energy by $W-E_1$.
Since there is no cost required for falling down into the state 2, we may assume that the transition rate from the state 1 to 2 is fully determined by the relative height $W-E_1$ of the wall.
This means that we can write $c(1\to2)=\psi(W-E_1)$ with some function $\psi(h)$.
Similarly we have $c(2\to1)=\psi(W-E_2)$ for the opposite transition.
By requiring the detailed balance condition \eqref{e:db2}
\begin{equation}
\frac{c(1\to2)}{c(2\to1)}=\frac{\psi(W-E_1)}{\psi(W-E_2)}=e^{\beta(E_1-E_2)},
\label{e:ccppeb}
\end{equation}
we find that the only possible choice is $\psi(h)=c\,e^{-\beta h}$ with some constant $c$.
Thus we find that 
\begin{equation}
c(1\to2)=\varepsilon\,e^{\beta E_1},\quad
c(2\to1)=\varepsilon\,e^{\beta E_2},
\label{e:c12c21}
\end{equation}
with $\varepsilon=c\,e^{-\beta W}$.
Note that the assumption of very high energy barrier led us to the essentially unique choice of the transition rates\footnote{
Also note that the rates used here do {\em not}\/ fit into the general form assumed in 
section~\ref{s:sp1}.}.

Now we come  back to the task of realizing a weak contact between two sublattices of driven lattice gas defined on $\La_1$ and $\La_2$.
We assume that, for each $j=0,\pm1,\ldots,\pm(\ell-1)/2$, a particle can hop between the site\footnote{
Note that both the sites are in the bulk of each subsystem since we impose periodic boundary conditions.
We have chosen this definition since our main purpose is to characterize bulk properties of nonequilibrium steady states rather than to investigate properties of realistic contact.
See section~\ref{s:faqwall}.
In equilibrium, the second law guarantees that it does not make any difference (in macroscopic scale) whether one puts boundary sites or bulk sites into contact.
We still do not know whether nonequilibrium steady states possess similar robustness.
} $(j,0)$ in $\La_1$ and the corresponding site $(j,0)$ in $\La_2$.
We also assume that the particle must go over a very high energy barrier to execute such a hop.
Then we see that the transition rates should be of the form \eqref{e:c12c21}, where $E_1$, $E_2$ should be interpreted as local energies.
Then we get\footnote{
Going back to the definition \eqref{e:etaxy} of $\etabx$, one sees immediately
that the values (which can be 0 or 1) of $\etax$ and $\zetax$
are simply exchanged
in the process $(\etab,\zetab)\to(\etabx,\zetabx)$,
no matter what $\etab$ and $\zetab$ are.
}
\begin{eqnarray}
c[(\etab,\zetab)\to(\etabx,\zetabx)]&=&
\varepsilon\,\etax(1-\zetax)\,\exp[\beta\{H_{\La_1}(\etab)-H_{\La_1}(\etabx)+u_1\}]
\ret
&&+
\varepsilon\,(1-\etax)\zetax\,\exp[\beta\{H_{\La_2}(\zetab)-H_{\La_2}(\zetabx)+u_2\}],
\label{e:cetazeta}
\end{eqnarray}
for $\xb=(j,0)$ with $j=0,\pm1,\ldots,\pm(\ell-1)/2$.
We set $c[(\etab,\zetab)\to(\etab',\zetab')]=0$ for configurations not of the form \eqref{e:cetazeta}. 
The first term in \eqref{e:cetazeta} represents a hop from $\La_1$ to $\La_2$ and the second term from $\La_2$ to $\La_1$.
Note that we do not have terms involving the electric field $E$ in \eqref{e:cetazeta} since we imagine that particles hop in the direction orthogonal to the field.
Indeed for any nonvanishing $c[(\etab,\zetab)\to(\etabx,\zetabx)]$ we have the detailed balance condition (see \eqref{e:db2})
\begin{equation}
\frac{c[(\etab,\zetab)\to(\etabx,\zetabx)]}{c[(\etabx,\zetabx)\to(\etab,\zetab)]}
=
\exp[\beta\{
H_{\rm tot}^{(u_1,u_2)}(\etab,\zetab)
-
H_{\rm tot}^{(u_1,u_2)}(\etabx,\zetabx)
\}].
\label{e:DB12}
\end{equation}

We investigate the steady state of this model with the total particle
number $|\etab|+|\zetab|=2N$ where $N$ is a constant.
When $\varepsilon=0$, the two sublattices decouple, and hence
the distribution 
$p^{(T,E)}_{\La_1,N_1}(\etab)\,p^{(T,E)}_{\La_2,N_2}(\zetab)$
with any fixed $N_1$, $N_2$ such that $N_1+N_2=2N$ is stationary.
We want to see how this situation is modified in the lowest order of
$\varepsilon>0$.

In the limit $\varepsilon\to0$, where particle exchanges between 
$\La_1$ and $\La_2$ are infinitesimally rare,
the states within $\La_1$ and $\La_2$ first reach their steady states.
When a particle hops between $\La_1$ and $\La_2$, again the 
states in the two sublattices relax to the steady states with the new
particle numbers.
Such a separation of time scale enables us to study stochastic
dynamics of particle exchange between $\La_1$ and $\La_2$
separately from the relaxation process in each sublattice.

We fix the total particle number $2N$, and denote by $N_1$ the number of particles in $\La_1$.
Since the number of particles in $\La_2$ is automatically known to be $N_2=2N-N_1$, we only specify $N_1$ in what follows.
Since a hop between the two sublattices takes place in the 
distribution 
$p^{(T,\nu)}_{\La_1,N_1}(\etab)\,p^{(T,\nu)}_{\La_2,N_2}(\zetab)$,
the effective transition rate for a hop from $\La_1$ to $\La_2$ is obtained
by averaging the sum of the rates \eqref{e:cetazeta} as
\begin{eqnarray}
&&\tilde{c}(N_1\to N_1-1)
\ret
&&
=
\bkt{
\sum_{j=-(\ell-1)/2}^{(\ell-1)/2}
\eta_{(j,0)}(1-\zeta_{(j,0)})\,
c[(\etab,\zetab)\to(\etab^{(j,0)},\zetab^{(j,0)})]
}^{(T,E)}_{\La_1,N_1;\La_2,N_2}
\ret
&&
=\varepsilon
\hspace{-6pt}\sum_{j=-(\ell-1)/2}^{(\ell-1)/2}\hspace{-6pt}
\bkt{
\eta_{(j,0)}(1-\zeta_{(j,0)})
\exp[\beta\{
H_{\La_1}(\etab)-H_{\La_1}(\etab^{(j,0)})+u_1
\}]
}^{(T,E)}_{\La_1,N_1;\La_2,N_2},
\label{e:ctil}
\end{eqnarray}
where the average is taken over 
$p^{(T,E)}_{\La_1,N_1}(\etab)\,p^{(T,E)}_{\La_2,N_2}(\zetab)$.
By noting that the average separates into those for each sublattice, and using the  translation invariance, we find
\begin{equation}
\tilde{c}(N_1\to N_1-1)
=\varepsilon\,\ell\,(1-\frac{N_2}{V})\,g(N_1)\,e^{\beta\,u_1},
\label{e:cgg1}
\end{equation}
where $V=\ell h$ and
\begin{equation}
g(N)=\bkt{
\etax\,\exp\sqbk{\beta\{\HL(\etab)-\HL(\etabx)\}}
}^{(T,E)}_{\La,N}
=\langle\,
\etax\,\exp[-\beta J\sum_{\yb;|\yb-\xb|=1}\eta_{\yb}]\,
\rangle^{(T,E)}_{\La,N},
\label{e:gm}
\end{equation}
where we substituted the Hamiltonian \eqref{e:Hgee}.
Here $\xb$ is one of $(j,0)$, but it can actually be an arbitrary site in $\La$ because of the translation invariance.

Similarly we have 
\begin{equation}
\tilde{c}(N_1\to N_1+1)
=\varepsilon\,\ell\,(1-\frac{N_1}{V})\,g(N_2)\,e^{\beta\,u_2},
\label{e:cgg2}
\end{equation}
for a hop from $\La_2$ to $\La_1$.

\subsubsection{Chemical potential of lattice gases}
\label{s:mulg}
To find the chemical potential let $\tilde{N}_1$ and $\tilde{N}_2$ be 
the average particle numbers in the steady state of the whole system.
Since hops between the two sublattices balance with each other,
we can assume that
\begin{equation}
\tilde{c}(\tilde{N}_1\to\tilde{N}_1-1)
=
\tilde{c}(\tilde{N}_1-1\to\tilde{N}_1).
\label{e:cct}
\end{equation}
Then by using \eqref{e:cgg1} and \eqref{e:cgg2}, we get
\begin{equation}
\frac{1}{\beta}\log\frac{g(\tilde{N}_1)}{1-(\tilde{N}_1-1)/V}+u_1
=
\frac{1}{\beta}\log\frac{g(\tilde{N}_2+1)}{1-\tilde{N}_2/V}+u_2.
\label{e:ggu}
\end{equation}
By comparing this equality with our definition \eqref{e:mumu} of the
chemical potential, we are led to define
\begin{equation}
\mu(\rho)=\frac{1}{\beta}\log\frac{g(\rho\ell h)}{1-\rho+V^{-1}}
\simeq\frac{1}{\beta}\log\frac{g(\rho\ell h)}{1-\rho},
\label{e:mudef}
\end{equation}
where $\rho=N/(\ell h)$ is the density.
We assume that $\mu(\rho)$ defined by \eqref{e:mudef} has a sensible 
infinite volume limit $\ell,h\to\infty$ (i.e., $V\to\infty$).
Then \eqref{e:ggu} precisely coincides with the desired balance equation \eqref{e:mumu}.

Recalling the definitions \eqref{e:gm} of $g(N)$, 
we stress that we have obtained a concrete formula \eqref{e:mudef} for the chemical
potential which involve expectation values in the steady state.
Therefore one can in principle compute $\mu(\rho)=\mu(T,E;V,N)$
by using systematic approximations or a computer.
See section~\ref{s:pert}.
Note that our formula \eqref{e:mudef} for the chemical potential involves
correlation functions of only local quantities.
Therefore it is not affected by the long range power law correlations
which are found universally in nonequilibrium steady states of lattice gases.
This can be clearly seen in our perturbative calculation in  section~\ref{s:pert}.
See section~\ref{s:faqlrc} for more discussions about long range correlations.

Note also that in \eqref{e:mudef} we have written down $\mu(\rho)$ without any
ambiguities.
To be precise, the relation \eqref{e:mumu} only determines the 
$\rho$ dependence of the chemical potential.
So there is a freedom to add any function $\mu_0(T,E)$ to \eqref{e:mudef}.
But the simplicity of the formula  \eqref{e:mudef} suggests that this is 
the ``right'' choice.
Indeed a consideration about $\mu$-walls
in section~\ref{s:mmw} indicates that \eqref{e:mudef}
is the complete formula for the chemical potential.

\subsubsection{Pressure of lattice gases}
\label{s:wp}
We shall define pressure for the driven lattice gas.
Since lattice gas models do not have the notion of momentum, the standard definition in terms of mechanical forces do not apply.
Instead we here 
obtain pressure from the mechanical work required to change the volume (area).

\begin{figure}
\centerline{\epsfig{file=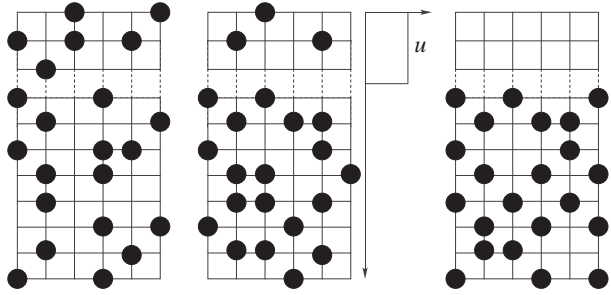,width=8cm}}
\caption[dummy]{
The setting used to determine the pressure in the lattice gas.
By changing the potential $u$ applied to the smaller subsystem from 0 to $\infty$,
we can effectively reduce the volume (area) of the system by $\ell\tilde{h}$.
From the mechanical work required to perform this change,
we define the pressure.
}
\label{f:lgp}
\end{figure}

We consider two lattices with the same width $\ell$
and different heights $h$ and $\tilde{h}$.
We assume $h\gg\tilde{h}$.
We place the smaller lattice on top of the larger one, and put them into a weak contact with each other
as in Fig.~\ref{f:lgp}.
We apply a uniform potential $u$ only to the smaller subsystem,
and  keep $T$ and $E$ constant over the whole system.

We first set $u=0$ and let the whole system reach its steady state.
Since the conditions are the same everywhere,
we get a uniform local steady state with density $\rho$
over the whole system.
Then we slowly vary $u$ from 0 to $\infty$.
When $u=\infty$, there are no particles in the smaller subsystem, and we have
a uniform local steady state with $T$, $E$ and the density 
$\rho\{1+(\tilde{h}/h)\}\simeq\rho$ in the larger subsystem.

Let $\tilde{\rho}(u)$ be the density in the smaller subsystem
when the potential is $u$.
By the balance condition \eqref{e:mumu}, this density can be 
determined by
\begin{equation}
\mu(\rho)=\mu(\tilde{\rho}(u))+u,
\label{e:mmru}
\end{equation}
where we set the density in the larger subsystem to
$\rho$.
This is allowed since the deviation is of order $\rho\tilde{h}/h$,
and we can make $\tilde{h}/h$ as small as we want.
By differentiating \eqref{e:mmru} by $u$, we get
\begin{equation}
\frac{d\tilde{\rho}(u)}{du}=-\frac{1}{\mu'(\tilde{\rho}(u))}.
\label{e:drdu}
\end{equation}

Now we evaluate the mechanical work required to change the potential
$u$.
Since the total number of particles in the smaller subsystem is 
$\tilde{\rho}(u)\,\ell\,\tilde{h}$, the work needed to
change the potential from $u$ to $u+\Delta u$ is equal to
$\Delta W=\Delta u\,\tilde{\rho}(u)\,\ell\,\tilde{h}+O((\Delta u)^2)$.
The total work is obtained by summing this up as
\begin{equation}
W=\ell\,\tilde{h}\int_0^\infty du\,\tilde{\rho}(u)
=\ell\,\tilde{h}\int_0^\rho d\tilde{\rho}\,\tilde{\rho}\,\mu'(\tilde{\rho}),
\label{e:W}
\end{equation}
where we used \eqref{e:drdu} to change the variable.

Defining the pressure $p$ by the standard relation $W=p\,\Delta V$
with $\Delta V=\ell\,\tilde{h}$ being the volume (area) of the smaller subsystem,
\eqref{e:W} implies
\begin{equation}
p(\rho)=\int_0^\rho d\tilde{\rho}\,\tilde{\rho}\,\mu'(\tilde{\rho}),
\label{e:pr}
\end{equation}
which is the final formula of pressure in the lattice gas.
Note that the Maxwell relation \eqref{e:Max} is obvious
from the formula \eqref{e:pr}.

\subsubsection{Free energy of lattice gases}
\label{s:fe}
Now that we have the formulae for the chemical potential
\eqref{e:mudef} and the pressure \eqref{e:pr}, we use the Euler equation
\eqref{e:Eu} in the form
\begin{equation}
f(\rho)=-p(\rho)+\rho\,\mu(\rho)=\int_0^\rho d\tilde{\rho}\,\mu(\tilde{\rho}),
\label{e:Euf}
\end{equation}
to define the specific free energy 
$f(\rho)=F(T,E;1,\rho)=F(T,E;V,N)/V$ with $\rho=N/V$.
The second equality in \eqref{e:Euf} is obtained by using the formula \eqref{e:pr}
and integrating by parts.

We have thus obtained concrete formulae for the chemical potential,
the pressure, and the free energy of a nonequilibrium steady state of 
the lattice gas.
The formulae only involve local correlation functions in the steady state distribution,
and may be computed in a concrete model.

\subsection{Steady state and density fluctuation}
\label{s:dfw}
We will now study the steady state distribution of a weakly coupled system
and show that Einstein's formula \eqref{e:pN12} for the density fluctuation holds
exactly in the present setting.
Recall that
in equilibrium statistical mechanics, the relation \eqref{e:pN12}
follows trivially from the canonical distribution (section~\ref{s:sm}).
We note that an analogous derivation is never possible 
here since we still do not  know
anything about general forms of steady state distribution in
nonequilibrium systems.
Nevertheless the same relation can be proved by examining the effective
stochastic process of particle exchange between the two subsystems.

Consider again the situation in section~\ref{s:wc}, where two identical systems
on $\La_1$ and $\La_2$ are weakly coupled with each other.
There are uniform potentials $u_1$ and $u_2$ on $\La_1$ and $\La_2$,
respectively.

We are interested in the steady state with the total particle number
$|\etab|+|\zetab|=2N$.
When $\varepsilon=0$, 
the distribution 
$p^{(T,E)}_{\La_1,N_1}(\etab)\,p^{(T,E)}_{\La_2,N_2}(\zetab)$
with any fixed $N_1$, $N_2$ such that $N_1+N_2=2N$ is 
stationary\footnote{
Recall that $p^{(T,E)}_{\La,N}(\etab)$ is the steady state distribution
for the system on $\La$, and is nonvanishing only when $|\etab|=N$.
}.
If we take into account the effect of $\varepsilon>0$ in the lowest order,
the above degeneracy is lifted and we get a unique steady 
state distribution of the form
\begin{equation}
p^{(T,E;u_1,u_2)}_{\La_1,\La_2;2N}(\etab,\zetab)=
\tilde{p}^{(T,E;u_1,u_2)}_{\La_1,\La_2;2N}(|\etab|,|\zetab|)\,
p^{(T,E)}_{\La_1,|\etab|}(\etab)\,p^{(T,E)}_{\La_2,|\zetab|}(\zetab),
\label{e:pss}
\end{equation}
where $\tilde{p}^{(T,E;u_1,u_2)}_{\La_1,\La_2;2N}(N_1,N_2)$
is nonvanishing only when $N_1+N_2=2N$.
We again specify only $N_1$, and abbreviate $\tilde{p}^{(T,E;u_1,u_2)}_{\La_1,\La_2;2N}(N_1,N_2)$ simply as $\tilde{p}(N_1)$.
It  is the probability of finding $N_1$ particles in $\La_1$ and
$2N-N_1$ particles in $\La_2$, and is normalized as
\begin{equation}
\sum_{N_1}\tilde{p}(N_1)=1.
\label{e:pNN1}
\end{equation}
We will now evaluate $\tilde{p}(N_1)$.

We substitute the steady state distribution \eqref{e:pss} into the general
condition \eqref{e:std} of stationarity, and then take partial sums over
$\etab$ and $\zetab$ with fixed $|\etab|$ and
$|\zetab|$.
By recalling the definition \eqref{e:ctil} of the effective transition rate
for hop between the two sublattices, this gives
\begin{equation}
\sum_{\sigma=\pm1}
\{
-\tilde{c}(N_1\to N_1+\sigma)\,\tilde{p}(N_1)
+\tilde{c}(N_1+\sigma\to N_1)\,
\tilde{p}(N_1+\sigma)\}
=0,
\label{e:ptss}
\end{equation}
for any $N_1$. 

To solve \eqref{e:ptss}, we try the ansatz
\begin{equation}
\tilde{p}(N_1)=q^{(u_1)}(N_1)\,q^{(u_2)}(2N-N_1),
\label{e:pqq}
\end{equation}
with a function $q^{(u)}(N)$.
We also assume that each term in the sum in the left-hand side of
\eqref{e:ptss} vanishes.
This is a kind of detailed balance condition.
By using \eqref{e:cgg1} and \eqref{e:cgg2}, we find that this condition
is satisfied if
\begin{equation}
e^{\beta u}\,g(N)\,q^{(u)}(N)=(1-\frac{N-1}{V})\,q^{(u)}(N-1),
\label{e:egq}
\end{equation}
which means
\begin{equation}
q^{(u)}(N)=
q^{(u)}(0)\,e^{-\beta u N}\prod_{M=1}^N\frac{1-(M-1)/V}{g(M)}
={\rm const.}\,
\exp\sqbk{
-\beta u N-\beta\sum_{M=1}^N\mu(\frac{M}{V})
},
\label{e:qeg}
\end{equation}
where we used \eqref{e:mudef}.
Note that 
\begin{equation}
\sum_{M=1}^N\mu(\frac{M}{V})
\simeq
V\int_0^\rho d\tilde{\rho}\,\mu(\tilde{\rho})
=V\,f(\rho),
\label{e:imuf}
\end{equation}
where we used \eqref{e:Euf} and wrote $\rho=N/V$.
Thus from \eqref{e:qeg} and \eqref{e:pqq}, we get the desired estimate
\begin{equation}
\tilde{p}^{(T,E;u_1,u_2)}_{\La_1,\La_2;2N}(N_1,N_2)
=
{\rm const.}\,
\exp[-\beta V\{u_1\rho_1+u_2\rho_2+f(\rho_1)+f(\rho_2)\}],
\label{e:pNNe}
\end{equation}
where $\rho_1=N_1/V_1$ and $\rho_2=N_2/V_2$.

If one sets $u_1=u_2=0$, \eqref{e:pNNe} becomes
\begin{equation}
\tilde{p}^{(T,E)}_{\La_1,\La_2;2N}(N_1,N_2)
=
{\rm const.}\,
\exp[-\beta\{F(T,E;V,N_1)+F(T,E;V,N_2)\}],
\label{e:pte}
\end{equation}
which is precisely Einstein's formula \eqref{e:pN12} for density fluctuation.
It should be stressed that this is an exact relation, which covers
both small fluctuations and large deviations.
At least in the setting of weakly coupled lattice gases, we have shown 
rigorously that the density fluctuation in nonequilibrium steady states
is exactly governed by the SST free energy.

It is easy to extend the fluctuation formula to the case where 
more than two subsystems are weakly coupled with each other.

\subsection{Minimum work principle}
\label{s:mwp}
Let us continue and study the steady state distribution in more detail.
This will lead us to the minimum work principle
for steady states.

By substituting \eqref{e:pNNe} to the full expression \eqref{e:pss}
of the steady state distribution, we get the  final 
formula for the steady state distribution
\begin{equation}
p^{(T,E;u_1,u_2)}_{\La_1,\La_2;2N}(\etab,\zetab)=
c_0\,e^{\beta\,\Phi(u_1,u_2)
-\beta\{u_1|\etab|+u_2|\zetab|+F(|\etab|)+F(|\zetab|)\}
}\,
p^{(T,E)}_{\La_1,|\etab|}(\etab)\,p^{(T,E)}_{\La_2,|\zetab|}(\zetab),
\label{e:pfin}
\end{equation}
for $\etab$ and $\zetab$ such that $|\etab|+|\zetab|=2N$,
where $F(N)$ is a shorthand for $F(T,E;V,N)$.
We wrote the (unknown) normalization factor as\footnote{
We may set $c_0=1$.
But with the freedom of adjusting $c_0$, we do not have to worry about additive constant in $\Phi(u_1,u_2)$.
} $c_0\exp[\beta\,\Phi(u_1,u_2)]$, where $c_0$ does not depend on $u_1$ and $u_2$.
We now investigate the function $\Phi(u_1,u_2)$, and show that it is nothing but a Legendre transformation of the free energy $F(N)$.

We fix all the parameters except $u_1$, $u_2$, and denote the average over 
the distribution \eqref{e:pfin} as $\bkt{\cdots}_{u_1,u_2}$.
Let us write
\begin{equation}
N_1(u_1,u_2)=\bkt{|\etab|}_{u_1,u_2},\quad
N_2(u_1,u_2)=\bkt{|\zetab|}_{u_1,u_2}.
\label{e:N1N2}
\end{equation}

From the normalization condition
\begin{equation}
\sumtwo{\etab,\zetab}{(|\etab|+|\zetab|=2N)}
p^{(T,E;u_1,u_2)}_{\La_1,\La_2;2N}(\etab,\zetab)
=1,
\label{e:pnorm}
\end{equation}
and the expression \eqref{e:pfin}, we find
\begin{equation}
\Phi(u_1,u_2)=-\frac{1}{\beta}\log
\bkt{
e^{-\beta(u_1|\etab|+u_2|\zetab|)}
}_{0,0}.
\label{e:Phib}
\end{equation}
By differentiating \eqref{e:Phib}, we get
\begin{equation}
\partialf{\Phi(u_1,u_2)}{u_1}=N_1(u_1,u_2),\quad
\partialf{\Phi(u_1,u_2)}{u_2}=N_2(u_1,u_2),
\label{e:Phidif}
\end{equation}
which can be regarded as differential equations for determining $\Phi(u_1,u_2)$.

We now claim that the solution of \eqref{e:Phidif} is given by
\begin{equation}
\Phi(u_1,u_2)=\sum_{i=1}^2\{
F(N_i(u_1,u_2))+u_i\,N_i(u_1,u_2)
\}.
\label{e:Phian}
\end{equation}
To see this, we simply differentiate \eqref{e:Phian} by $u_1$ to get
\begin{equation}
\partialf{\Phi(u_1,u_2)}{u_1}=N_1(u_1,u_2)
+\sum_{i=1}^2\sqbk{
\partialf{N_i(u_1,u_2)}{u_1}
\{\mu(\frac{N_i(u_1,u_2)}{V})+u_i\}
},
\label{e:Phiu}
\end{equation}
where we used \eqref{e:muF}.
Note that the balance condition \eqref{e:mumu} implies that 
$\mu(N_i(u_1,u_2)/V)+u_i$ is independent of $i=1,2$.
Then since $N_1(u_1,u_2)+N_2(u_1,u_2)=2N$ is constant,
the second term in the right-hand side of \eqref{e:Phiu} vanishes.
We thus get the first equation in \eqref{e:Phidif}.
The second equation follows similarly.

We have thus determined the steady state distribution \eqref{e:pfin}
including the normalization function.
The formula \eqref{e:Phian} shows that $\Phi(u_1,u_2)$ is obtained
from $F(T,E;V,N)$ by a  Legendre transformation.

At this stage we shall see
what we get from the general second law of Markov
processes described in Appendix~\ref{s:M2nd}.

We consider a time-dependent Markov process where the potentials $u_1(t)$,
$u_2(t)$ become time-dependent in the present model.
The pair $(u_1,u_2)$ correspond to the parameter $\alpha$
in Appendix~\ref{s:M2nd}.
The system is initially in the steady state corresponding to
$(u_1(0),u_2(0))$, and we vary $u_1(t)$, $u_2(t)$ in an arbitrary manner
for $0\le t\le t_{\rm f}$.

Using \eqref{e:pfin}, we see that the function $\varphi^{(\alpha)}$
of \eqref{e:phia}, which play a fundamental role in the second law,
becomes
\begin{equation}
\varphi^{(u_1,u_2)}(\etab,\zetab)
=
\beta\{u_1|\etab|+u_2|\zetab|-\Phi(u_1,u_2)\}
-\log p^{(T,E;0,0)}_{\La_1,\La_2;2N}(\etab,\zetab).
\label{e:phiu}
\end{equation}
Therefore the second law reads
\begin{equation}
\int_0^{t_{\rm f}}dt\cbk{
\frac{du_1(t)}{dt}\bkt{|\etab|}_t
+\frac{du_2(t)}{dt}\bkt{|\zetab|}_t}
\ge
\Phi(u_1(t_{\rm f}),u_2(t_{\rm f}))-\Phi(u_1(0),u_2(0)).
\label{e:mwpP}
\end{equation}
Since $\bkt{|\etab|}_t$ and $\bkt{|\zetab|}_t$ are the number of particles
in $\La_1$ and $\La_2$, respectively, the left-hand side of \eqref{e:mwpP}
is precisely the mechanical work $W$ needed to change the potentials
$u_1(t)$, $u_2(t)$.
Thus \eqref{e:mwpP} becomes
\begin{equation}
W\ge\Phi(u_1(t_{\rm f}),u_2(t_{\rm f}))-\Phi(u_1(0),u_2(0)),
\label{e:WPP}
\end{equation}
which is the minimum work principle for steady states.
Note that $\Phi$ rather than $F$ appears on the right-hand side
because we are controlling the potentials $u_1$, $u_2$, rather than
the volume.
If we set $u_1(0)=u_2(0)=0$ and $u_1(t_{\rm f})=0$, $u_2(t_{\rm f})=\infty$, 
then \eqref{e:WPP} becomes
\begin{equation}
W\ge F(T,E;V,2N)-2\,F(T,E;V,N)=F(T,E;V,2N)-F(T,E;2V,2N),
\label{e:WFF}
\end{equation}
which is precisely of the form \eqref{e:mwpV}
of the conjectured minimum work principle.

It is an easy exercise to extend the minimum work principle
\eqref{e:WPP} to the cases where $n$ subsystems are weakly coupled with
each other.
Then one is allowed to change the potentials 
$u_1(t),\ldots,u_n(t)$ in an arbitrary manner.

The present minimum work principle, although valid for any 
$u_1(t)$, $u_2(t)$, may not be too exciting since weak coupling 
ensures that each subsystem is in a steady state for any $t$.
But see the following remark.

\Rem
The essences of the minimum work principle \eqref{e:WPP}
may be the following property that we shall call
``weak canonicality.''
Consider a Markov process for a lattice gas, and let
$p_0(\etab)$ be the stationary distribution for the case where no potential is
applied to the system.
We then apply a potential $u(\cdot)$ to the system and 
denote the new stationary distribution as $p_{u(\cdot)}(\etab)$.

We say that the system has ``weak canonicality'' for a class of
potentials $\calU$ if
\begin{equation}
p_{u(\cdot)}(\etab)\simeq
\exp\sqbk{\beta\Phi[u(\cdot)]-\beta\sum_{\xiL}u(x)\,\etax}
p_0(\etab),
\label{e:wcan}
\end{equation}
for any $u(\cdot)\in\calU$, where the function $\Phi[u(\cdot)]$ ensure the normalization.

If \eqref{e:wcan} holds we can easily show a minimum work principle
\begin{equation}
W\ge\Phi[u_{t_{\rm f}}(\cdot)]-\Phi[u_{0}(\cdot)],
\label{e:mwpgen}
\end{equation}
for a time-dependent Markov process where $u_t(\cdot)$ varies
within $\calU$.
We can also show the expected formula for density fluctuation and a version of fluctuation-response relation based on \eqref{e:wcan}.

We therefore regard it an important task to study which models possess
weak canonicality for which classes of potentials.
So far we only know of some simple examples.

\subsection{Linear response}
\label{s:DLGLR}
We can also develop a linear response theory for time-dependent phenomena which take place when the potentials $u_1$, $u_2$ become time-dependent.
As we have already remarked in section~\ref{s:WCSL}, it may not be too surprising that we can study linear transport phenomena, since we are dealing only with transport through the weak contact.
Nevertheless it might be of some importance that a linear response theory in a highly nonequilibrium system can be constructed unambiguously.
Recent numerical study \cite{Hayashi05} indicates the validity of the fluctuation response relation  in DLG when a weak probing field orthogonal to the strong driving field is applied.
This finding is similar to the result we present here, but (theoretically speaking) is more delicate.

Since the potential affects only hopping between $\La_1$ and $\La_2$, we only need to treat slow stochastic dynamics of particle exchange.
We again fix the total number of particles to $2N$, and only specify $N_1$, the number of particles in $\La_1$.

For simplicity\footnote{
Extension to models with varying $u_1$ is automatic.
} we set $u_1=0$ and $u_2=\lambda\,h(t)$, where $h(t)$ is an arbitrary (not too pathological) function such that $h(t)=0$ for $t<0$.
$\lambda$ is a small parameter in which we shall expand.
We denote by $\tilde{p}_t(N_1)$ the probability that there are $N_1$ particles in $\La_1$ (and hence $2N-N_1$ particles in $\La_2$) at time $t$.
It satisfies the master equation 
\begin{equation}
\frac{d}{dt}\tilde{p}_t(N_1)=\sum_{\sigma=\pm1}
\{-\tilde{p}_t(N_1)\,\tilde{c}_{\lambda}(N_1\to N_1+\sigma)+
\tilde{p}_t(N_1+\sigma)\,
\tilde{c}_{\lambda}(N_1+\sigma\to N_1)\},
\label{e:ptmaster}
\end{equation}
for any $t\ge0$ and $N_1$, where $\tilde{c}_{\lambda}(N_1\to N_1+\sigma)$ is the transition rate \eqref{e:cgg1} or \eqref{e:cgg2} with $u_1=0$ and $u_2=\lambda\,h(t)$.
The initial distribution $\tilde{p}_0(N_1)$ is taken to be the stationary distribution (studied in section~\ref{s:dfw}) with $u_1=u_2=0$.
We wish to solve \eqref{e:ptmaster} in the lowest order of $\lambda$.

From  \eqref{e:cgg1} and \eqref{e:cgg2}, we find that
\begin{equation}
\tilde{c}_{\lambda}(N_1\to N_1-1)
=\tilde{c}_{0}(N_1\to N_1-1)
\label{e:clc01}
\end{equation}
and
\begin{eqnarray}
\tilde{c}_{\lambda}(N_1\to N_1+1)
&=&\tilde{c}_{0}(N_1\to N_1+1)\,e^{\lambda\,\beta\,h(t)}
\ret
&=&
\tilde{c}_{0}(N_1\to N_1+1)+
\lambda\,\beta\,h(t)\,\tilde{c}_{0}(N_1\to N_1+1)+O(\lambda^2).
\label{e:clc02}
\end{eqnarray}
By substituting \eqref{e:clc01} and \eqref{e:clc02} into the master equation \eqref{e:ptmaster}, we get
\begin{eqnarray}
&&\frac{d}{dt}\tilde{p}_t(N_1)=\sum_{N_1'}\Gamma(N_1,N_1')\,\tilde{p}_t(N_1')
\ret
&&+\lambda\,\beta\,h(t)\,
\{-\tilde{p}_t(N_1)\,\tilde{c}_0(N_1\to N_1+1)+
\tilde{p}_t(N_1-1)\,\tilde{c}_0(N_1-1\to N_1)\}
+O(\lambda^2),
\label{e:ptmaster2}
\end{eqnarray}
where we have defined $\Gamma(N_1,N_1')$ by 
\begin{equation}
\sum_{N_1'}\Gamma(N_1,N_1')\,q(N_1')
=\sum_{\sigma=\pm1}
\{-q(N_1)\,\tilde{c}_{0}(N_1\to N_1+\sigma)+
q(N_1+\sigma)\,
\tilde{c}_{0}(N_1+\sigma\to N_1)\},
\label{e:Gammadef}
\end{equation}
for any $q(\cdot)$.

The time evolution operator $U_t(N_1,N_1')$ for $u_1=u_2=0$ is defined as the solution of 
\begin{equation}
\frac{d}{dt}U_t(N_1,N_1')=\sum_{N_1''}\Gamma(N_1,N_1'')\,
U_t(N_1'',N_1'),
\label{e:Utdef}
\end{equation}
with the initial condition $U_0(N_1,N_1')=\delta_{N_1,N_1'}$.
Then it is standard\footnote{
A quick derivation.
Here we write $N$ instead of $N_1$.
Write \eqref{e:ptmaster2} as $d\tilde{p}_t(N)/dt=\sum_{N'}\{\Gamma(N,N')+\lambda\,\chi_t(N,N')\}\,\tilde{p}_t(N)+O(\lambda^2)$, and write the solution as $\tilde{p}_t(N)=\tilde{p}_0+\lambda\,q_t(N)+O(\lambda^2)$.
Then by noting that $\sum_{N'}\Gamma(N,N')\,\tilde{p}_0(N')=0$, one finds that $q_t(N)$ satisfies the differential equation $dq_t(N)/dt=\sum_{N'}\Gamma(N,N')\,q_t(N')+\lambda\sum_{N'}\chi_t(N,N')\,\tilde{p}_0(N')$.
It is easy to check that the solution is written as $q_t(N)=\int_0^tds\sum_{N',N''}U_{t-s}(N,N')\,\chi_s(N',N'')\,p_0(N'')$.
} that the perturbative solution of the master equation \eqref{e:ptmaster2} is given by
\begin{eqnarray}
&&\tilde{p}_t(N_1)=
\tilde{p}_0(N_1)+\lambda\,\beta\int_0^tds\sum_{N_1'}
U_{t-s}(N_1,N_1')\,h(s)\times
\ret
&&
\times
\{-\tilde{p}_0(N_1')\,\tilde{c}_0(N_1'\to N_1'+1)+
\tilde{p}_0(N_1'-1)\,\tilde{c}_0(N_1'-1\to N_1')\}+O(\lambda^2)
\ret
&&=
\tilde{p}_0(N_1)-\lambda\,\beta\int_0^tds\sum_{N_1'}\sum_{\sigma=\pm1}
U_{t-s}(N_1,N_1')\,h(s)\,\sigma\,\tilde{p}_0(N_1'+\sigma)\,\tilde{c}_0(N_1'+\sigma\to N_1')+O(\lambda^2),
\label{e:ptsol1}
\end{eqnarray}
where we used the detailed balance condition $\tilde{p}_0(N_1')\,\tilde{c}_0(N_1'\to N_1'+1)=\tilde{p}_0(N_1'+1)\,\tilde{c}_0(N_1'+1\to N_1')$ to get the final expression.

We define the ``current operator'' by
\begin{equation}
j_{N_1,N_1'}=\tilde{c}_0(N_1'\to N_1)\,(N_1'-N_1),
\label{e:jNNdef}
\end{equation}
which measure the number of particle (which number is either 0, 1 or $-1$) that have passed from $\La_1$ to $\La_2$ at a given instant.
Then the solution \eqref{e:ptsol1} can be written in the form
\begin{equation}
\tilde{p}_t(N_1)=
\tilde{p}_0(N_1)-\lambda\,\beta\int_0^tds\sum_{N_1',N_1''}
U_{t-s}(N_1,N_1')\,h(s)\,j_{N_1',N_1''}\,\tilde{p}_0(N_1'')
+O(\lambda^2),
\label{e:DLGLR}
\end{equation}
which is the basic equation of our linear response theory.

The average $\sbkt{\hat{N}_1(t)}_\lambda=\sum_{N_1}N_1\,\tilde{p}_t(N_1)$ of the number of particles in $\La_1$ at time $t$ can be evaluated by using \eqref{e:DLGLR} as
\begin{equation}
\sbkt{\hat{N}_1(t)}_\lambda
=
N
-\lambda\,\beta\int_0^tds\sum_{N_1,N_1',N_1''}N_1\,
U_{t-s}(N_1,N_1')\,h(s)\,j_{N_1',N_1''}\,\tilde{p}_0(N_1'')
+O(\lambda^2),
\label{e:N1avDLG}
\end{equation}
where we noted that the average in the steady state with $u_1=u_2=0$ satisfies $\sum_{N_1}N_1\,\tilde{p}_0(N_1)=N$.
We can further rewrite \eqref{e:N1avDLG} as
\begin{equation}
\sbkt{\hat{N}_1(t)}_\lambda
=
N
-\lambda\,\beta\int_0^tds\,h(s)\,
\sbkt{\hat{N}_1(t)\,\hat{j}(s)}_0
+O(\lambda^2),
\label{e:N1avDLG2}
\end{equation}
where the definition of the time-dependent correlation function $\sbkt{\hat{N}_1(t)\,\hat{j}(s)}_0$ may be clear by comparing \eqref{e:N1avDLG2} with \eqref{e:N1avDLG}.
This is the desired fluctuation-response relation for the particle exchange process in the weakly coupled DLG.

When $h(t)$ is a step function with $h(t)=0$ for $t<0$ and $h(t)=1$ for $t\ge0$, then \eqref{e:N1avDLG2} simplifies as
\begin{eqnarray}
\sbkt{\hat{N}_1(t)}_\lambda
&=&
N
-\lambda\,\beta\int_0^tds\,
\sbkt{\hat{N}_1(t)\,\hat{j}(s)}_0
+O(\lambda^2)
\ret
&=&
N+\lambda\,\beta\,
\sbkt{\hat{N}_1(t)\,\{\hat{N}_1(t)-\hat{N}_1(0)\}}_0
+O(\lambda^2).
\label{e:N1avDLG3}
\end{eqnarray}

\subsection{$\mu$-wall in the driven lattice gas}
\label{s:mmw}
Let us comment on how the notion of $\mu$-walls introduced in
section~\ref{s:muw} can be implemented in the driven lattice gas.

The answer seems to be very simple.
The weak coupling scheme of section~\ref{s:wc} can be applied to
situations where the subsystems on $\La_1$ and $\La_2$ have 
different values of nonequilibrium parameters.
If we set $E=0$ for the subsystem on $\La_2$,
a contact between a steady state and an equilibrium state
is  realized.
We may assume that this weak coupling realizes a $\mu$-wall.
(But see the remark at the end of this section.)

To see that this defines $\mu(T,E;V,N)$ consistently
(see the discussion at the end of section~\ref{s:muw}),
it suffices to show that the formula \eqref{e:mudef} gives the
standard chemical potential for equilibrium states.
This fact can be checked directly for general cases, but the easiest
way is  to treat the free system with $\HL(\cdot)=0$
at equilibrium.
Then \eqref{e:mudef} gives
\begin{equation}
\mu(\rho)=\frac{1}{\beta}\log\frac{\rho}{1-\rho},
\label{e:mufree}
\end{equation}
which coincides with the result from equilibrium statistical mechanics.
Since it is guaranteed that two systems in a weak contact have the
same values of $\mu(\rho)$, it follows that \eqref{e:mudef}
gives the equilibrium chemical potential in an arbitrary
equilibrium system.

We therefore conclude that the formula \eqref{e:mudef} gives a consistent chemical
potential $\mu(T,E;V,N)$ for a local steady state $(T,E;V,N)$
including its dependence on $T$ and $E$.
Similarly the formula \eqref{e:Euf} defines the corresponding free energy $F(T,E;V,N)$
without ambiguity.

It should be recalled, however, that we concluded in section~\ref{s:muw} that a contact between an equilibrium state and a nonequilibrium steady state cannot be realized in the $(T,\phi;V,N)$ formalism.
If we stick onto the interpretation that $E$ represents the electric field, we must again conclude that our contact is not realistic.
Nevertheless the fact that we can construct (at least theoretically) a consistent $\mu$-wall may be of importance and interest.

\subsection{Perturbative estimate of the SST free energy}
\label{s:pert}
We found that \eqref{e:mudef} can be regarded as the definition of SST chemical potential for the steady state in the driven lattice gas.
It is worth trying to evaluate this  and other thermodynamic quantities explicitly.
Since exact calculations are so far impossible, we here calculate the leading contributions in the limit of low density and high temperature.

First by substituting the definition \eqref{e:gm} of $g(N)$ into the chemical potential \eqref{e:mudef}, and expanding the exponential, we get
\begin{equation}
\mu(\rho)=
\frac{1}{\beta}\log
\frac{\rho-\beta J\sum_{\yb;|\yb-\xb|=1}\bkt{\etax\etay}+O(\rho^3)}
{1-\rho},
\label{e:muexp}
\end{equation}
where we noted that $\bkt{\etax}=\rho$.
In the present section we consider a uniform system on $\La$, and suppress the suffixes $T$, $E$, $\La$, and $N$ because they are always fixed.
We thus need to evaluate the two-point correlation function $\bkt{\etax\etay}$.

We evaluate the leading nonequilibrium correction to  $\bkt{\etax\etay}$ by using the standard procedure.
One first notes that, for any function $h(\etab)$ of the configuration, there is an identity
\begin{equation}
\sum_{\etab}p(\etab)\sum_{\bkt{\ub,\vb}}c(\etab\to\etab^{\ub,\vb})
\{h(\etab^{u,v})-h(\etab)\}=0,
\label{e:pcf}
\end{equation}
which follows from the master equation \eqref{e:maeq}.
$p(\etab)$ is the steady state distribution.
We here set $h(\etab)=\etax\etay$.
We substitute the transition rate (\ref{e:CTE1}) into (\ref{e:pcf}) and keep only those terms which have order $\rho^2$.
After a straightforward (but a little tedious) calculation, we get
\begin{eqnarray}
&&\sumtwo{\ub\in\La}{|\ub-\xb|=1,\ \ub\ne\yb}
\bigl\{-\phi(\beta\{J_{\xb,\yb}-J_{\ub,\yb}+E(x_1-u_1)\})\,\tilde{G}(\xb,\yb)
\ret
&&\hspace{70pt}
+\phi(\beta\{J_{\ub,\yb}-J_{\xb,\yb}+E(u_1-x_1)\})\,\tilde{G}(\ub,\yb)\bigr\}
\ret
&+&\sumtwo{\vb\in\La}{|\vb-\xb|=1,\ \vb\ne\xb}
\bigl\{-\phi(\beta\{J_{\xb,\yb}-J_{\xb,\vb}+E(y_1-v_1)\})\,\tilde{G}(\xb,\yb)
\ret
&&\hspace{70pt}
+\phi(\beta\{J_{\xb,\vb}-J_{\xb,\yb}+E(v_1-y_1)\})\,\tilde{G}(\xb,\vb)\bigr\}=O(\rho^3),
\label{e:pGpG}
\end{eqnarray}
where $\tilde{G}(\xb,\yb)=\bkt{\etax\etay}$, and $J_{\xb,\yb}=J$ if $|\xb-\yb|=1$ and $J_{\xb,\yb}=0$ otherwise.
Since $\sum_{\xb}\etax=N$, the two point function $\tilde{G}(\xb,\yb)$ must satisfy $\sum_{\xb,\yb}\tilde{G}(\xb,\yb)=N(N-1)$.
This normalization condition and the equation (\ref{e:pGpG}) determines $\tilde{G}(\xb,\yb)$ (to the order $\rho^2$).

By using the translation invariance, we can write $\tilde{G}(\xb,\yb)=G(\xb-\yb)+O(\rho^3)$ with a function $G(\xb)$ of $\xb\in\La\backslash\{\ob\}$.
Here $\ob=(0,0)$ is the origin.
From (\ref{e:pGpG}), we see that $G(\xb)$ satisfies the equation\footnote{
This is the same equation as one gets for the driven lattice gas with only two particles as studied in \cite{Tasaki04a}.
}
\begin{equation}
\sumtwo{\yb\in\La}{|\xb-\yb|=1,\ \yb\ne\ob}
\{-\tilde{c}(\xb\to\yb)\,G(\xb)+\tilde{c}(\yb\to\xb)\,G(\yb)\}=0,
\label{e:ctG}
\end{equation}
for any $\xb\in\La$.
Here the effective transition rate is given by
\begin{equation}
\tilde{c}(\xb\to\yb)=
\phi(\beta\{J_{\xb}-J_{\yb}+E(x_1-y_1)\})
+\phi(\beta\{J_{\xb}-J_{\yb}+E(y_1-x_1)\}),
\label{e:ctpJE}
\end{equation}
with $J_{\xb}=J$ if $|\xb|=1$ and $J_{\xb}=0$ otherwise.

Let us denote the unit vectors as $\eb_1=(1,0)$ and $\eb_2=(0,1)$.
We introduce $\calU=\{\eb_1,-\eb_1,\eb_2,-\eb_2\}$, which is the set of sites neighboring to the origin $\ob$.
Now unless $\xb$ or $\yb$ is in $\calU$, we have $\tilde{c}(\xb\to\yb)=t\equiv\phi(\beta E)+\phi(-\beta E)$ if $\xb-\yb=\pm\eb_1$, and  $\tilde{c}(\xb\to\yb)=s\equiv2\phi(0)$  if $\xb-\yb=\pm\eb_2$.
The effective hopping rate $\tilde{c}(\xb\to\yb)$ become irregular only around the origin.
More precisely, for $\xb\in\calU$, we have  $\tilde{c}(\xb\to\xb\pm\eb_1)=\phi(\beta\{J+E\})+\phi(\beta\{J-E\})$,  $\tilde{c}(\xb\pm\eb_1\to\xb)=\phi(-\beta\{J+E\})+\phi(-\beta\{J-E\})$,  $\tilde{c}(\xb\to\xb\pm\eb_2)=2\phi(\beta\,J)$,  and  $\tilde{c}(\xb\pm\eb_2\to\xb)=2\phi(-\beta\,J)$.

By explicitly separating the equilibrium behavior in $G(\xb)$, we write
\begin{equation}
G(\xb)=G_0\,e^{\beta\,J_{\xb}}(1+\psi_{\xb}),
\label{e:Gbp}
\end{equation}
where $\psi_{\xb}$ is the nonequilibrium correction which vanishes if $E=0$, and $G_0$ is the normalization constant.
By substituting \eqref{e:Gbp} into \eqref{e:ctG}, we get
\begin{equation}
\sumtwo{\yb\in\La}{|\xb-\yb|=1,\ \yb\ne\ob}
\{-t(\xb\to\yb)(1+\psi_{\xb})+t(\yb\to\xb)(1+\psi_{\yb})\}=0,
\label{e:txyp}
\end{equation}
where the hopping rates are written as
\begin{equation}
t(\xb\to\yb)=t^{(0)}(\xb\to\yb)+\Delta t(\xb\to\yb),
\label{e:ttdelta}
\end{equation}
with $t^{(0)}(\xb\to\xb\pm\eb_1)=t$ and  $t^{(0)}(\xb\to\xb\pm\eb_2)=s$ for any $\xb\in\La$.
As for the correction near the origin, we have, for $x\in\calU$,
\begin{eqnarray}
\Delta t(\xb\to\xb\pm\eb_1)
&=&
e^{\beta J}\{\phi(\beta\{J+E\})+\phi(\beta\{J-E\})\}-\{\phi(\beta E)+\phi(-\beta E)\},
\ret
\Delta t(\xb\pm\eb_1\to\xb)
&=&
\phi(-\beta\{J+E\})+\phi(-\beta\{J-E\})-\{\phi(\beta E)+\phi(-\beta E)\},
\ret
\Delta t(\xb\to\xb\pm\eb_2)
&=&
2e^{\beta J}\phi(\beta J)-2\phi(0),
\ret
\Delta t(\xb\pm\eb_2\to\xb)
&=&
2\phi(-\beta J)-2\phi(0),
\label{e:dt1234}
\end{eqnarray}
and $\Delta t(\xb\to\yb)=0$ for other combinations.

Note that  $\Delta t(\xb\to\yb)\to0$  as $\beta\to0$.
We thus neglect the cross terms $\Delta t\,\psi$ in \eqref{e:txyp} to get the lowest order contribution for $\psi_{\xb}$ in the limit of high temperature.
The result is
\begin{equation}
\sumtwo{\yb\in\La}{|\xb-\yb|=1,\ \yb\ne\ob}
\{-t^{(0)}(\xb\to\yb)\psi_{\xb}+t^{(0)}(\yb\to\xb)\psi_{\yb}\}=-Q_{\xb},
\label{e:PL}
\end{equation}
with
\begin{equation}
Q_{\xb}=\sumtwo{\yb\in\La}{|\xb-\yb|=1,\ \yb\ne\ob}
\{\Delta t(\yb\to\xb)-\Delta t(\xb\to\yb)\}.
\label{e:Qtt}
\end{equation}
Note that \eqref{e:PL} is nothing but the Laplace-Poisson equation on the two-dimensional lattice with charge distribution $Q_{\xb}$.

\begin{figure}
\centerline{\epsfig{file=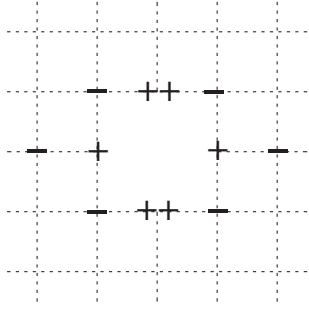,width=4cm}}
\caption[dummy]{
The nonequilibrium correction $\psi_{\xb}$ is the solution of the Laplace-Poisson equation with the charge distribution $Q_{\xb}$.
The symbols $+$, $-$ represents the charge $Q_0$ and $-Q_0$.
Note that this is a collection of quadrupoles.
}
\label{f:Qdis}
\end{figure}

The charge  $Q_{\xb}$ can be calculated using the explicit forms \eqref{e:dt1234} of $\Delta t(\xb\to\yb)$ and the (local detailed balance) condition $\phi(h)=e^{-h}\phi(-h)$, which is \eqref{e:phicond}.
After a little calculation one finds that $Q_{\pm\eb_1}=Q_0$, $Q_{\pm2\eb_1}=-Q_0$, $Q_{\pm\eb_2}=2Q_0$,  $Q_{\pm\eb_1\pm\eb_2}=Q_{\pm\eb_1\mp\eb_2}=-Q_0$, and $Q_{\xb}=0$ otherwise, where
\begin{eqnarray}
Q_0&=&
\phi(-\beta\{J-E\})+\phi(-\beta\{J+E\})-e^{\beta J}
\{\phi(\beta\{J+E\})+\phi(\beta\{J-E\})\}
\ret
&=&
(1-e^{\beta E})\phi(-\beta\{J-E\})+(1-e^{-\beta E})\phi(-\beta\{J+E\}).
\label{e:Q0}
\end{eqnarray}
As can be seen from Fig.~\ref{f:Qdis}, the charge distribution $Q_{\xb}$ is a collection of  quadrupoles.

Clearly one has $Q_0=0$ if $E=0$.
By expanding in $\beta$, we find that the leading behavior of $Q_0$ is
\begin{equation}
Q_0=\frac{\beta^3}{2}E^2J(4\phi''(0)-\phi(0)),
\label{e:Q0exp}
\end{equation}
where we assumed that $\phi(h)$ is differentiable, and used the fact that \eqref{e:phicond} implies $\phi'(0)=-\phi(0)/2$.

Thus the nonequilibrium correction $\psi_{\xb}$ can be obtained by solving the Laplace-Poisson equation \eqref{e:PL} with the charge distribution $Q_{\xb}$ of Fig.~\ref{f:Qdis}.
Note also that, to get the lowest order behavior in $\beta$, we can set $t=s=2\phi(0)$.
Since the charge distribution is a collection of quadrupoles, the long range behavior of $\psi_{\xb}$ is given by
\begin{equation}
\psi_{\xb}\approx\frac{(x_1)^2-(x_2)^2}{|\xb|^4},
\label{e:psiLRC}
\end{equation}
which is the famous $1/r^d$ long range correlation.
See \cite{DorfmanKirkpatrickSengers94,Tasaki04a,Sasa04}.
But we are here interested only in the very short range behavior, i.e., the values of $\psi_{\xb}$ for $\xb\in\calU$.
Clearly for $\xb\in\calU$, the solution of \eqref{e:PL} is given by $\psi_{\xb}\simeq{\rm const.}Q_0/t$ with  positive constants.
It is not an easy task to determine the constants, but they are certainly numerical constants (which do not depend on any of the model parameters) of order 1.

We therefore conclude that the leading correction in the nearest neighbor correlation function is given as
\begin{equation}
\sum_{\yb;|\yb-\xb|=1}\bkt{\etax\etay}\simeq
\sum_{\yb;|\yb-\xb|=1}\bkt{\etax\etay}^{\rm eq}+
C(4\frac{\phi''(0)}{\phi(0)}-1)\beta^3E^2J\rho^2,
\label{e:correcorr}
\end{equation}
where $\bkt{\cdots}^{\rm eq}$ denotes the expectation in the corresponding equilibrium, and $C$ is a positive numerical constant which does not depend on any of the model parameters (and may be computed numerically if necessary).

By substituting \eqref{e:correcorr} into \eqref{e:muexp} and further expanding, we finally get
\begin{equation}
\mu(T,E;\rho)\simeq\mu(T,0;\rho)-C(4\frac{\phi''(0)}{\phi(0)}-1)\beta^3E^2J^2\rho.
\label{e:muexp2}
\end{equation}
By substituting this into \eqref{e:Euf}, we get
\begin{equation}
f(T,E;\rho)\simeq f(T,0;\rho)-\frac{C}{2}(4\frac{\phi''(0)}{\phi(0)}-1)\beta^3E^2J^2\rho^2,
\label{e:fexp}
\end{equation}
which is the concrete form of SST free energy for the driven lattice gas.

One of our postulates of SST stated in section~\ref{s:ftn} is that the free energy $f(T,E;\rho)$ is a concave function of $E$, i.e., $\partial^2f(T,E;\rho)/\partial E^2\le0$.
It is apparent from \eqref{e:fexp} that the validity of this postulate depends crucially on the choice of the function $\phi(h)$.
As for the most standard heat-bath rule with $\phi(h)=(1+e^h)^{-1}$, we have $\phi''(0)=0$, which means that $f(T,E;\rho)$ does not satisfy the conjectured concavity.

\section{The ``second law'' for a general time-dependent Markov process}
\label{s:M2nd}
Let us state and prove the ``second law'' in a Markov process
which was used in sections~\ref{s:sp3} and \ref{s:mwp}.

\subsection{Main results}
\label{s:M2nds}
Let $\calS$ be a finite state space, and $c^{(\alpha)}(s\to s')$
be arbitrary transition rates which ensure ergodicity.
(See section~\ref{s:sp1} for the general setup.)
By $p^{(\alpha)}_\infty(s)$ we denote the stationary distribution for the
time-independent Markov process with constant transition rates
$c^{(\alpha)}(s\to s')$.
More precisely it satisfies
\begin{equation}
\sumtwo{s'\in\calS}{(s'\ne s)}
\{-c^{(\alpha)}(s\to s')\,p^{(\alpha)}_\infty(s)
+c^{(\alpha)}(s'\to s)\,p^{(\alpha)}_\infty(s')\}=0,
\label{e:stda}
\end{equation}
for any $s\in\calS$.
(See \eqref{e:std}.)
Let us define
\begin{equation}
\varphi^{(\alpha)}(s)=-\log p^{(\alpha)}_\infty(s).
\label{e:phia}
\end{equation}

Fix an arbitrary differentiable function $\alpha(t)$ with
$0\le t\le t_{\rm f}$, and consider a time-dependent Markov 
process with the transition rates $c^{(\alpha(t))}(s\to s')$.
The  probability distribution $p_t(s)$
at time $t$  satisfies the master equation
\begin{equation}
\frac{d}{dt}p_t(s)=\sumtwo{s'\in\calS}{(s'\ne s)}
\{-c^{(\alpha(t))}(s\to s')\,p_t(s)+c^{(\alpha(t))}(s'\to s)\,p_t(s')\},
\label{e:tdmea}
\end{equation}
for any $s\in\calS$.
We set the initial condition as
\begin{equation}
p_0(s)=p^{(\alpha(0))}_\infty(s).
\label{e:inita}
\end{equation}
We denote the average over $p_t(s)$ as
$\bkt{g(s)}_t=\sum_{s\in\calS}g(s)\,p_t(s)$.

Then the well known ``second law'' is
\begin{equation}
\int_0^{t_{\rm f}}dt\,\frac{d\alpha(t)}{dt}
\bkt{\left.\frac{d}{d\alpha}\varphi^{(\alpha)}(s)\right|_{\alpha=\alpha(t)}}_t
\ge0.
\label{e:M2nd}
\end{equation}

\subsection{Proof}
\label{s:M2ndp}
For completeness we prove the ``second law'' \eqref{e:M2nd}.
Although the standard proof makes use of relative entropy
(see, for example, section~2.9 of \cite{CoverThomas91}),
we here present a direct proof.

Only in the present proof, we make use of the probabilistic language
and denote by $\hat{s}(t)$ the random variables for the present Markov
process.
Thus $p_t(s)={\rm Prob}[\hat{s}(t)=s]$.

Fix a positive integer $N$, and let $\Delta t=t_{\rm f}/N$.
Let\footnote{
${\rm Prob}[A|B]={\rm Prob}[A\ \mbox{and}\ B]/{\rm Prob}[B]$
is the conditional probability of an event $A$ given that an event
$B$ is true.
}
\begin{equation}
T^t(s,s')={\rm Prob}[\hat{s}(t)=s\,|\,\hat{s}(t-\Delta t)=s'].
\label{e:Ttss}
\end{equation}
In physicists' language, $T^t(s,s')$ is the time-evolution kernel from
$t-\Delta t$ to $t$, since it satisfies
\begin{equation}
p_t(s)=\sum_{s'\in\calS}T^t(s,s')\,p_{t-\Delta t}(s'),
\label{e:pTp}
\end{equation}
where $p_t(s)$ is the solution of the master equation \eqref{e:tdmea} with
an arbitrary initial condition.
Note that
\begin{equation}
\sum_{s'\in\calS}T^t(s,s')\,p^{(\alpha(t))}_\infty(s')
=
p^{(\alpha(t))}_\infty(s)+O((\Delta t)^2).
\label{e:Ttpa}
\end{equation}

For $n=0,1,\ldots,N$, let $t_n=n\,\Delta t$.
Define a random quantity
\begin{equation}
\hat{Q}=\prod_{n=0}^{N-1}
\frac{
p^{(\alpha(t_{n+1}))}_\infty(\hat{s}(t_n))
}{
p^{(\alpha(t_{n}))}_\infty(\hat{s}(t_n))
}.
\label{e:Qpp}
\end{equation}
By using \eqref{e:Ttpa} repeatedly, we find that
\begin{eqnarray}
\bkt{\hat{Q}}&=&
\sum_{s_0,s_1,\ldots,s_{N-1}\in\calS}
\cbk{\prod_{n=1}^{N-1}T^{t_n}(s_n,s_{n-1})}
p^{\alpha(t_0)}_\infty(s_0)
\prod_{n=0}^{N-1}
\frac{
p^{(\alpha(t_{n+1}))}_\infty(s_n)
}{
p^{(\alpha(t_{n}))}_\infty(s_n)
}
\ret
&=& 
\sum_{s_1,\ldots,s_{N-1}\in\calS}
\cbk{\prod_{n=2}^{N-1}T^{t_n}(s_n,s_{n-1})}
p^{\alpha(t_1)}_\infty(s_1)
\prod_{n=1}^{N-1}
\frac{
p^{(\alpha(t_{n+1}))}_\infty(s_n)
}{
p^{(\alpha(t_{n}))}_\infty(s_n)
}+O((\Delta t)^2)
\ret
&=&\cdots
\ret
&=&
\sum_{s_{N-1}\in\calS}p^{\alpha(t_N)}_\infty(s_{N-1})
+N\,O((\Delta t)^2)
\ret
&=&
1+O(\Delta t).
\label{e:Q=1}
\end{eqnarray}
Then since $\log x\le x-1$, we have
$\bkt{\log\hat{Q}}\le\bkt{\hat{Q}-1}=O(\Delta t)$.
But from \eqref{e:Qpp} and \eqref{e:phia}, we have
\begin{eqnarray}
\bkt{\log\hat{Q}}&=&
-\sum_{n=0}^{N-1}\bkt{
\varphi^{(\alpha(t_{n+1}))}(\hat{s}(t_n))
-\varphi^{(\alpha(t_{n}))}(\hat{s}(t_n))
}
\ret
&=&
-\sum_{n=0}^{N-1}\bkt{
\varphi^{(\alpha(t_{n+1}))}(s)
-\varphi^{(\alpha(t_{n}))}(s)
}_{t_n}
\ret
&=&
-\sum_{n=0}^{N-1}\cbk{
\Delta t\,\frac{d\alpha(t_n)}{dt}
\bkt{\left.\frac{d}{d\alpha}
\varphi^{(\alpha)}(s)\right|_{\alpha=\alpha(t_n)}}_{t_n}
+O((\Delta t)^2)
}.
\label{e:logQ}
\end{eqnarray}
By letting $N\to\infty$, we get the desired \eqref{e:M2nd}.

\bigskip\bigskip

It is a pleasure to thank Yoshi Oono for
introducing us to the present subject and
for various indispensable advices and suggestions.
We wish to thank
Kumiko Hayashi for related collaboration with one of us (Sasa), Hisao Hayakawa for valuable comments about boundary effects, 
Akira Yoshimori for useful discussions related to fluctuation-response relations,
and Kyozi Kawasaki for valuable comments on the manuscript.
We also wish to thank
Ichiro Arakawa,
Hiroshi Ezawa,
Makoto Fushiki,
Nobuyasu Ito,
Christopher Jarzynski,
Arisato Kawabata,
Kim Hyeon-Deuk,
Tohru Koma,
Teruhisa Komatsu,
Joel Lebowitz,
Raphael Lefevere,
Hiroshi Mano,
Tsuyoshi Mizuguchi,
Satoru Nasuno,
Ooshida Takeshi,
Kazuya Saito,
Tomohiro Sasamoto,
Masaki Sano,
Ken Sekimoto,
Akira Shimizu,
Herbert Spohn,
Masaru Sugiyama,
and
Hirofumi  Wada
for various useful discussions and comments during the years through which  we have developed the  present work.

The work of one of us (Sasa)
was supported by grants from the Ministry of Education, Science, 
Sports and Culture of Japan, No. 16540337.



\begin{thebibliography}{99}

\bibitem{Wightman79}
A. S. Wightman, 
{\em Convexity and the Notion of Equilibrium State in 
Thermodynamics and Statistical Mechanics}\/,
Introduction to {\em Convexity in the Theory of Lattice Gases}\/ by
Robert B. Israel
(Princeton University Press, 1979)

\bibitem{OonoPaniconi98}
Y. Oono, and M. Paniconi,
{\em Steady state thermodynamics}\/,
Prog. Theor. Phys. Suppl. {\bf 130}, 29--44 (1998).

\bibitem{LiebYngvason99}
E. H. Lieb and J. Yngvason,
{\em The physics and mathematics of the second law of thermodynamics}\/,
Phys. Rep. {\bf 310}, 1--96 (1999).

\bibitem{Gallavotti99}
G. Gallavott,
{\em Statistical mechanics: a short treatise}\/
(Springer, 1999).




\bibitem{Einstein05a}
A. Einstein,
{\em A new determination of molecular dimensions}\/
(University of Zurich dissertation, 1905).

\bibitem{Einstein05b}
A. Einstein,
{\em On the movement of small particles suspended in stationary liquids required by molecular-kinetic theory of heat}\/,
Annalen der Physik {\bf 17}, 549--560 (1905).

\bibitem{Onsager31a}
L. Onsager,
{\em Reciprocal relations in irreversible processes. I}\/,
Phys. Rev. {\bf 37}, 405--426 (1931).

\bibitem{Onsager31b}
L. Onsager,
{\em Reciprocal relations in irreversible processes. II}\/,
Phys. Rev. {\bf 38}, 2265--2279 (1931).


\bibitem{Nyquist28}
H. Nyquist,
{\em Thermal agitation of electric charge in conductors}\/,
Phys. Rev. {\bf 32}, 110--113 (1928).

\bibitem{KuboTodaHashitsume85}
R. Kubo, M. Toda, and N. Hashitsume,
{\em Statistical Physics II}\/ (Springer, 1985).

\bibitem{Nakano93}
H. Nakano,
{\em Linear response theory --- a historical perspective}\/,
Int. J. Mod. Phys.  B {\bf 7}, 2397--2467 (1993).


\bibitem{Prigogine67}
I. Prigogine,
{\em Introduction to thermodynamics of irreversible processes}\/
(Interscience, 1967).

\bibitem{Hashitsume52}
N. Hashitsume,
{\em A statistical theory of linear dissipative systems}\/,
Prog. Theor. Phys. {\bf 8}, 461--478 (1952).

\bibitem{OnsagerMachlup53} 
L. Onsager and S. Machlup, 
{\em Fluctuations and irreversible processes}\/,
Phys. Rev. {\bf 91},  1505--1512  (1953).

\bibitem{Hashitsume56}
N. Hashitsume,
{\em A statistical theory of linear dissipative systems, II}\/,
Prog. Theor. Phys. {\bf 15}, 369--413 (1956).

\bibitem{Ono61}
S. Ono, {\em Variational principles in thermodynamics and 
statistical mechanics of irreversible processes}\/,
Adv. Chem. Phys. {\bf 3},  267-321 (1961).

\bibitem{GrrotMazur62}
S. R. de Groot and P. Mazur, 
{\em Non-equilibrium thermodynamics}\/
(North-Holland, 1962). 


\bibitem{YamadaKawasaki68}
T. Yamada and K. Kawasaki, 
{\em Nonlinear Effects in the shear viscosity of critical mixture}\/,
Prog. Theor. Phys. {\bf 38}, 1031--1051, (1967).

\bibitem{KawasakiGunton73}
K. Kawasaki and J. D. Gunton, 
{\em Theory of nonlinear transport processes:
nonlinear shear viscosity and normal stress effects}\/, 
Phys. Rev. A {\bf 8}, 2048--2064 (1973). 

\bibitem{DorfmanKirkpatrickSengers94}
J.R. Dorfman, T. R. Kirkpatrick, and J. V. Sengers, 
{\em Generic long-range correlations in molecular fluids}\/,
Ann. Rev. Phys. Chem.
{\bf 45}, 213--239 (1994). 

\bibitem{Zwanzig61}
R. Zwanzig, 
{\em Memory Effects in irreversible thermodynamics}\/, 
Phys. Rev. {\bf 124}, 983--992, (1961).

\bibitem{Mori65}
H. Mori, 
{\em Transport, collective motion, and Brownian motion}\/,
Prog. Theor. Phys. {\bf 33}, 423--455, (1965).

\bibitem{McLennan90}
J. A. Mclennan,
{\em Introduction to nonequilibrium statistical mechanics}
(Prentice Hall, 1990). 

\bibitem{Zubarev74}
D. N. Zubarev, 
{\em Nonequilibrium statistical thermodynamics}
(Consultants Bureau, 1974). 

\bibitem{Crooks00} 
G. E. Crooks, 
{\em Path-ensemble averages in systems driven far from equilibrium}\/,
Phys. Rev. E {\bf 63}, 2361--2366 (2000).

\bibitem{WadaSasa03}
H. Wada and S. Sasa, 
{\em Anomalous pressure in fluctuating shear flow}\/,
Phys. Rev. E {\bf 67}, 065302(R)  (2003).

\bibitem{SpohnLebowitz78}
H. Spohn and J. L. Lebowitz,
{\em Irreversible thermodynamics for quantum systems weakly coupled to thermal reservoirs}\/,
Adv. Chem. Phys. {\bf 38}, 109--142 (1978).

\bibitem{JaksicPillet02}
V. Jak\v{s}i\'{c} and C.-A. Pillet,
{\em Non-equilibrium steady states of finite quantum systems coupled to thermal reservoirs}\/,
Comm. Math. Phys. {\bf 226}, 131--162 (2002).

\bibitem{HuLiZhao00}
B. Hu, B. Li, and H. Zhao,
{\em heat conduction in one-dimensional nonintegrable systems}\/,
Phys. Rev. E {\bf 61}, 3828--3831 (2000).

\bibitem{EckmannPilletReyBelle99}
J.-P. Eckmann, C.-A. Pillet, and L. Rey-Bellet,
{\em Non-equilibrium statistical mechanics of anharmonic chains coupled to two heat baths at different temperatures}\/,
Comm. Math. Phys. {\bf 201}, 657--697 (1999).

\bibitem{EckmannHairer00}
J.-P. Eckmann and M. Hairer,
{\em Non-equilibrium statistical mechanics of strongly anharmonic chains of oscillators}\/,
Comm. Math. Phys. {\bf 219}, 523--565 (2001).

\bibitem{LefevereSchenkel04}
R. Lefevere and S. Schenkel,
{\em Perturbative analysis of anharmonic chains of oscillators out of equilibrium}\/,
J. Stat. Phys. {\bf 115}, 1389--1421 (2004).



\bibitem{ChapmanCowling39}
S. Chapman and T. G.  Cowling, {\em The mathematical theory of non-uniform gases}\/
(Cambridge University Press, 1939).


\bibitem{Lanford75}
O. E. Lanford, {\em Time evolution of large classical systems}\/,
ed. J. Moser, Lecture Notes in Physics {\bf 38}, 1--111 (1975). 

\bibitem{KimHayakawa03a}
Kim, H.-D. and H. Hayakawa, 
{\em Kinetic theory of a dilute gas system under steady heat conduction}\/,
J. Phys. Soc. Jpn. {\bf 72}, 1904-1916 (2003);
{\bf 73}, 1609 (2003).

\bibitem{KimHayakawa03b}
Kim, H.-D. and H. Hayakawa, 
{\em Test of information theory on the Boltzmann equation}\/,
J. Phys. Soc. Jpn. {\bf 72}, 2473-2476 (2003).

\bibitem{HohenbergHalperin77}
P. C. Hohenberg and B. I. Halperin, 
{\em Theory of dynamic critical phenomena}\/,
Rev. Mod. Phys. {\bf 49}, 435--479 (1977).

\bibitem{OnukiKawasaki79}
A. Onuki and K. Kawasaki, 
{\em Nonequilibrium steady state of critical fluids under shear flow:
a renormalization group approach}\/, 
Annals of Physics {\bf 121}, 456--528 (1979).



\bibitem{Spohn91}
H. Spohn, {\em Large scale dynamics of interacting particles}\/ (Springer, 1991).

\bibitem{Spohn83}
H. Spohn,
{\em  Long range correlations for stochastic lattice gases in a nonequilibrium steady-state}\/,
 J. Phys. {\bf A 16}, 4275--4291 (1983).

\bibitem{PraehoferSpohn02}
M. Praehofer and H. Spohn,
{\em Current fluctuations for the totally asymmetric simple exclusion process}\/,
in  ``In and out of equilibrium'', ed. V. Sidoravicius, Progress in Probability Vol. 51, 185--204 
(Birkhauser, 2002), cond-mat/0101200.



\bibitem{KatzLebowitzSpohn84} 
S. Katz, J. L. Lebowitz, and H. Spohn, 
{\em Nonequilibrium steady states of stochastic lattice gas models of fast ionic conductors}\/,
J. Stat. Phys. {\bf 34}, 497--537 (1984).


\bibitem{SchimttmannZia95}
B. Schimttmann and R. K. P. Zia, 
{\em Statistical mechanics of driven diffusive systems}\/
(Academic Press, 1995).

\bibitem{LefevereTasaki04}
R. Lefevere and H. Tasaki,
{\em High-temperature expansion for nonequilibrium steady states in
driven lattice gases}\/,
Phys. Rev. Lett.~{\bf 94}, 200601 (2005).


\bibitem{EynkLebowitzSpohn96}
G. L. Eyink, J. L. Lebowitz, and H. Spohn, 
{\em Hydrodynamics and fluctuations outside of local equilibrium: driven diffusive systems}\/,
J. Stat. Phys. {\bf 83},
385--472 (1996).

\bibitem{AlexanerEyink98}
F. J. Alexander and G. L. Eyink,
{\em Shape-dependent thermodynamics and nonlocal hydrodynamics in a non-Gibbsian steady state of a drift-diffusion system}\/,
Phys. Rev. E {\bf 57} R6229--R6232 (1998).

\bibitem{HayashiSasa03}
K. Hayashi and S. Sasa,
{\em Thermodynamic relations in a driven lattice gas: numerical experiments}\/,
Phy. Rev. E {\bf 68}, 035104(R) (2003).


\bibitem{EvansCohenMorris93}
D. J. Evans, E. G. D. Cohen, and G. P. Morrriss,
{\em Probability of second law violations in steady flow}\/,
Phys. Rev. Lett. {\bf 71}, 2401--2404 (1993).

\bibitem{GallavottCohen95}
G. Gallavott and E. G. D. Cohen,
{\em Dynamical ensemble in stationary states}\/,
J. Stat. Phys. {\bf 80}, 931--970 (1995).

\bibitem{Kurchan98}
J. Kurchan, 
{\em Fluctuation theorem for stochastic dynamics}\/, 
J. Phys. A: Math. Gen. {\bf 31}, 3719--3729 (1998). 

\bibitem{LebowitzSpohn99}
J. L. Lebowitz and H. Spohn,
{\em A Gallavott-Cohen-type symmetry in the large deviation functional for stochastic dynamics}\/,
J. Stat. Phys. {\bf 95}, 333--365 (1999).

\bibitem{Maes99}
C. Maes,
{\em The fluctuation theorem as a Gibbs property}\/,
J. Stat. Phys. {\bf 95}, 367--392 (1999).


\bibitem{LebowitzMaesSpeer90}
J. L. Lebowitz, C. Maes, and E. R. Speer,
{\em Statistical mechanics of probabilistic cellular automata}\/,
J. Stat. Phys. {\bf 59}, 117--170 (1990).
 


\bibitem{DerridaLebowitzSpeer01}
 B. Derrida, J. L. Lebowitz, and E. R. Speer,
{\em Free energy functional for nonequilibrium systems: an exactly solvable case}\/,
Phys. Rev. Lett.  {\bf 87}, 150601 (2001).
 
\bibitem{DerridaLebowitzSpeer02a}
 B. Derrida, J. L. Lebowitz, and E. R. Speer,
{\em Large deviation of the density profile in the steady state of the open symmetric simple exclusion process}\/,
J. Stat. Phys. {\bf 107}, 599--634 (2002).


\bibitem{DerridaLebowitzSpeer02b}
 B. Derrida, J. L. Lebowitz, and E. R. Speer,
{\em Exact free energy functional for a driven diffusive open stationary nonequilibrium system}\/,
 Phys. Rev. Lett.  {\bf 89}, 030601 (2002)
 
\bibitem{DerridaLebowitzSpeer03}
B. Derrida, J. L. Lebowitz, and E. R. Speer,
{\em Exact large deviation functional of a stationary open driven diffusive system: the asymmetric exclusion process}\/,
J. Stat. Phys. {\bf 110}, 775--810 (2003).


\bibitem{BodineauDerrida04}
T. Bodineau and B. Derrida,
{\em Current fluctuations in nonequilibrium diffusive systems: an additivity principle}\/,
Phys. Rev. Lett.  {\bf 92}, 180601 (2004)

\bibitem{Bertinietal01} 
L. Bertini, A. De Sole, D. Gabrielli, G. Jona-Lasinio and C. Landim
{\em Fluctuations in stationary nonequilibrium states 
of irreversible processes}\/,
Phys. Rev. Lett. {\bf 87},  040601 (2001). 

\bibitem{Bertinietal02} 
L. Bertini, A. De Sole, D. Gabrielli, G. Jona-Lasinio and C. Landim,
{\em Macroscopic fluctuation theory for stationary non-equilibrium states}\/,
J. Stat. Phys. {\bf 107},  635--675 (2002). 






\bibitem{Landauer75}
R. Landauer,
{\em Inadequacy of entropy and entropy derivatives in characterizing the steady state}\/,
Phys. Rev. A {\bf 12}, 636--638 (1975).

\bibitem{Landauer94}
R. Landauer,
{\em Statistical physics of machinery: forgotten middle-ground}\/,
Physica A {\bf 194}, 551--562 (1993).


\bibitem{JouCasasVazquezLebon88}
D. Jou, J. Casas-V\'{a}zquez, and G. Lebon,
{\em Extended irreversible thermodynamics}\/,
Rep. Prog. Phys. {\bf 51}, 1105--1179 (1988).

\bibitem{JouCasasVazquezLebon01}
D. Jou, J. Casas-V\'{a}zquez, and G. Lebon,
{\em Extended irreversible thermodynamics}\/
(Springer, 2001).


\bibitem{MullerRuggeri98}
I. M\"{u}ller and T. Ruggeri,
{\em Rational extended thermodynamics}\/
(Springer, 1998).

\bibitem{Keizer87}
J. Keizer,
{\em Statistical thermodynamics of nonequilibrium processes}\/,
especially chapter 8
(Springer,1987).

\bibitem{Eu98}
B. C. Eu, 
{\em Nonequilibrium statistical mechanics}\/
(Kluwer, 1998).

\bibitem{Eu89a}
B. C. Eu,
{\em Nonequilibrium thermodynamic function for sheared fluids}\/,
Physica {\bf A 160}, 53--86 (1989).

\bibitem{Eu89b}
B. C. Eu,
{\em Shear-induced melting point depression}\/,
Physica {\bf A 160}, 87--97 (1989).

\bibitem{Eu88}
B. C. Eu,
{\em Entropy for irreversible processes}\/,
Chem. Phys. Lett. {\bf 143}, 65--70 (1988).

\bibitem{EvansHanley80a}
D. J. Evans and H. J. M. Hanley,
{\em Shear induced phase transitions in simple fluids}\/,
Phys. Lett. {\bf 79A}, 178--180 (1980).

\bibitem{EvansHanley80b}
D. J. Evans and H. J. M. Hanley,
{\em A thermodynamics of steady homogeneous flow}\/,
Phys. Lett. {\bf 80A}, 175--177 (1980).

\bibitem{DominguezJou95}
R. Domin\'{i}nguez and D. Jou,
{\em Thermodynamic pressure in nonequilibrium gases}\/,
Phys. Rev. E {\bf 51}, 158--163 (1995).

\bibitem{CasasVazquezjou94}
J. Casas-V\'{a}zquez and D. Jou,
{\em Nonequilibrium temperature versus local-equilibrium temperature}\/,
Phys. Rev. E {\bf 49}, 1040--1048 (1994).





\bibitem{Callen85}
H. B. Callen,
{\em Thermodynamics and an introduction to thermostatistics}\/
(Wiley, 1985).



\bibitem{LandauLifshitz80}
L. D. Landau and E. M. Lifshitz,
{\em Statistical mechanics, 3rd edition, part 1}\/
(Butterworth-Heinemann, 1980).


\bibitem{PuszWoronowicz78}
W. Pusz and S. L. Woronoicz,
{\em Passive states and KMS states for general quantum systems}\/,
Comm. Math. Phys. {\bf 58}, 273--290 (1978).

\bibitem{Lenard78}
A. Lenard, 
{\em Thermodynamic proof of the Gibbs formula for elementary quantum systems}\/,
J. Stat. Phys. {\bf 19}, 575--586 (1978).

\bibitem{Tasaki98}
H. Tasaki, 
{\em From quantum dynamics to the canonical distribution:  general picture and a rigorous example}\/,
Phys. Rev. Lett. {\bf 80}, 1373--1376 (1998).


\bibitem{Feller68}
W. Feller,
{\em An introduction to probability theory and its applications, vol. 1}\/
(Wiley, 1968).

\bibitem{Tasaki04a}
H. Tasaki,
{\em  A remark on the choice of stochastic transition rates in driven nonequilibrium systems}\/,
preprint, cond-mat/0407262.


\bibitem{Kubo81}
R. Kubo,
{\em H-theorems for Markoffian processes}\/,
in {\em Perspectives in statistical physics}\/, H. J. Ravech\'{e} ed., 101--110 (North-Holland, 1981).

\bibitem{Yosida68}
K. Yosida,
{\em Functional analysis}\/ (Springer, 1968).

\bibitem{Hayashi05}
K. Hayashi,
{\em Fluctuation-dissipation relations outside the linear response regime in a two-dimensional driven lattice gas along the direction transverse to the driving force}\/,
Phys. Rev. E {\bf 72}, 047105 (2005).

\bibitem{HatanoSasa01}
T. Hatano and S. Sasa, 
{\em Steady state thermodynamics of Langevin systems}\/,
Phys. Rev. Lett. {\bf 86}, 3463--3466 (2001).

\bibitem{Trepagnieretal04}
E.H. Trepagnier, C. Jarzynski, F. Ritort, G. E. Crooks,
C. J. Bustamante, and J. Liphardt, 
{\em Experimental test of Hatano and Sasa's nonequilibrium
steady-state equality}\/, 
Proc. Natl. Acad. Sci. USA {\bf 101}, 15038--15041 (2004).


\bibitem{Ugawa05}
H. Ugawa,
{\em Extended hydrodynamics from Enskog's equation: The bidimensional case}\/,
Physica A {\bf 354}, 77--87 (2005).

\bibitem{ButlerHarrowell02}
S. Butler and P. Harrowell,
{\em Factors determining crystal-liquid coexistence under shear}\/,
Nature {\bf 415}, 1008--1011 (2002).

\bibitem{NishinoHayakawa05}
T. H. Nishino, H. Hayakawa,
{\em Knudsen effect in a non-equilibrium gas}\/,
J. Phys. Soc. Jpn. {\bf 74}, 2655--2658 (2005), {\bf 74}, 3398 (2005). 
cond-mat/0506491.

\bibitem{MillsPhillips02}
C. T. Mills and  L. F. Phillips,
{\em Onsager heat transport at the aniline liquid-vapour interface}\/,
Chem. Phys. Lett. {\bf 366} 279--283 (2003).

\bibitem{MillsPhillips03}
C. T. Mills and  L. F. Phillips,
{\em Distillation of a cool liquid onto warmer surface}\/,
Chem. Phys. Lett. {\bf 372} 615--619 (2003).

\bibitem{Sasa04}
S. Sasa,
{\em Long range spatial correlation between two Brownian particles under external driving}\/,
 Physica {\bf D 205}, 233--241 (2005).


\bibitem{CoverThomas91}
T. A. Cover and J. A. Thomas,
{\em Elements of Information Theory}\/
(Wiley, 1991).




\end{thebibliography}
\end{document}